\documentclass[prd,aps,twocolumn,a4paper,showkeys,nofootinbib]{revtex4-1}

\usepackage{graphicx,psfrag}
\usepackage{mathrsfs}
\usepackage{amsmath,amsfonts,amssymb}
\usepackage{comment}

\usepackage{hyperref}
\usepackage{enumitem}

\usepackage{longtable}
\usepackage{multirow}

\usepackage[utf8]{inputenc}
\usepackage[T1]{fontenc}

\allowdisplaybreaks

\newcommand{\be}{\begin{equation}}
\newcommand{\ee}{\end{equation}}
\newcommand{\bea}{\begin{eqnarray}}
\newcommand{\eea}{\end{eqnarray}}
\newcommand{\bel}{\begin{align}}
\newcommand{\eel}{\end{align}}

\newcommand{\orcid}[1]{\href{https://orcid.org/#1}{
\includegraphics[width=10pt]{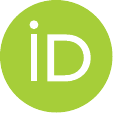}
}}

\def\l{\ell}
\def\lm{{\ell m}}
\def\p{\partial}
\def\F{{\cal F}}

\def\oe{{(o/e)}}

\def\GMc2{{\rm G M_{\odot} c^{-2}}}

\def\MM{\bar{F}}

\def\kt2{\kappa^\text{T}_2}

\def\sYlm{{}_{-2}Y_{\lm}}
\def\Hschw{{\hat{H}_{\rm Schw}}}

\usepackage{pifont} 

\newcommand{\TEOB}{\texttt{TEOBResumS}}
\newcommand{\DALI}{\texttt{TEOBResumS-Dal\'i}}
\newcommand{\GRA}{\texttt{GR-Athena++}}
\newcommand{\RWZ}{\texttt{RWZHyp}}

\usepackage{color}
\definecolor{cyan}{rgb}{0,0.9,0.9}
\definecolor{orange}{rgb}{0.9,0.5,0}
\definecolor{magenta}{rgb}{1,0,1}
\definecolor{purple}{rgb}{0.8,0.4,0.8}
\definecolor{gray}{rgb}{0.8242,0.8242,0.8242}

\begin{document}

\title{Scattering and dynamical capture of two black holes: 
synergies between numerical and analytical methods}

\author{Simone \surname{Albanesi}\orcid{0000-0001-7345-4415}$^{1,2,3,4}$}
\author{Alireza \surname{Rashti}\orcid{0000-0003-3558-7684}$^{5,6}$}
\author{Francesco \surname{Zappa}\orcid{0000-0002-0351-6518}$^{1}$}
\author{Rossella \surname{Gamba}\orcid{0000-0001-7239-0659}$^{1,5,7}$}
\author{William \surname{Cook}\orcid{0000-0003-2244-3462} $^{1}$}
\author{Boris \surname{Daszuta}\orcid{0000-0001-6091-2827} $^{1}$}
\author{Sebastiano \surname{Bernuzzi}\orcid{0000-0002-2334-0935}$^{1}$}
\author{Alessandro \surname{Nagar}$\orcid{0000-0001-7998-2673}^{3,8}$}
\author{David \surname{Radice}\orcid{0000-0001-6982-1008}$^{5,6,9}$}

\affiliation{${}^1$Theoretisch-Physikalisches Institut, Friedrich-Schiller-Universit{\"a}t Jena, 07743, Jena, Germany}
\affiliation{${}^2$Niels Bohr International Academy, Niels Bohr Institute, Blegdamsvej 17, 2100 Copenhagen, Denmark}
\affiliation{${}^3$INFN sezione di Torino, Torino, 10125, Italy}
\affiliation{${}^4$Dipartimento di Fisica, Università di Torino, Torino, 10125, Italy}
\affiliation{${}^5$Institute for Gravitation and the Cosmos, The Pennsylvania State University, University Park, PA 16802, USA}
\affiliation{${}^6$Department of Physics, The Pennsylvania State University, University Park, PA 16802, USA}
\affiliation{${}^7$Department of Physics, University of California, Berkeley, CA 94720, USA}

\affiliation{${}^8$Institut des Hautes Etudes Scientifiques, 35 Route de Chartres, Bures-sur-Yvette, 91440, France}
\affiliation{${}^9$Department of Astronomy and Astrophysics, The Pennsylvania State University, University Park, PA 16802, USA}


\begin{abstract}
We study initially unbound systems of two black holes using numerical relativity (NR) simulations performed with \GRA{}.
We focus on regions of the parameter space close to the transition from scatterings to dynamical captures,
considering equal mass and spin-aligned configurations, as well as unequal mass and nonspinning ones.
The numerical results are then used to validate the effective-one-body (EOB) model
\DALI{} for dynamical captures and scatterings. 
We find good agreement for the waveform phenomenologies, scattering angles, mismatches, and energetics
in the low energy regime ($E_0\lesssim 1.02\,M$). In particular, 
mismatches weighted with the zero-detuned, high-power noise spectral density of Advanced LIGO
are typically below or around the $1\%$ level, 
with only a few cases -- corresponding to spinning binaries -- slightly above the $3\%$ threshold,
thus suggesting the usability of \DALI{} for current data analysis of low-energy scatterings and dynamical captures.
We also discuss dynamical captures in the test-mass limit by solving numerically the Zerilli equation with the 
time domain code \RWZ{}. The latter analysis provides valuable insights into both the analytical 
noncircular corrections of the EOB waveform and the integration of NR Weyl scalars.
\end{abstract}


\maketitle

\section{Introduction} 
\label{sec:intro}

With the experimental advancements in gravitational wave (GW) observatories 
and the development of new data analysis techniques 
by the 
LIGO-Virgo-KAGRA (LVK) collaboration~\cite{LIGOScientific:2018mvr,LIGOScientific:2020ibl,LIGOScientific:2021djp}, 
GW astronomy has become a 
standard tool to explore the universe.
This prompts the necessity to provide accurate descriptions 
of the dynamics of astrophysical compact binaries, which are the sources of all the GW signals detected so far. 
This urgency is amplified by the high sensitivity that will be reached with next-generation detectors, 
such as Einstein Telescope~\cite{Punturo:2010zz,Maggiore:2019uih}, 
Cosmic Explorer~\cite{Reitze:2019iox} and LISA~\cite{LISA:2017pwj}. 
Although the majority of compact binaries circularize by the time their gravitational
signals enter the sensitivity band of our detectors~\cite{Peters:1963ux}, certain dense astrophysical environments, such as
globular clusters, can host compact two-body systems with elliptic or 
hyperbolic-like orbits~\cite{Samsing:2013kua,Rodriguez:2016kxx,Belczynski:2016obo,Samsing:2017xmd,
Rodriguez:2018pss,Mukherjee:2020hnm,DallAmico:2021umv,DallAmico:2023neb}. 
The event GW190521 might have been precisely generated by a black hole binary (BBH)
of this nature~\cite{Romero-Shaw:2020thy,Gayathri:2020coq,Gamba:2021gap}.

In this work we focus on initially unbound two-black-hole systems.
Configurations of this kind can become bound due to dissipative effects linked to the emission of GWs, 
ultimately leading to the merger of the two objects. Such systems are known as {\it dynamical captures}.
Crucially, depending on the energy and angular momentum considered,
the dynamically captured black holes can undergo many highly eccentric orbits before merging. 
On the other hand, if the GW emission due to the mutual interaction is not strong
enough to make the system bound, the two compact objects will simply scatter.
The transition from scatterings to captures is particularly delicate to study, since small variations
in the initial data can lead to very different phenomenologies~\cite{Gold:2012tk}. 
This aspect is particularly relevant if these systems are studied numerically, since
numerical relativity (NR) simulations are computationally expensive, and 
the outcome of the simulation cannot be easily predicted by just looking at the initial data.
It is therefore preferable to also have (semi-)analytical models that can 
quickly provide predictions on the phenomenologies of these systems.
Analytical methods that have been proved to be successful in describing
coalescing BBHs include 
the post-Newtonian (PN)~\cite{Einstein:1938yz,Einstein:1940mt,Blanchet:1989ki,Damour:2014jta,Schafer:2018kuf}
and post-Minkowskian (PM)~\cite{Bel:1981be,Damour:2016gwp,Bern:2019nnu,Damour:2020tta,Dlapa:2022lmu}
frameworks, whose synergies are essential to build effective-one-body (EOB) 
models~\cite{Buonanno:1998gg,Buonanno:2000ef,Damour:2000we,Buonanno:2006ui,Nagar:2006xv,Damour:2007xr,Damour:2008gu,Damour:2012ky,Nagar:2018zoe,Cotesta:2018fcv,Pompili:2023tna,Nagar:2023zxh}.
In this work we consider the last avatar of \DALI{}~\cite{Nagar:2024dzj}, 
a PN-based EOB model that has been already used 
to study both elliptic and hyperbolic 
systems, providing reliable results despite being numerically informed only on quasi-circular
NR simulations~\cite{Chiaramello:2020ehz,Nagar:2020xsk,Nagar:2021gss,Albanesi:2021rby,
Nagar:2021xnh,Placidi:2021rkh,Albanesi:2022ywx,Albanesi:2022xge,Bonino:2022hkj,
Hopper:2022rwo,Nagar:2023zxh,Andrade:2023trh,Nagar:2024dzj}. 
Since the model is based on PN-approximate results, numerical data for generic orbits 
are also needed for its validation. 
The interplay between analytical and numerical methods in studying BBHs is 
indeed well-established, and it is therefore not surprising that 
this synergy retains its importance also for initially unbound orbits, 
where its relevance may even be heightened.
We thus explore the hyperbolic regime close to the scattering-capture transition by performing
NR simulations with \GRA{}~\cite{Daszuta:2021ecf,Cook:2023bag}. 
After surveying the parameter space and highlighting
the challenges encountered, we compare the numerical and analytical 
results.
We also study the metric perturbations generated by dynamical captures in the test-mass limit,
that are governed by the Regge-Wheeler and Zerilli (RWZ)
equations with a test particle source term~\cite{Regge:1957td,Zerilli:1970se,Nagar:2005ea,Martel:2005ir}.
In particular, we solve the Zerilli equation
by means of the time-domain code \RWZ{}~\cite{Bernuzzi:2010ty,Bernuzzi:2011aj,Bernuzzi:2012ku}. 
These waveforms can be employed to assess the reliability of the EOB 
analytical prescriptions in a more controlled environment, as well as to 
elucidate the numerical integration procedures used in the NR case, 
highlighting any associated systematic issues.

The paper is structured as follows. In Sec.~\ref{sec:numrel_simulations} 
we present our new numerical simulations:
we discuss the initial data in Sec.~\ref{sbsec:initial_data}, the 
time evolution performed with \GRA{} in Sec.~\ref{sbsec:gra}, and the post-processing
in Sec.~\ref{sbsec:postproc}. The phenomenologies 
are examinated in Sec.~\ref{sbsec:phenom}, with a focus
on the transition from scattering to capture. Remnant
properties for merging configurations are investigated in Sec.~\ref{sbsec:remnants}.
Subsequently, in Sec.~\ref{sec:eob_nr} we introduce the EOB model \DALI{} and
the comparisons with the numerical results, discussing, in particular:
i) the phenomenologies across the parameter space in Sec.~\ref{sbsec:eobnr_parspace},
ii) the scattering angles in Sec.~\ref{sbsec:eobnr_q1_nospin_scat}, iii)
the mismatches in Sec.~\ref{sbsec:eobnr_mm}, and iv) the energetics in Sec.~\ref{sbsec:eobnr_energetics}. 
Finally, in Sec.~\ref{sec:testmass} we study the test mass limit, focusing 
on nonspinning configurations. 
The dynamics and the relevance of the radiation reaction 
in this regime are analyzed in Sec.~\ref{sbsec:testmass_dynamics}, while the RWZ
numerical perturbations are presented in Sec.~\ref{sbsec:rwz}. The latter waveforms 
are also employed to validate the EOB inspiral waveform in Sec.~\ref{sbsec:rwz_eob}. 
Throughout this paper we use geometrized unit, $G=c=1$.

\section{Numerical relativity simulations} 
\label{sec:numrel_simulations}
\begin{figure}[t]
  \centering 
    \includegraphics[width=0.48\textwidth]{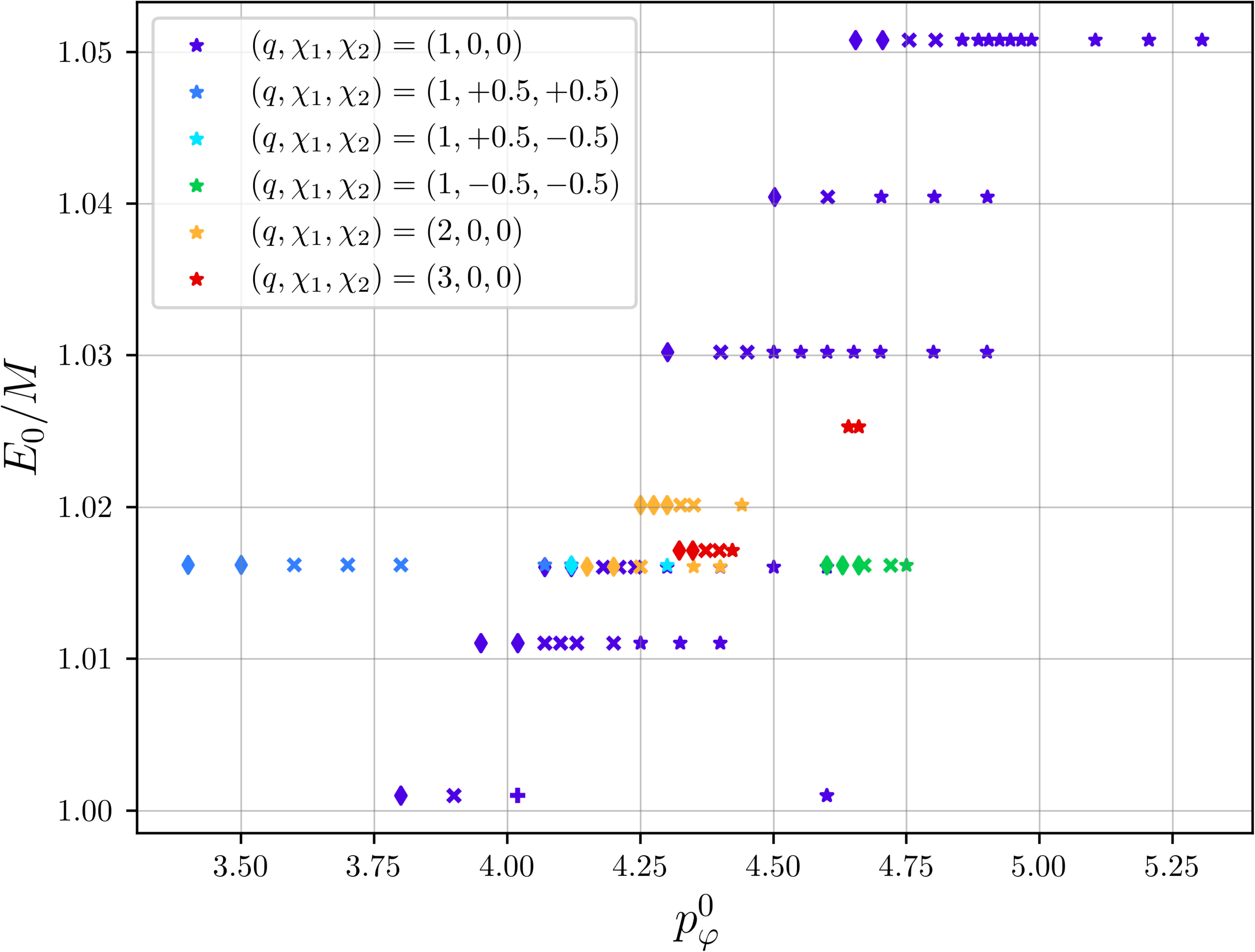}
    \caption{Comparable mass configurations considered in this work. 
    Different markers indicate different phenomenologies:
    stars for scatterings, diamonds for direct plunges, crosses for double encounters, and a plus
    for the only triple encounter of our
    dataset. Colors are related to $(q,\chi_1,\chi_2)$ combinations, see legend. 
    }
  \label{fig:parspace_NR}
\end{figure}

We begin our survey of the parameter space of initially unbound systems by presenting 
84 new NR simulations with varying mass ratios
and spins. 
More precisely, in Sec.~\ref{sbsec:initial_data} we discuss how we choose the initial data for these runs,
that are then time-evolved as described in Sec.~\ref{sbsec:gra}. 
While most of the configurations in this paper correspond to equal mass nonspinning systems, we also
simulate some cases with spin and higher mass-ratios. The complete list
of simulations considered in this work is shown in Fig.~\ref{fig:parspace_NR} and reported in
Tables~\ref{tab:runs_q1_nospin} and~\ref{tab:runs_q1_spin_q23}.
In Sec.~\ref{sbsec:phenom} we discuss their phenomenologies by means of guage-invariant quantities 
computed as explained in Sec.~\ref{sbsec:postproc}. 

\subsection{Initial data}
\label{sbsec:initial_data}
When simulating systems close to the scattering-capture transition, 
small variations in the initial data can lead to completely different outcomes; 
it is thus difficult to guess, a priori, the phenomenology.
To predict the evolution across the parameter space,
we employ the state-of-the-art EOB model 
\DALI{}~\cite{Nagar:2020xsk,Nagar:2021gss,Nagar:2021xnh,Nagar:2023zxh,Nagar:2024dzj}.
After fixing the initial ADM energy to a value greater than the 
total rest mass of the system ($E_0>M$), 
we systematically explore various initial orbital angular momenta $L_0$ with the EOB model, 
selecting specific sets that define configurations close to the threshold between scatterings and captures.
Note that the EOB prediction is not guaranteed to be fully
consistent with the outcome of the numerical simulation, 
since the model is PN-approximate. This is however not an issue, 
since we only use \DALI{} to understand which region of the parameter space should 
be explored with numerical simulations.
In practice, we fix $r_{\rm EOB}=100\,M$, and for each pair of values of energy
and angular momentum, we compute the corresponding Cartesian EOB 
initial positions and momenta. We then compute the Cartesian positions and momenta using
a 2PN EOB/ADM transformation~\cite{Buonanno:1998gg,Bini:2012ji}. 
These quantities
are then used to initialize the NR simulation. The energy and the angular momenta have the same values
in both ADM and EOB charts,
so in terms of $(E,L,r)$, the only quantity that actually needs the 2PN transformation is the initial separation $r$.
For the values of energy and angular momentum considered in this work, $r_{\rm EOB}=100\,M$ translates in
an ADM separation $D$ that is still close to $100\,M$ (precise values are reported in 
Tables~\ref{tab:runs_q1_nospin} and~\ref{tab:runs_q1_spin_q23}). 
While in practice these small variations do not drastically change the outcome of the numerical simulation, since
the GW emission at these large separations is indeed small, 
we still employ the transformation to compute $D$ to be as consistent as possible.  

The reliability of the EOB/ADM transformation
can be tested by comparing the initial ADM energy and angular momentum of the simulation
with the original values used in the EOB chart. As anticipated, they should be equal, but given 
the finite PN-accuracy of the transformation, this is not guaranteed.
As expected, the agreement tends to decrease for higher initial velocities
due to the PN approximation employed in the coordinate transformation.
For example, for an initial EOB energy $1.011\,M$, the relative difference with the 
corresponding ADM energy at the beginning of the numerical simulation is $3\cdot 10^{-5}$;
for initial energy $1.050\,M$ this relative difference grows to $8\cdot 10^{-4}$.

The Cartesian momenta and separation
are used together with the individual black hole spins to 
initialize a stand-alone version of the thorn 
\texttt{TwoPunctures}~\cite{Brandt:1997tf, Ansorg:2004ds,Daszuta:2021ecf}.
The total rest mass of the system is computed as $M=M_1+M_2$, where
$M_i$ are the ADM masses of the punctures as defined in Eq.~(83) of
Ref.~\cite{Ansorg:2004ds}; we choose $M_1>M_2$ and $M=1$ for all the configurations considered. 
The mass ratio is denoted as $q=M_1/M_2$.
We always consider systems whose spins are (anti-)aligned with the orbital angular momentum,
so that the only non vanishing component of the spin vectors is the $z$ one, 
simply denoted as $S_i$, or with its dimensionless 
counterpart, $\chi_i \equiv S_i/M_i^2$.
The total angular momentum of the system is given by
$J = L + S_1 + S_2$. It is convenient to also define the 
reduced angular momentum $j\equiv J/(\nu M^2)$ and the canonical orbital angular 
momentum $p_\varphi \equiv L/(\nu M^2)$, 
where $\nu = q/(q+1)^2$ is the symmetric mass ratio.
We often use the dimensionless energy defined by $\hat{E} \equiv E/M$. 

When reporting the NR initial data, we do not remove the contribution of the junk radiation.
We checked that the relative differences in energy and angular momentum
before and after junk radiation are below the $10^{-4}$ threshold, and they are 
thus negligible for the purposes of this work. 
The junk contributions are estimated from the waveform and the corresponding
integrated energy and angular momentum fluxes 
evaluated at $t=200\,M$, computed as detailed in Sec.~\ref{sbsbsec:postproc_waves} and~\ref{sbsbsec:postproc_energetics}.
Similar small values were also found for
scattering configurations with initial separation $D=100\,M$ in Ref.~\cite{Damour:2014afa},
where the waveform and the fluxes where computed with time domain integration.

\subsection{Numerical methods}
\label{sbsec:gra}
Given some initial data prescribed as discussed in the previous section, we perform
the time evolution by means of {\GRA}~\cite{Daszuta:2021ecf,Cook:2023bag}, that employs the 
Z4c formulation of the Einstein field equations~\cite{Bernuzzi:2009ex,Ruiz:2010qj,Weyhausen:2011cg,Hilditch:2012fp}. 
For these new runs, we employ the same moving puncture gauge conditions and gauge parameters 
as those used in Refs.~\cite{Daszuta:2021ecf,Andrade:2023trh}.
Further, similarly to the \GRA{} runs considered in Ref.~\cite{Andrade:2023trh}, we employ 
6th order finite-difference methods for the spatial derivatives, and a 8th order
Kreiss-Oliger operator, but with dissipation factor $\epsilon = 0.1$.
For the time-integration, we use a 4th order 
Runge-Kutta algorithm and a Courant-Friedrichs-Lewy factor of 0.25. 
The typical edge of the grid considered is $L=3072\,M$. 
Different choices have also been explored, without finding any substantial difference in the results.
We evolve only the $z>0$ portion of the space and we complete the rest of the grid exploiting bitant symmetry.
We use outflow boundary conditions.

\GRA{} utilizes an oct-tree structure, with the grid initially organized as a mesh divided into meshblocks. 
These meshblocks have the same number of grid points, but may differ in physical size. In the case of a cubic 
initial mesh and cubic meshblocks, three parameters govern the grid setup: the number of 
grid points on the edges of the unrefined initial mesh $N_M$, the number of grid points on the edges of 
meshblocks $N_B$, and the number of physical refinement levels $N_L$. The grid structure is ultimately dictated 
by an adaptive mesh refinement (AMR) criterion. When satisfied, this criterion (de)refines a given MeshBlock, 
resulting in a smaller or larger block with double or half the resolution, respectively. For these simulations 
we consider the AMR criterion with 
$L_2$-norm as described in Ref.~\cite{Rashti:2023wfe}. In all of the simulations considered in this work,
we use $N_B=16$,
while we choose $N_L$ according to the mass-ratio. More specifically, we typically use $N_L=\lbrace 11,12,13 \rbrace$ 
for $q=\lbrace 1,2,3 \rbrace$, respectively. Finally, we use different values for $N_M$ depending on the configuration
considered, typically $N_M=\lbrace 128,192,256\rbrace$.  
The resolution at punctures $\delta x_p$ is given by Eq.~(40) of Ref.~\cite{Daszuta:2021ecf}.  
The information on the runs is summarized in Tables~\ref{tab:runs_q1_nospin} and~\ref{tab:runs_q1_spin_q23}, where we 
report the information for the highest resolution of each physical configuration
analyzed in this work.

\subsection{Post-processing}
\label{sbsec:postproc}
In this section we address the computation of gauge-invariant quantities derived 
from the numerical simulations previously discussed. In particular, we examine 
gravitational waveforms, energetics, remnant properties, and 
scattering angles.

\subsubsection{Numerical waveforms}
\label{sbsbsec:postproc_waves}
During the numerical time-evolution of the system, \GRA{} extracts the Weyl scalar
$\psi_4$ projected on a null tetrad at finite distance and integrating over discrete 2-spheres with geodesic grids,
~\cite{Daszuta:2021ecf}
and different radii. The $2\%$ discrepancy in the \GRA{} amplitude that was observed
in previous comparisons with other NR codes~\cite{Daszuta:2021ecf,Andrade:2023trh},
which was related to a bug on the symmetrization of the stencils used to compute the Weyl finite-differences in the ghost zones, has now been fixed.
The Weyl scalar for each extraction radius $R$ is then decomposed in multipoles using 
the spin-weighted spherical harmonics $\sYlm(\Theta,\Phi)$. 
We consider $R=100\,M$, if not specified otherwise.
The gravitational strain is also decomposed on the same basis, 
\be
h_+ - i h_\times = D_L^{-1}\sum_{\l=2} \sum_{m=-\l}^{\l} \sYlm(\Theta, \Phi) h_\lm,
\ee
where $D_L$ is the luminosity distance, and the waveform multipoles $h_\lm$ are
related to the Weyl scalar by the asymptotic relation 
\be 
\ddot{h}_\lm = \psi_4^{\lm}.
\ee
Since we are interested in the waveform at infinity, we need 
to extrapolate $\psi_4^\lm$ before performing the double-time integration.
We thus apply the procedure 
proposed in Refs.~\cite{Lousto:2010qx,Nakano:2015pta,Nakano:2015rda,Daszuta:2021ecf},
\begin{equation}
\label{eq:extrapolation}
\lim_{r \rightarrow \infty} r \psi_4^\lm \simeq A 
\bigg( \bar{r} \psi_4^\lm- \frac{(\ell-1)(\ell+2)}{2 \bar{r}} \int dt \, \bar{r} \psi_4^\lm \bigg) \, ,
\end{equation}
where $A=1-2M/\bar{r}$ is the Schwarzschild metric potential, and $\bar{r}=R (1+M/(2R))^2$
is the areal radius. 
This quantity is also used to compute the tortoise coordinate 
\be
\bar{r}_* = \bar{r} + 2 M \log \left( \frac{\bar{r}}{2M}-1 \right)\, ,
\ee
and the retarded time $u \equiv t - \bar{r}_*$. 
From the extrapolated scalar,
obtaining $h_\lm$ appears straightforward at a first glance.
It is however well-known that the presence of numerical noise can induce 
drifts in the signal~\cite{Reisswig:2010di}, making the integration 
not trivial.
The standard procedure used in the quasi-circular case is the fixed 
frequency integration (FFI), that consists in performing a frequency-domain integration 
with a low-frequency cut-off $f_0$~\cite{Reisswig:2010di}. Clearly, if $f_0$ is too low, the unphysical drifts
are not removed from the waveform. In the quasi-circular or eccentric cases,
the physical range of frequencies is finite, and a cut-off frequency below the lowest physical frequency
is typically enough to remove the noise-generated drifts, without cutting-off physical contributions. 
However, for highly-eccentric configurations, the waveform frequency at apastron $\omega_{\rm ap}$
can be so low that a cut-off $f_0<\omega_{\rm ap}$ does not yield reliable waveforms.
The situation is even more delicate for hyperbolic-like orbit, where
the frequency spectrum is not bounded. As a consequence, for highly non-circular configurations,
the FFI inevitably removes physical contributions from the waveform. For this reason, direct 
time-domain integration (TDI) is in principle preferable. 
However, as mentioned before, a direct TDI can lead to evident drifts. A typical solution
to overcome this problem is to remove the drift, after each integration step,
with a polynomial fit (typically linear or constant),
that is usually performed only on the last portion of the signal.
However, this method can lead to unreliable waveforms for
some configurations, especially for signal with long time durations, such 
as dynamically-bounded binaries with multiple encounters. 
For this reason, in this work we decide to use the FFI with $f_0=0.007 M^{-1}$ for all the equal mass
cases, and $f_0=0.010 M^{-1}$ for the unequal mass ones, 
even if this procedure inevitably
removes the low-frequency part of waveforms. In particular, the portion of the waveform generated
when the two black holes are farther apart will be strongly suppressed\footnote{In principle, the FFI also removes the
zero-frequency tail signal that dominates the quasi-normal-ringing at late time after merger. 
However, the tail is not typically resolved in finite-distance extracted Weyl scalars due
to the finite-resolution used. }.

The drifts caused by the numerical noise are not the only ones present in highly non-circular waveforms,
since the motion of the punctures toward the extraction zone can lead to drifts in $\psi_4^\lm$. 
While this could be in principle cured by just considering farther extraction zones, waveforms extracted 
at large radii are negatively impacted by numerical dissipation. 
This issue is particularly relevant for bounded 
configurations with large apastra and scatterings. Higher mass ratios make the problem even worse. 
This drift in the scattering case affects both TDI and FFI integrations: 
in the first case this drift spoils the late-time fit, 
while in the latter having a non-periodic $\psi_4^\lm$ can induce spectral leakage due to the 
fast Fourier transform used\footnote{The signal from dynamical captures is not strictly periodic, 
but $\psi_4^\lm$ is small enough both at the beginning of the simulation and approximately $\sim 200\,M$ 
after merger, so spectral leakage is not an issue in practice.}. 
However, the latter issue can be easily solved by applying 
time-windows that artificially send the initial and last portions of the signal to zero.
The windowing is particularly relevant for the last part of scattering waveforms. 

We conclude by mentioning that, for some configurations, we also consider Cauchy Characteristic Extraction (CCE).
We dumped the metric quantity needed for the characteristic extraction, that was then performed as post-processing 
step employing the public code \texttt{PITTNull}~\cite{Bishop:1998uk,Babiuc:2010ze}. 
Comparisons between extrapolated finite-distance and CCE waveforms
are reported in Appendix~\ref{app:cce_waves}, where we argue that
improved high-pass frequency filters
should be considered in future works.
For this reason, in the main body of this work we use waveforms extracted at finite distance
and then extrapolated according to Eq.~\eqref{eq:extrapolation}.

\subsubsection{Energetics}
\label{sbsbsec:postproc_energetics}

In our analysis, it is crucial to accurately differentiate between scattering 
events and captures. Visual inspection of the punctures' tracks can be misleading, 
as it is challenging to distinguish, within the simulation time, proper scatterings from systems
that will merge in the distant future. 
The most reliable method to differentiate these outcomes is to examine the energy 
after the first encounter, when the two black holes have reached a reasonable large
separation.  If the energy satisfies the condition $E<M$, 
the two objects will inevitably merge; otherwise, they will scatter.
The energetics are thus a fundamental tool in our analysis. 

These curves can be computed once that the waveform has been extracted. The 
energy and angular momentum instantaneous fluxes are indeed given by
\begin{subequations}
\label{eq:fluxes}
\begin{align}
\label{eq:fluxE}
\dot{E} &= \frac{1}{16 \pi} \sum_{\l=2} \sum_{m=-\l}^{\l} | \dot{h}_{\lm} |^2  ,\\
\label{eq:fluxJ}
\dot{J} &= -\frac{1}{16 \pi} \sum_{\l=2} \sum_{m=-\l}^{\l} m \Im \left( \dot{h}_{\lm} h_{\lm}^* \right),
\end{align}
\end{subequations}
where the asterisk denotes a complex conjugation and $\Im$ is the imaginary part operator.
In this work we consider all the multipoles up to $\ell=5$.
Note that $\dot{E}$ involves only the News $\dot{h}_{\lm}$, while the computation of 
$\dot{J}$ also requires the strain. As a consequence, energy fluxes are more accurate
than angular momentum ones, since they require one fewer integration step.
We can then integrate Eqs.~\eqref{eq:fluxes} and compute the total energy and angular
momentum radiated by the system as a function of time. We can then subtract these quantities
to the initial values $(E_0, J_0)$ and obtain
the time-evolution of the energy and
angular momentum, $E(t)$ and $J(t)$. 
We can then calculate the reduced binding energy 
\be
\hat{E}_b = \frac{E(t) - M}{\mu},
\ee
where $\mu = \nu M$ is the reduced mass of the system. 
Finally, combining the binding energy with the reduced angular momentum $j=J/(\nu M^2)$, we obtain 
the gauge-invariant energetic curves $\hat{E}_b(j)$.

In the case of coalescing binaries, the energetic curves can be used to estimate
the properties of the remnant. Indeed, balance equations
imply that the  mass $M_f$ and spin $a_f$ of the remnant are given by 
the final values of energy and angular momentum.

\subsubsection{Scattering angles}
\label{sbsbsec:postproc_scats}
In the case of systems with final positive binding energy, 
we can compute the gauge-invariant scattering angle by extrapolating 
the tracks of the black holes~\cite{Damour:2014afa}.
More precisely, we convert the relative motion of the two punctures in polar coordinates $(r,\varphi)$,
and we extrapolate the incoming and outgoing trajectories, $\varphi_{\rm in}(r)$ and $\varphi_{\rm out}(r)$,
using $1/r$-polynomials. The constant terms of the two polynomials give the asymptotic incoming and outgoing
angles, $\varphi_{\rm in}^\infty$ and $\varphi_{\rm out}^\infty$, that can be used to compute the 
scattering angle
\be
\chi \equiv \varphi_{\rm out}^\infty - \varphi_{\rm in}^\infty - \pi.
\ee
The portions of the tracks that we consider for the fits are $r\in [25,80]\,M$ and $r\in [25\,M,r_f]$, 
for the ingoing and outgoing motion, respectively. Here $r_f$ is the separation reached 
by the two punctures at the end of the simulation. 
For the ingoing part, we consider an initial radius that is smaller 
than the initial separation; this is due to the fact that the two punctures have zero initial
velocities\footnote{This is a direct consequence of the typical gauge choice
adopted in moving punctures gauges, where the shift vector $\beta^i$ is
set to zero at the beginning of the simulation. This should not be confused with the initial ADM linear
momenta.}.
Similar to previous works~\cite{Damour:2014afa,Hopper:2022rwo,Damour:2022ybd,Rettegno:2023ghr}, 
the polynomial extrapolation is performed with a
least-squares fitting method that uses a singular-value decomposition (SVD).
Singular values smaller than $10^{-13}$ times the maximum
singular value are dropped.
Typically, with the aforementioned SVD, the angle tends to plateau around polynomial order $n = 6$, 
yielding consistent values for higher orders; this resulting value is assigned to $\chi$.
The error associated 
to this extrapolation is computed as the difference between the highest and lowest angles
obtained with the different polynomial extrapolations, i.e.,
$\Delta \chi_{\rm fit} = \chi_n^{\rm max} - \chi_n^{\rm min}$.

Another source of error is the finite resolution employed in the simulations. We test the relevance 
of this error for a few scattering
configurations by considering two resolutions, $N_M=\lbrace 128, 256 \rbrace$. 
We then compute the error $\Delta \chi_{\rm res}$ 
as the difference between the two angles obtained. 
However, for all the cases considered we find $\Delta \chi_{\rm res}< 0.3 {}^\circ$,
while $\Delta \chi_{\rm fit}\sim 1-3^\circ$. Since the total error is computed 
as a square sum of the individual errors, $\Delta \chi_{\rm res}$ is in practice negligible.
We thus considered most of the scattering configurations at resolution $N_M=128$, assigning to each of them 
a conservative resolution error $\Delta \chi_{\rm res} = 0.3^\circ$.

We conclude this discussion by recalling that the velocities of the punctures are strictly linked
to the shift vector $\beta^i$~\cite{Campanelli:2005dd}.
Since the tracks are gauge-dependent quantities, one should pay extra care
when using them to compute the gauge-invariant scattering angle.
While the relevance of the initial value for the lapse function $\alpha$ has been tested in Ref.~\cite{Rettegno:2023ghr},
tests on the Gamma-driver, and thus on $\beta_i$, should be performed to precisely quantify the relevance
of the gauge-choices on the numerical scattering angle. However,
a remarkable agreement between the numerical and the analytical agreement has 
been found in many works~\cite{Damour:2014afa,Hopper:2022rwo,Damour:2022ybd,Rettegno:2023ghr}
for most configurations, leading us to think, a posteriori, 
that the effects of the gauge choices are indeed small, at least
for the energy range considered.

\subsection{Phenomenology} 
\label{sbsec:phenom}
\begin{figure*}[t]
  \centering 
    \includegraphics[width=0.315\textwidth]{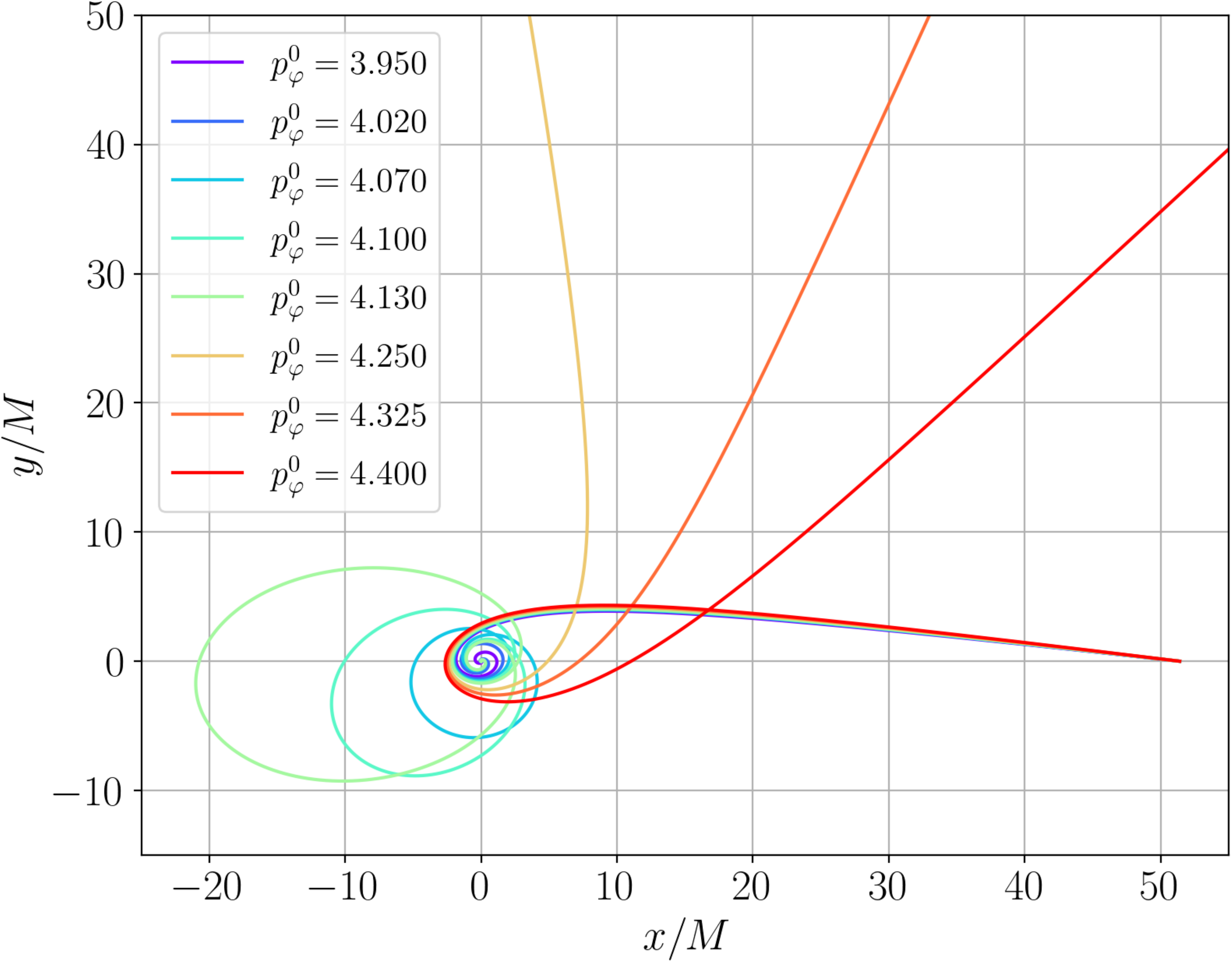}
    \includegraphics[width=0.340\textwidth]{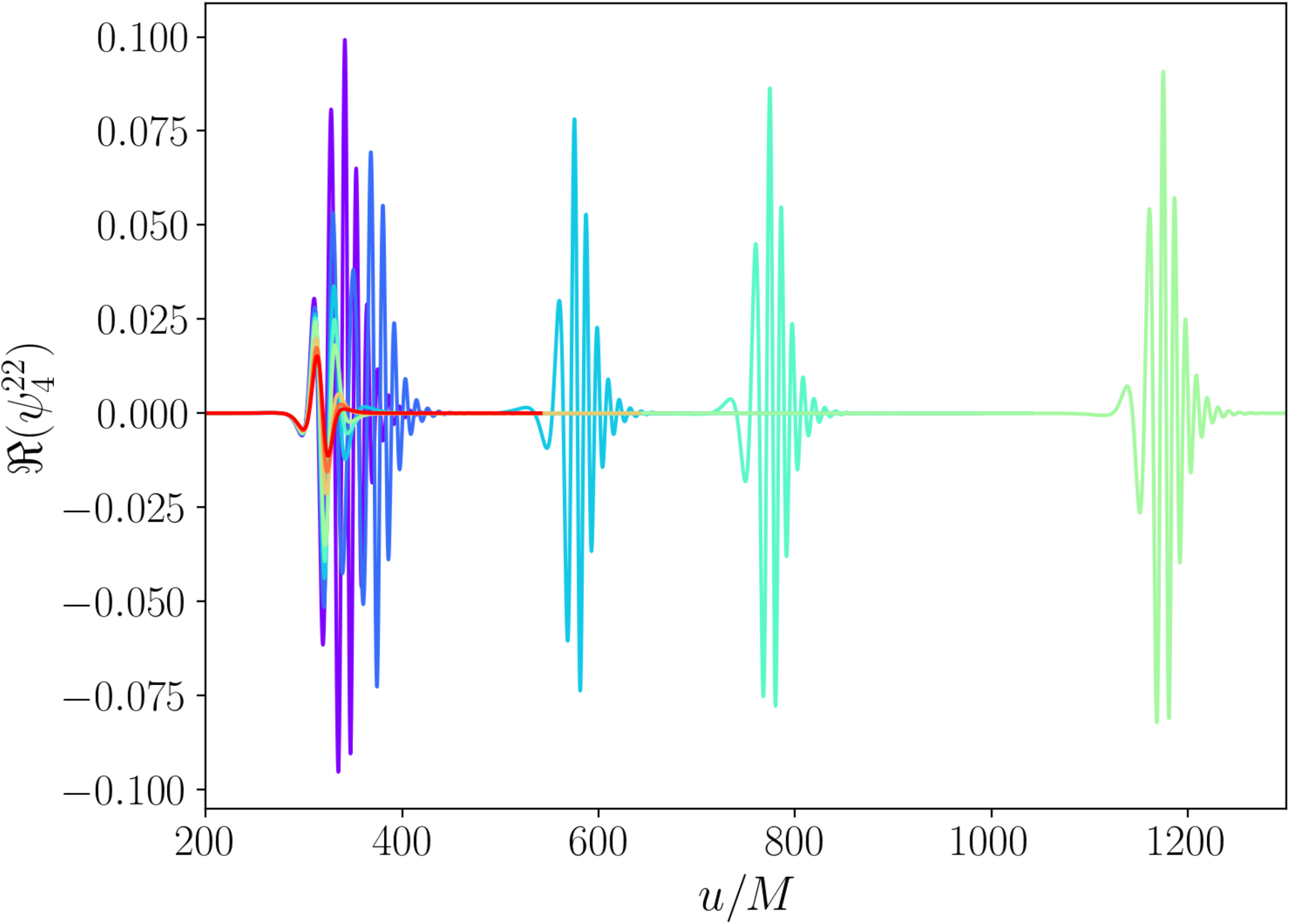}
    \includegraphics[width=0.330\textwidth]{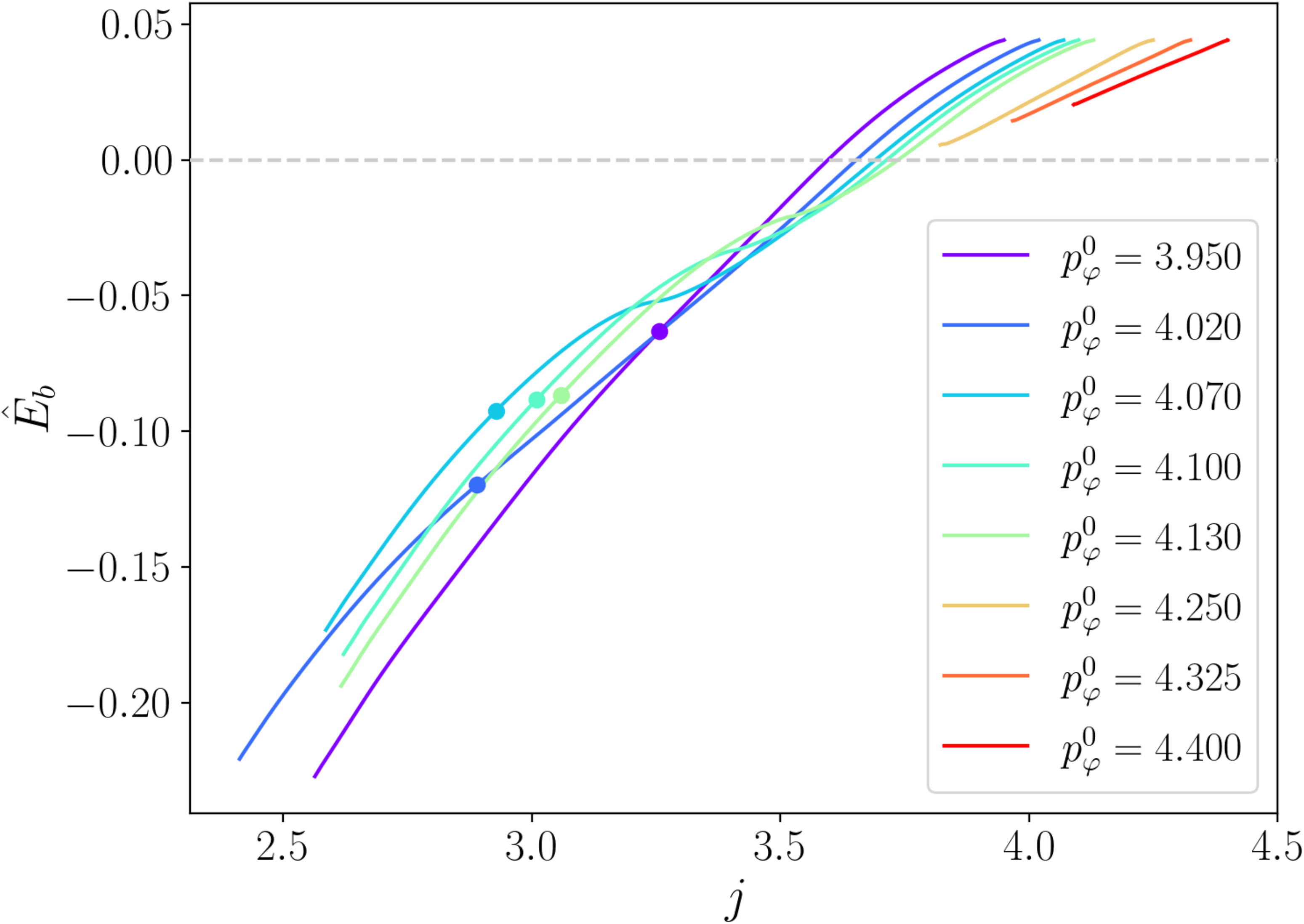}
    \caption{
    Scatterings and dynamical captures for equal mass systems with fixed initial energy 
    $E_0 \simeq 1.011\,M$ and different initial angular momenta $p_\varphi^0$;
    the latter are reported in the legends.
    We show the tracks of one puncture, the extrapolated (2,2) Weyl scalars, and the energetics. The circles
    in the $\hat{E}_b(j)$ curves mark the time of the merger, defined as the last peak of the (2,2) waveform amplitude
    (see Appendix~\ref{appendix:tmrg} for more details).
    The horizontal line in the right panel marks the parabolic limit. 
   }
  \label{fig:series_E1.011}
\end{figure*}
As anticipated, 
initially unbound systems can have a rich variety of phenomenologies. 
If the gravitational wave emission is not strong enough to make the system bound,
then the two black holes will simply scatter. 
However, if enough radiation
is emitted, then the system becomes bound and the two black holes are doomed to merge. 
The latter scenario can happen in essentially two situations, which correspond to
different regions of the orbital parameter space, i.e. the $(p_\varphi,E)$ plane. 
The first situation occurs when the system has a relatively low angular momentum 
or, equivalently, when the impact parameter $b= L/|p|$
is rather small, being $|p|$ the modulus of the linear momentum. 
In this case, the two black holes pass closer to each other,
prompting a significant emission of radiation.
The two black holes can either merge at their first encounter ({\it direct captures}), 
or they may initially rebound, 
only to complete a single orbit and ultimately merge during the subsequent close encounter.
We denote the latter systems as {\it double encounters}.
The second situation instead occurs when the initial energy is slightly 
above the parabolic limit $\hat{E}_b=0$.
In this case, even a small interaction can produce enough 
gravitational waves to bind the two black holes, that 
can undergo many encounters before merging, as long as enough angular momentum is provided.
These two regions of the parameter space are visualized 
for the equal mass nonspinning case
in Fig.~\ref{fig:parspace_TEOB_q1_nospin}, 
were we considered both numerical simulations 
and analytical predictions. This figure will be further discussed in Sec.~\ref{sec:eob_nr}

\subsubsection{Nonspinning equal mass systems}
\label{sbsbsec:phenom_q1_nospin}
We study the equal mass nonspinning case by considering several values of energy.
For each of them, we examine different angular momenta that result 
in scattering events, double encounters, and direct captures.
We start by discussing
eight simulations with fixed initial
energy\footnote{We typically write energies and angular momenta up to the third decimal. 
More precise values are reported in Tables~\ref{tab:runs_q1_nospin} and~\ref{tab:runs_q1_spin_q23}.} 
$E_0 \simeq 1.011\,M$ and initial separation $D\simeq102.8\,M$.
We report the tracks of the punctures in the left panel of Fig.~\ref{fig:series_E1.011}, while
in the middle panel of the same figure we show $\psi_{22}^4$ extrapolated to infinity through
Eq.~\eqref{eq:extrapolation}. In the right panel we draw the energetic curves $\hat{E}_b(j)$,
also highlighting the merger time, which
is defined as the last peak of the (2,2) waveform amplitude.
It is crucial to consider the last peak of the amplitude rather than the highest one,
since the latter might occur during encounters that do not promptly result in a coalescence;
see Appendix~\ref{appendix:tmrg} for more detail.

The three configurations with higher initial canonical angular momentum, 
$p_\varphi^0 \simeq \lbrace 4.400, 4.325, 4.250 \rbrace$, are scatterings.
As the initial angular momentum is decreased, the gravitational emission at the first encounter is
enhanced and the systems become bound. For $p_\varphi^0 \simeq \lbrace 4.130, 4.100, 4.070 \rbrace$,
the trajectories diverge after the first encounter, gradually increasing the separation between the two black holes.
However, upon reaching the apastron --the point of maximum separation-- 
they begin to move toward each other again and eventually merge.
As a consequence, the 
Weyl scalar clearly exhibits two amplitude peaks. For $p_\varphi^0\simeq4.020$, the two 
punctures perform a quasi-circular whirl before plunging and merging; this dynamics still 
produces two close peaks in the Weyl scalar, but no outgoing motion of the punctures is observed. 
Moreover, these two peaks unify after integration, thus becoming indistinguishable in the strain.
We thus classify this event as a direct capture.
Finally, the configuration with lowest angular momentum, $p_\varphi^0\simeq3.950$, also results in a direct capture.

We also mention that we evolve a configuration with same initial energy and $p_\varphi^0\simeq4.200$, 
that is not shown in Fig.~\ref{fig:series_E1.011}.
The reason for this choice is that, after the first encounter, the system becomes bound 
($E\sim 0.999\,M$ after the first encounter), but we did not complete the simulation till merger. 
However, knowing that this configuration is bound
allows us to put a more stringent bound on the interval in which the transition
from scattering to dynamical capture occurs.
Indeed, with this information, we can claim  that, at fixed initial energy $E_0 \simeq 1.011\,M$,
this transition occurs for a canonical angular momentum $p_\varphi^0$ 
in the interval $\left[ 4.200, 4.250 \right]$, where the first value corresponds to the capture with
highest simulated $p_\varphi^0$, and the second one corresponds to the scattering with lowest angular momentum. 
We thus identify the angular momentum at which the transition occurs as $p_\varphi^* = 4.225\pm 0.025$. 
We remark that we start our evolutions at large separation ($D\simeq 100\,M$), so that
the dissipative effects in the first part of the evolution are negligible, thus making 
these results mostly independent on the initial separation.

\begin{figure}[t]
  \centering 
    \includegraphics[width=0.49\textwidth]{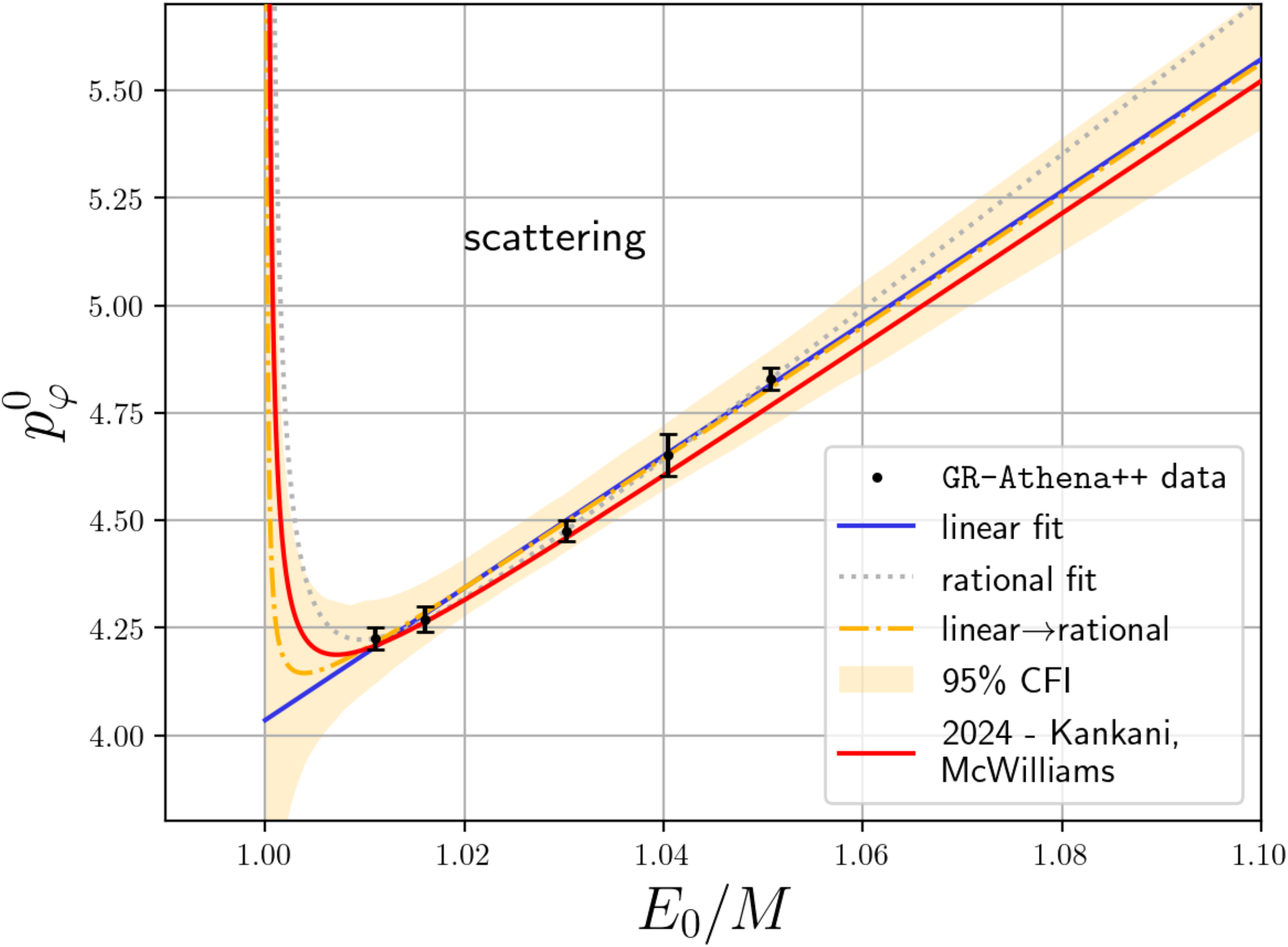}
    \caption{ 
  \label{fig:transition_fit} Canonical initial angular momentum that marks the transition from scattering
  to bound systems as a function of energy for $(q,\chi_1,\chi_2)=(1,0,0)$. 
  We show the data points obtained from the \GRA{} simulations (black),
  together with different fits (see text for more details).}
\end{figure}

We repeat this analysis for different energies, considering in total 
five values in the range $E_0 \in [1.011, 1.051]\,M$.
The corresponding $p_\varphi^*$, with associated error bars, are reported in Fig.~\ref{fig:transition_fit}. 
In the same figure, we also report the fit of the scattering-merger transition recently 
proposed in Ref.~\cite{kankani:2024may} (red). 
The latter is performed using energies up to $E=1.1\,M$, and assuming the ansatz
\be 
\label{eq:kankani_ansatz}
p_\varphi^*(\hat{E}) = \frac{a \hat{E}^2  +b \hat{E} + c}{\hat{E}-1},
\ee
where the coefficients are
$(a_K, b_K, c_K)= (15.4932, -27.0216,  11.5292)$
(note that here we use $p_\varphi$ rather than $L$, so that the coefficients of Ref.~\cite{kankani:2024may} 
are $\nu$-rescaled here). 
We use our simulations to perform a similar fit. However, our five datapoints approximately lay
on a line, so that directly regressing them using the ansatz of Eq.~\eqref{eq:kankani_ansatz} would provide a fit
that could not be safely extrapolated to higher energies (gray in Fig.~\ref{fig:transition_fit}).
While a linear fit is a more natural choice (blue in Fig.~\ref{fig:transition_fit}), 
this linear regression would certainly fail for energies close to the parabolic limit. 
We thus proceed as follows. We map the energies $\hat{E}\in[1, \hat{E}_0^{\rm max}]$ 
and the angular momenta $p_\varphi\in [p_\varphi^{0,\min}, p_\varphi^{0,\max}]$
in the interval $[0,1]$ with a linear rescaling. We then determine the angular coefficient $\bar{a}_l$ with a linear regression,
and subsequently perform a fit with the ansatz $y = (\bar{a}_l x^2 + \bar{b} x + \bar{c})/x$. 
Inverting the linear rescaling, we obtain 
a relation $p_\varphi^*(\hat{E})$ equivalent to Eq.~\eqref{eq:kankani_ansatz}.
The coefficients
found with this procedure are $\left( a,b,c \right) = \left( 15.3746, -26.7257, 11.3514\right)$.
The final result is also reported in orange in Fig.~\ref{fig:transition_fit}, 
together with the corresponding $95\%$ confidence interval.
The latter shows that our fit is compatible with the result of Ref.~\cite{kankani:2024may}.

However, our scattering-merger transition is systematically 
shifted at slightly higher angular momenta for $E\gtrsim 1.01$. 
Moreover, the fit proposed by the authors of Ref.~\cite{kankani:2024may} 
predicts $p_\varphi^* = 4.767$ for $\hat{E}=1.05079$, the highest energy considered in this work, while
from our numerical data we have $p_\varphi^*= 4.830\pm0.025$.
The small discrepancies might be linked to the different methods employed to compute the fluxes
from $\psi_4$, that are used to estimate the energy after the first encounter, i.e. 
the quantity that dictates the final state of the system (bound or unbound).
Indeed, Ref.~\cite{kankani:2024may} uses, for initially unbound configurations, 
a time-domain integration for $\psi_4$ extracted at $R=140\,M$, 
while we use an FFI applied to the Weyl scalar extracted at $R=100\,M$. Number of multipoles considered 
and resolutions employed might also play a role in this discrepancy. Finally,
notice that our simulations start from an initial separation $D\sim 100\,M$, while the ones
considered in Ref.~\cite{kankani:2024may} start from $D = 60\,M$, where the effects of
the junk radiation might be more relevant. Nonetheless, as mentioned above, the fit of Ref.~\cite{kankani:2024may}
still lays within our $95\%$ confidence interval. 

We conclude this discussion by remarking that we also run a series of simulations at initial
energy $E_0 = 1.001\,M$. However, since the fate of the configuration with highest 
orbital angular momentum ($p_\varphi^0\simeq4.600$)
is not clear and the other three configurations are captures, we do not employ these data in our fit. 
The uncertainties on the final state
are linked to the fact that different integration options lead to different energies after
the first close passage, some of them being slightly below the parabolic limit, and others slightly above.
In Figs.~\ref{fig:parspace_NR} and~\ref{fig:parspace_TEOB_q1_nospin}, we label this simulation as a 
scattering, basing the decision on our default integration setting. 
However, this classification should be interpreted with caution.

\subsubsection{Spinning equal mass systems}
\label{sbsbsec:phenom_q1_spin}
\begin{figure*}[t]
  \centering 
  	\includegraphics[width=0.310\textwidth]{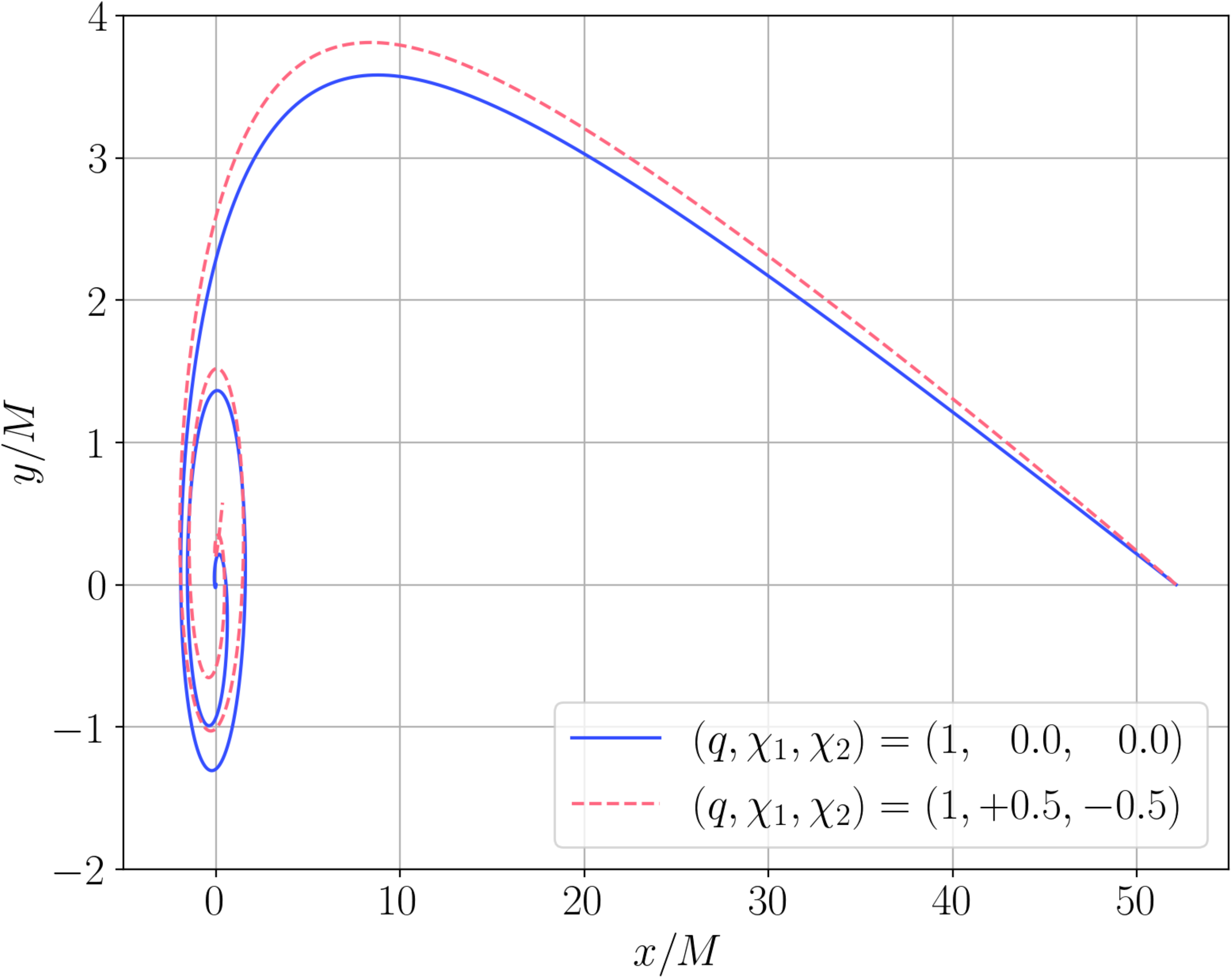}
    \includegraphics[width=0.340\textwidth]{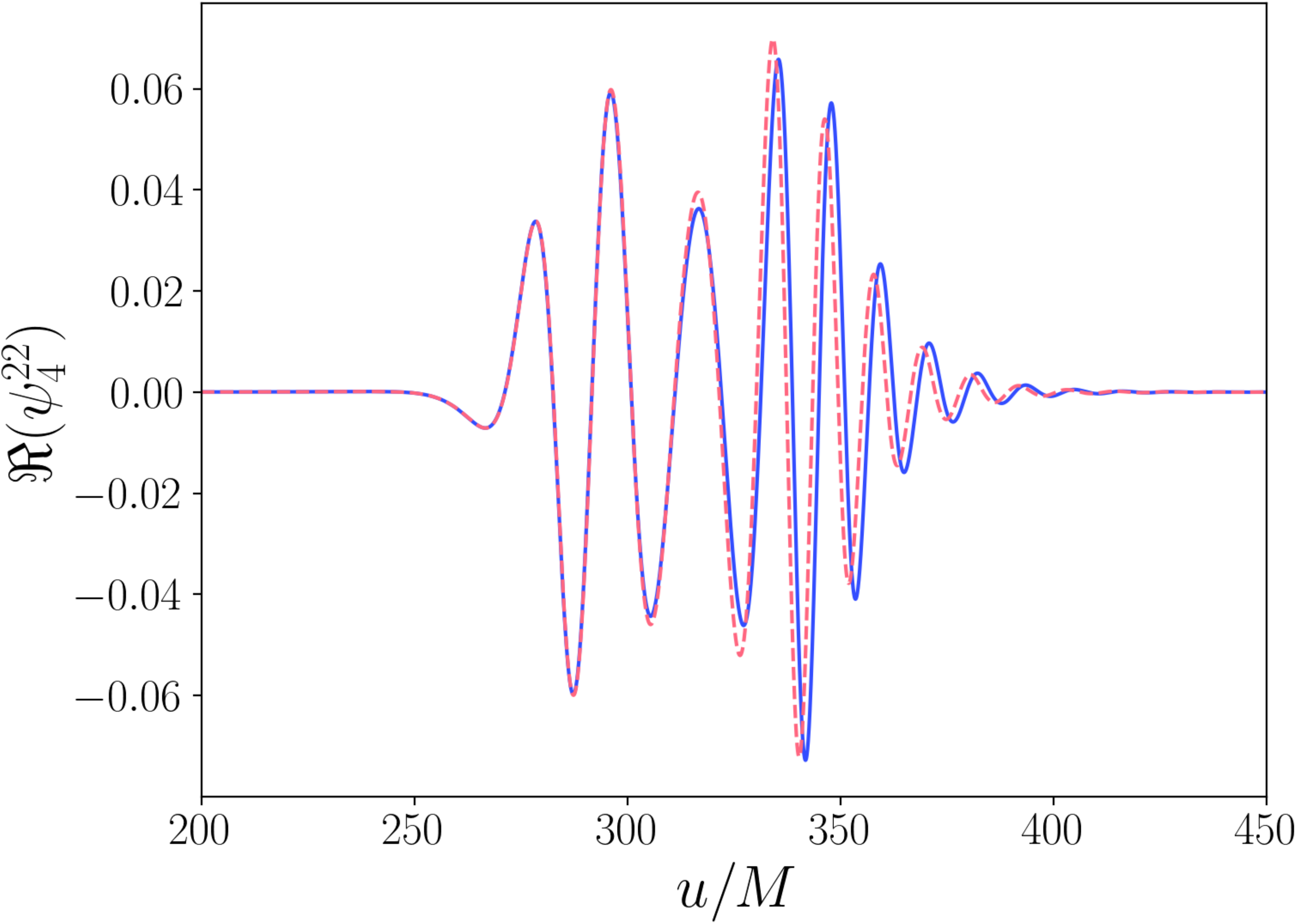}
    \includegraphics[width=0.325\textwidth]{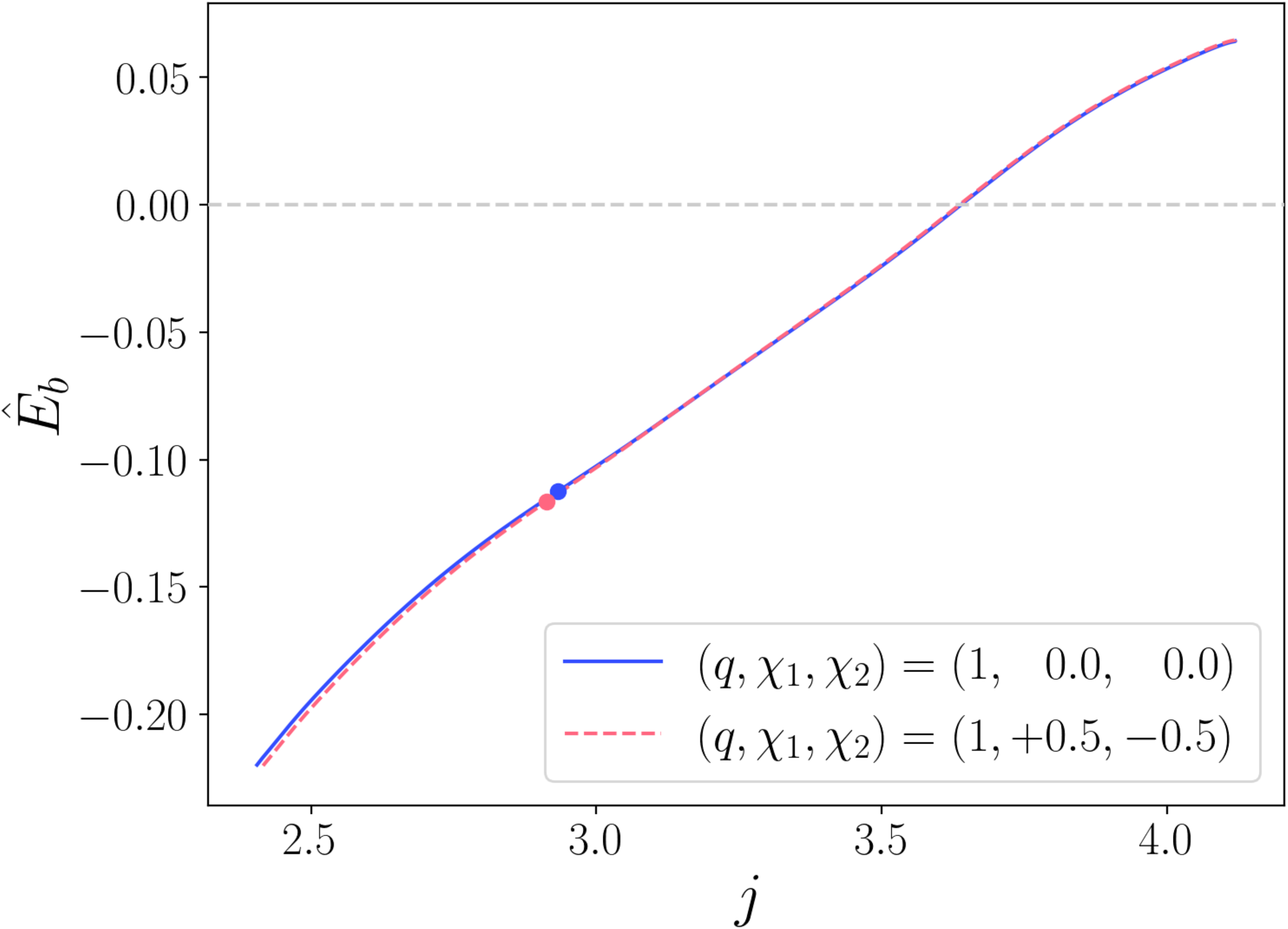}
    \caption{Tracks of one puncture, extrapolated Weyl scalars, and energetic curves for equal mass configurations with
    same initial separation, same angular momentum $p_\varphi^0=4.12001$,
    and same energy $E_0\simeq 1.016$ (within the $\sim0.01\%$ level), 
    but different spin configurations (solid blue for the 
    nonspinnig case, dashed red for $\chi_1=-\chi_2=0.5$). 
    The circles in the energetics mark the merger time.
    Note the different vertical and horizontal scales in the plot of the tracks.
    The two configurations are rather similar as predicted by PN theory, 
    since the leading order spin-orbit contribution is 
    vanishing for equal mass binaries with anti-aligned individual spins.
    }
  \label{fig:updown_spin}
\end{figure*}

We now turn our attentions to equal mass systems with aligned and anti-aligned spins. 
We consider spins with magnitude $|\chi_i|=0.5$ and different orientations. 

We start by studying configurations whose individual 
spins have opposite signs. 
According to PN theory, the main spin correction
vanishes for equal mass binaries with 
this spin configuration~\cite{Racine:2008qv,Santamaria:2010yb}. We thus expect the phenomenology
of systems with anti-aligned individual spins to be quite similar to the nonspinning case. 
An explicit example for dynamical capture described within
the EOB formalism is reported in Fig.~6 of Ref.~\cite{Nagar:2020xsk}
(see also Fig.~7 therein).

In order to verify this prediction, 
we consider a spinning configuration with $(q,\chi_1,\chi_2)=(1,+0.5,-0.5)$
and $(\hat{E}_0, p^0_\varphi) = (1.01619, 4.12001)$.
The corresponding Weyl scalar and energetic curve
are then compared against a nonspinning configuration with 
equivalent orbital parameters in Fig.~\ref{fig:updown_spin}. The 
initial energy in the latter case is $E_0=1.01607\,M$, with a $\sim 0.01\%$
discrepancy with respect to the spinning case.
%
The tracks of the punctures reported in the left panel of Fig.~\ref{fig:updown_spin} 
show that these two configurations have a quasi-circular whirl before the plunge. 
While the corresponding (2,2) waveforms are initially well aligned ($|\Delta \phi| \lesssim 0.015$ radians for
retarded time $u<300\,M$), 
the dephasing rapidly increases afterward, finally saturating at $|\Delta \phi|\sim1 $ radian for $u = 350\,M$.
This discrepancy in the phase is also clearly visible in the Weyl scalar reported in the middle
panel of Fig.~\ref{fig:updown_spin}. 
Since dephasing occurs only after $u = 300\,M$, we are prone to think that this phase
difference is linked to higher-order 
spin contributions rather than to the $0.01\%$ discrepancy in the initial energies. 
Indeed, significant differences in initial energies would influence the timing of the close encounter, 
causing a misalignment throughout the entire waveforms.
A similar conclusion can be also drawn by inspecting the energetic curves.

The analysis is repeated also for a larger angular momentum, $p_\varphi^0\simeq4.300$, 
which results in scattering in both the spinning and non-spinning configurations.
In this case, the comparison between the waveforms is even more striking,
since no evident dephasing is observed for the whole evolution 
($|\Delta \phi| \lesssim 0.015$ radians up to the close encounter,
reaching at most $|\Delta \phi| \sim 0.06$ radians afterward).
The result is not surprising since, as illustrated in Fig.~\ref{fig:updown_spin} for $p_\varphi^0\simeq4.120$, 
there is no evidence of strong dephasing in the waveform generated at the first close encounter;
instead, the dephasing becomes more pronounced during the quasi-circular whirl that occurs just 
before the plunge. 
Despite a visual discrepancy in the gauge-dependent tracks of the punctures for $p_\varphi^0\simeq4.120$
(that can be observed also for the previous case, as shown in the left panel of Fig.~\ref{fig:updown_spin}), 
the scattering angles 
are perfectly compatible, since they are 
$301.5 \pm 1.7 {}^\circ$ and $303.6 \pm 2.0 {}^\circ$ 
for the nonspinning and spinning cases, respectively.
These results, together with the ones for $p_\varphi^0\simeq4.120$, confirm the prediction of Ref.~\cite{Nagar:2020xsk}. 

\begin{figure*}[t]
  \centering 
    \includegraphics[width=0.32\textwidth]{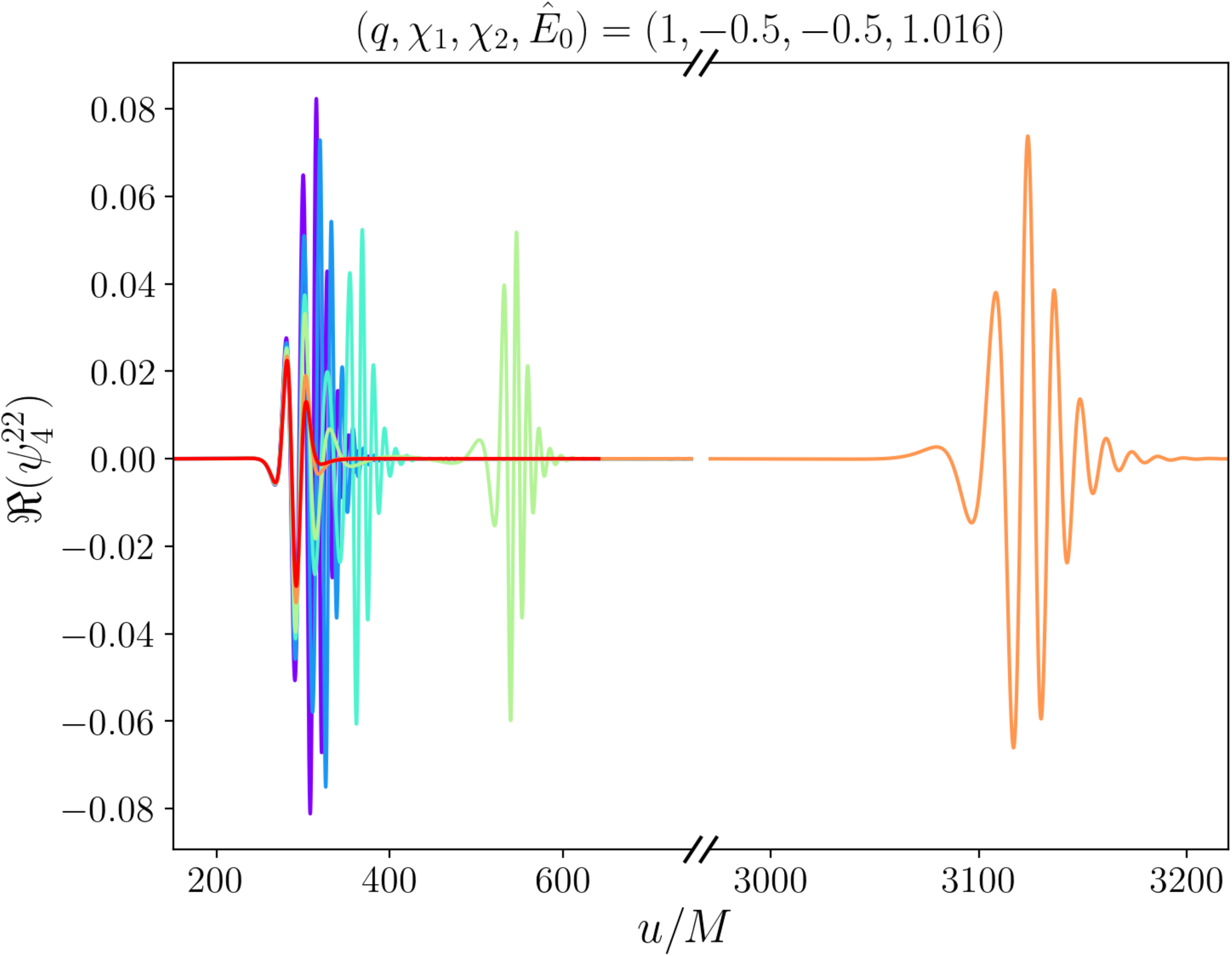}
    \includegraphics[width=0.32\textwidth]{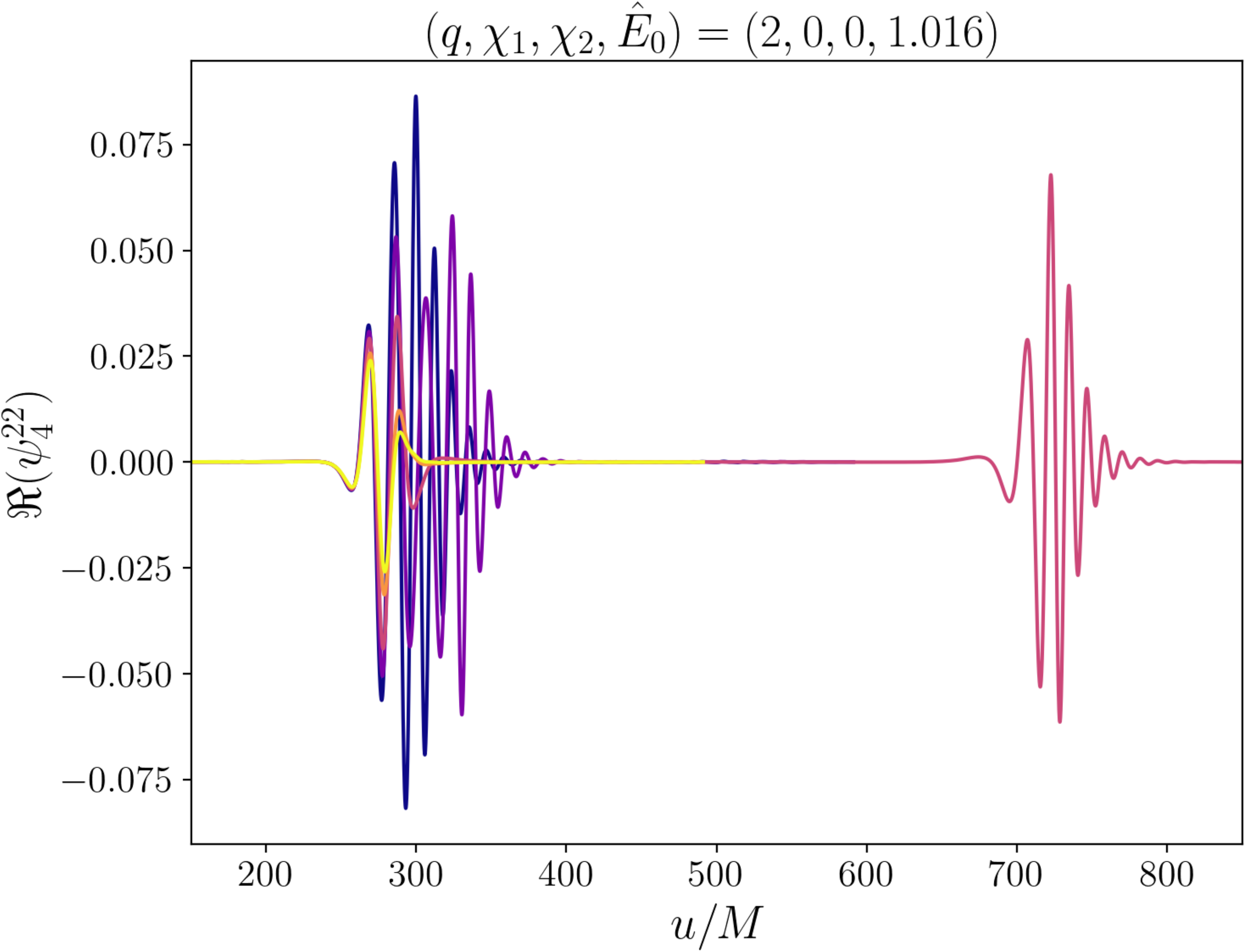}
    \includegraphics[width=0.32\textwidth]{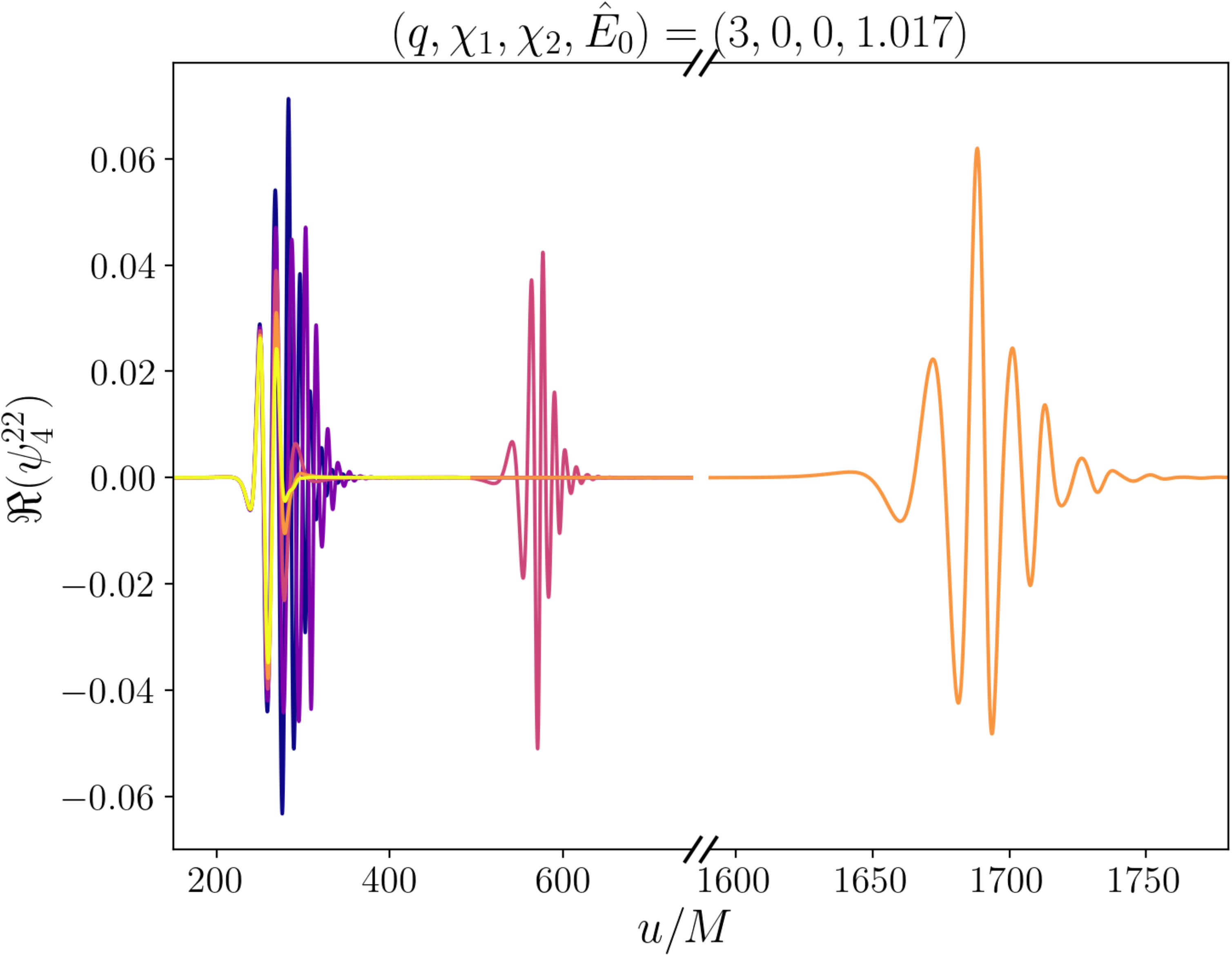}\\
    \includegraphics[width=0.32\textwidth]{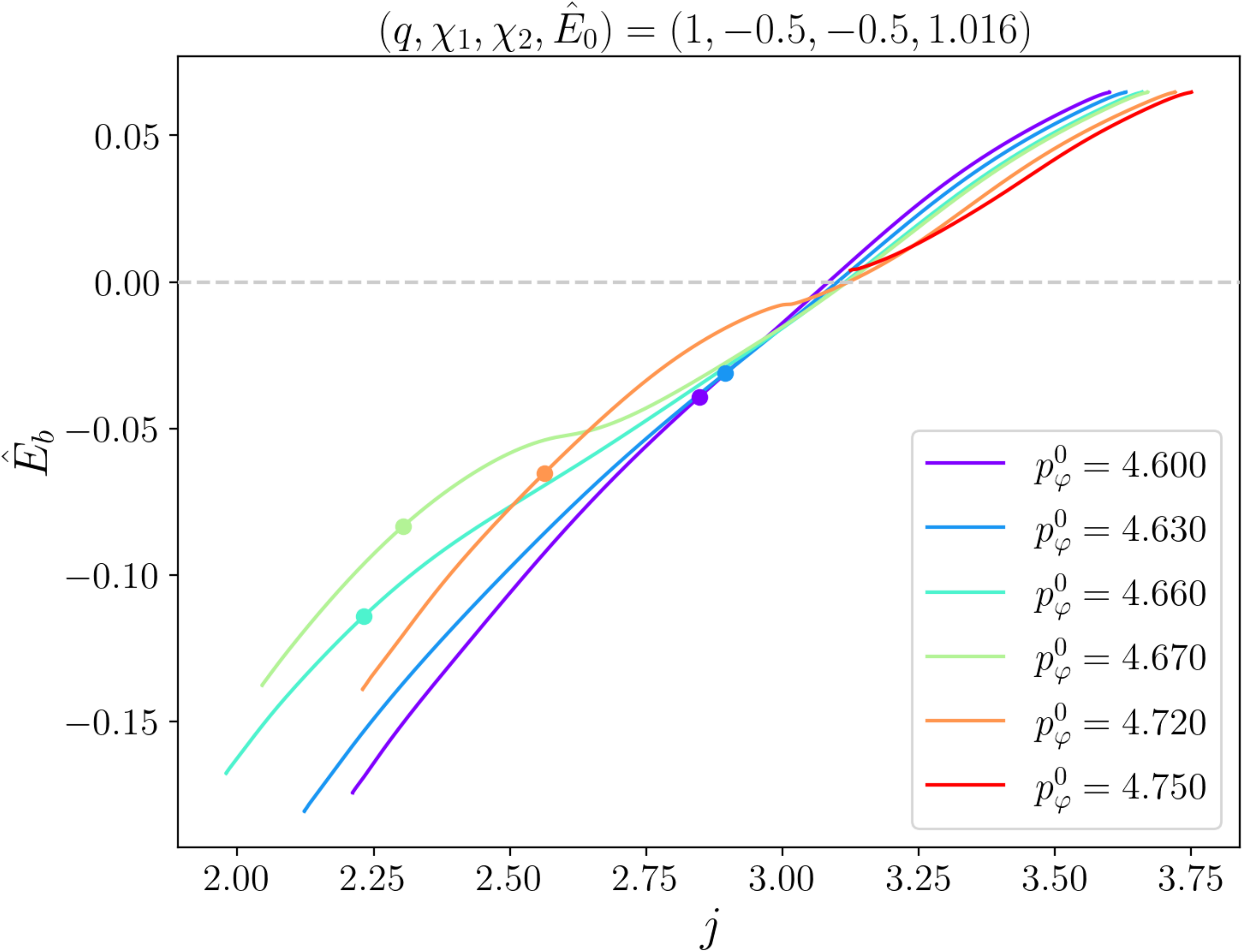}
    \includegraphics[width=0.32\textwidth]{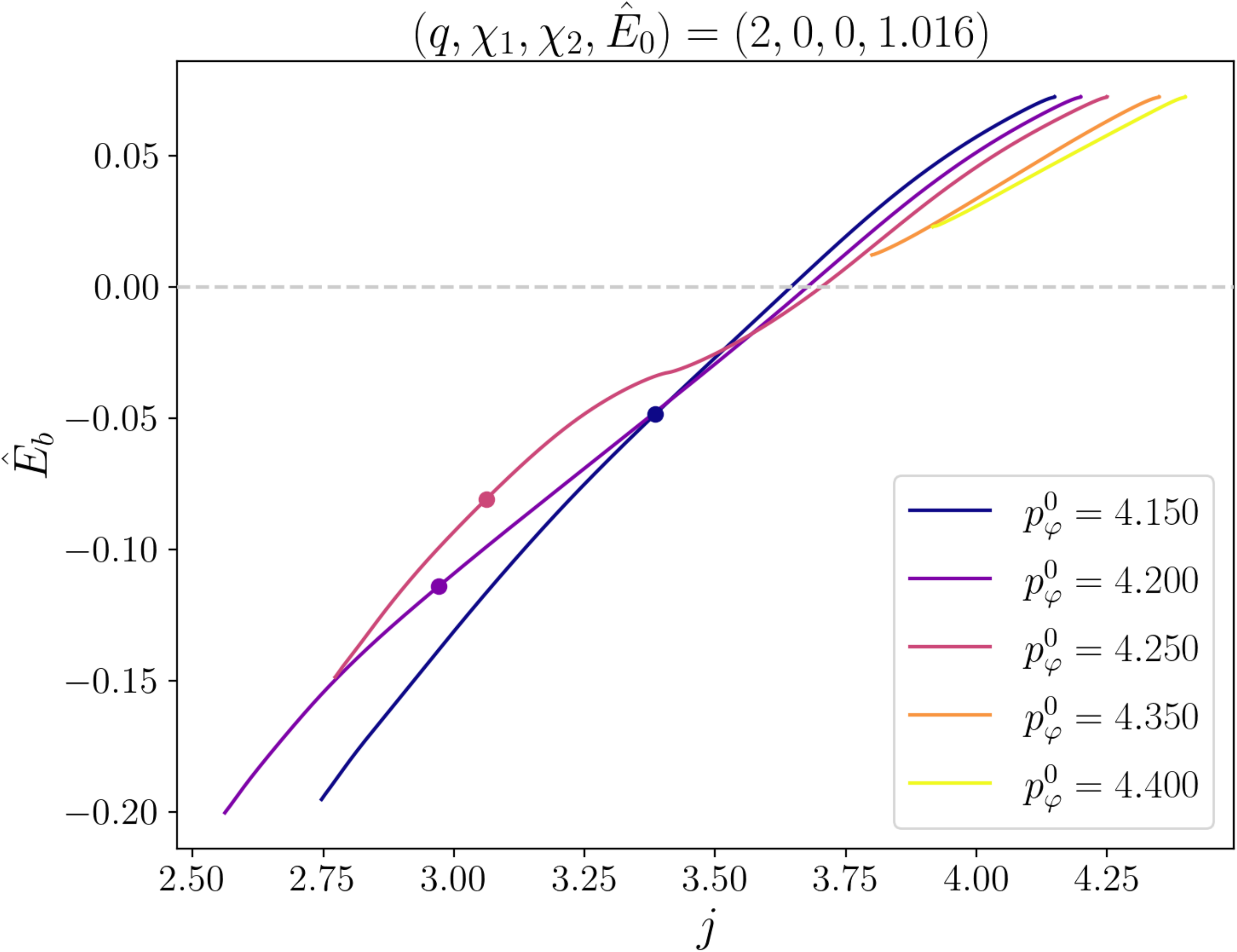}
    \includegraphics[width=0.32\textwidth]{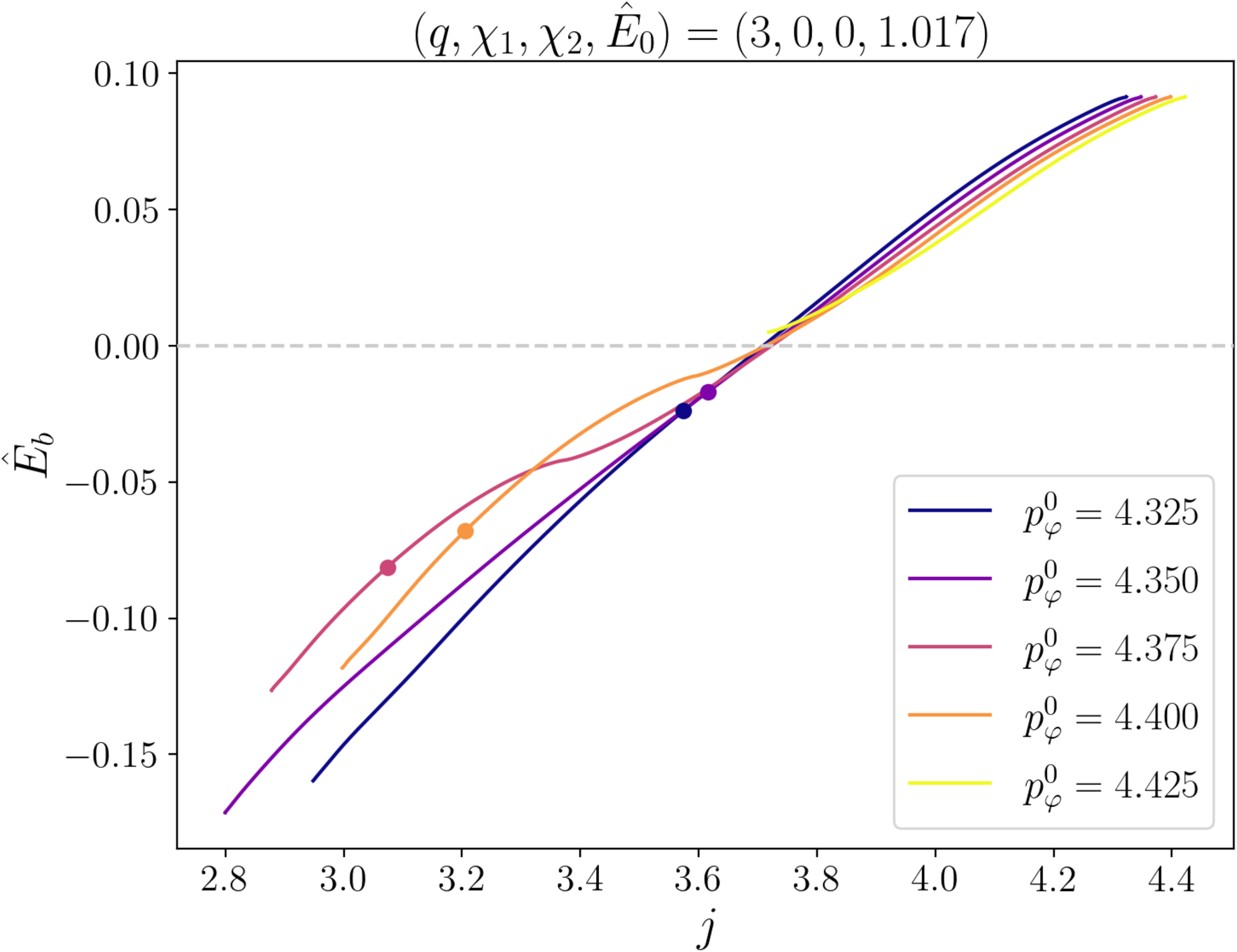}
    \caption{Real parts of $\psi_4^{22}$ and energetic curves for an equal mass energy series 
    with $\chi_i=-0.5$ (left) and two energy series for unequal mass nonspinning configurations. 
    The markers in the energetics highlight the merger time. 
    The angular momenta are shown in the legends of the bottom panels.}
  \label{fig:series_q123}
\end{figure*}

We now move to the case in which both spins are anti-aligned with
the orbital angular momentum, $\chi_1 = \chi_2 = -0.5$. 
In heuristic terms, the frame dragging of the two black holes opposes to the orbital motion,
so that, in order to avoid captures, we need to provide more orbital 
angular momentum (or, equivalently, higher impact parameters).
We thus simulate an energy series with $E_0 = 1.01619\,M$ and 
angular momenta $p_\varphi^0\in[4.600,4.750]$. 
The dominant multipoles of $\psi_4$ and the energetics computed with all 
the multipoles up to $\l=5$ are shown in the left panels of Fig.~\ref{fig:series_q123}.
For the configuration with $p_\varphi^0\simeq4.720$, we have a double encounter where the two black holes merge
at $u \sim 3115\,M$. This very long simulation poses different challenges to the integration
process. First of all, consider that the maximum separation reached after the first encounter 
is $D_{\rm ap} \sim 100\,M$, meaning that the two punctures have an outgoing motion up to 
a grid radius of $R\sim 50\,M$. This motion generates visible drifts in the waveform
extracted at radii $R\lesssim 120\,M$, that become negligible only for higher extraction radii.
Note that this issue might occur also for scatterings, if the outgoing motion is evolved long 
enough. However, this problematic behavior
can be easily fixed, at least for the cases investigated in this work,
by considering larger extraction radius; we thus consider $R=140\,M$ for this double encounter. 
Second, the long duration of the signal makes time-domain integration impracticable; 
even considering higher-order polynomials to remove the accumulated drift 
does not produce meaningful strains. On the contrary, FFI methods produce more reliable waveform
once that the aforementioned issue with the outgoing motion is solved.
For the energy considered ($E_0 = 1.01619\,M$), we find that the transition
from scattering to captures occurs
for $p_\varphi^* = 4.735 \pm 0.015$. In the nonspinning case, for a very similar energy ($E_0 = 1.01607\,M$)
we got $p_\varphi^* = 4.270 \pm 0.030$.
The nonspinning value is smaller than the spinning one, as expected. 
The larger error in the nonspinning case is only linked to the sampling of the parameter 
space; more targeted simulations would easily reduce the uncertainties. 

We repeat this analysis at the same energy, but for aligned spins, namely $\chi_1=\chi_2=0.5$.
We find that the transition from scattering to dynamical capture
occurs at $p_\varphi^* = 3.935 \pm 0.035$, i.e. at a lower orbital angular momentum 
with respect to the anti-aligned and nonspinning cases, as a priori expected.

\subsubsection{Nonspinning unequal mass systems}
\label{sbsbsec:phenom_q23_nospin}
Finally, we discuss nonspinning configurations with higher mass ratios. These 
configurations are more computationally expensive, since more resolution
has to be employed in order to correctly resolve the punctures. For this reason, we just focus on a few 
significant cases, and leave a more systematic exploration to future work. 
We consider three scattering-capture transitions at fixed energy, specifically
$E_0 \simeq\lbrace 1.016, 1.020 \rbrace \,M$ for $q=2$, and $E_0\simeq1.017\,M$ for $q=3$.
We further consider two spinning configurations with $q=3$ and $E_0 \simeq 1.025$. 

The (2,2) extrapolated Weyl scalars and the energetics for two series
are reported in the middle and right panels of Fig.~\ref{fig:series_q123}. 
The long duration of the $q=3$ configuration with $p_\varphi^0 \simeq 4.400$
poses integration issues that are similar to the ones discussed in the previous section.
Also in this case, considering $\psi_4$ extracted at $140\,M$ fixes the problems. 
The transition from scattering to captures occurs 
at $p_\varphi^* = 4.300 \pm 0.050$ for $q=2$ and $E\simeq 1.016\,M$, and 
at $p_\varphi^* = 4.410 \pm 0.012$ for $q=3$ and $E\simeq 1.017\,M$.
We recall that the width of the errors on $p_\varphi^*$ is just linked to the
sampling of the parameter space, and not to physical uncertainties.
The values obtained for these configurations are similar to
the equal mass nonspinning case, showing that the impact of the mass-ratio
on the scattering-capture threshold is smaller than the one related
to the spin.

\subsection{Properties of the remnants}
\label{sbsec:remnants}
\begin{figure*}[t]
  \centering 
    \includegraphics[width=0.325\textwidth]{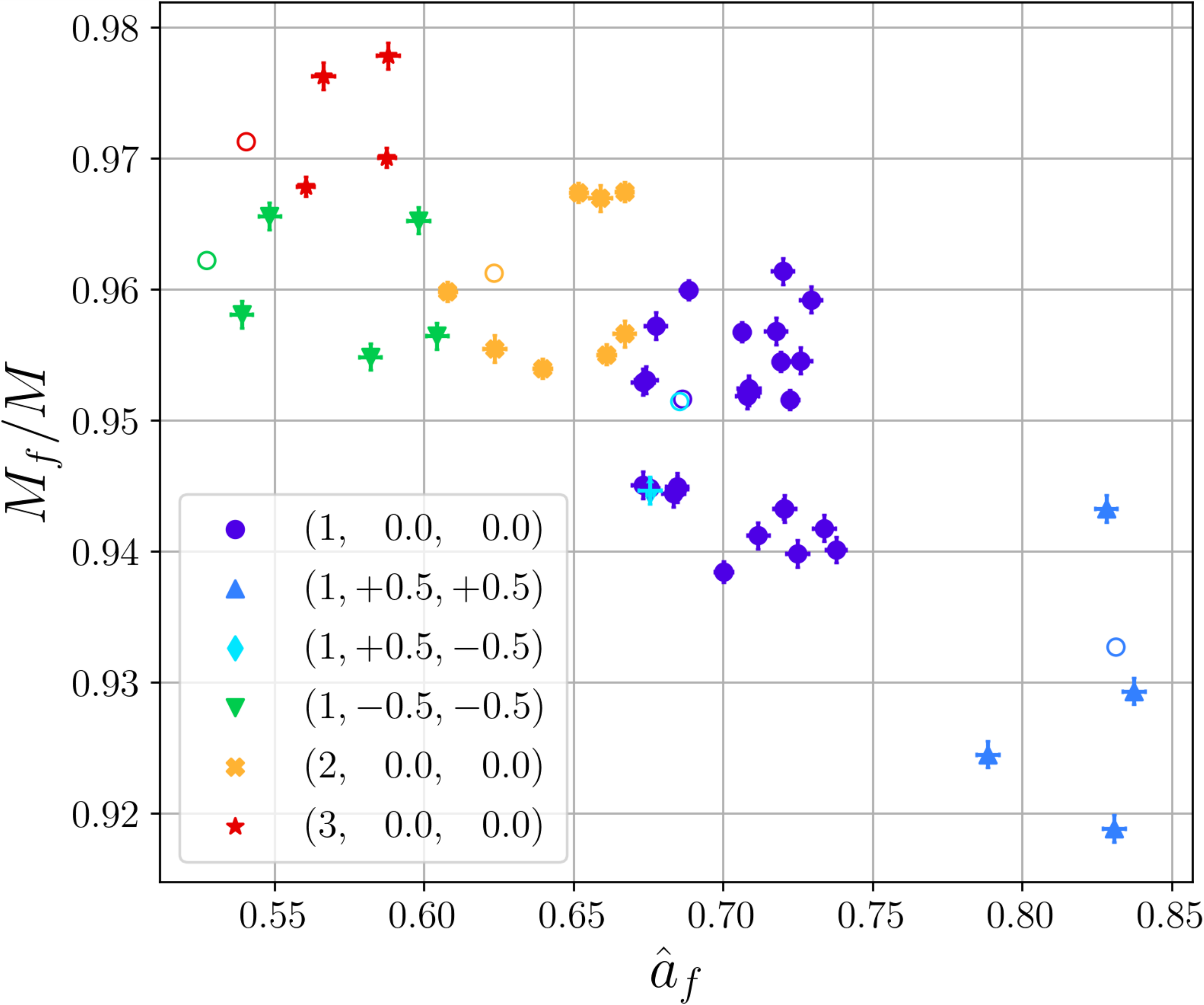}
    \includegraphics[width=0.320\textwidth]{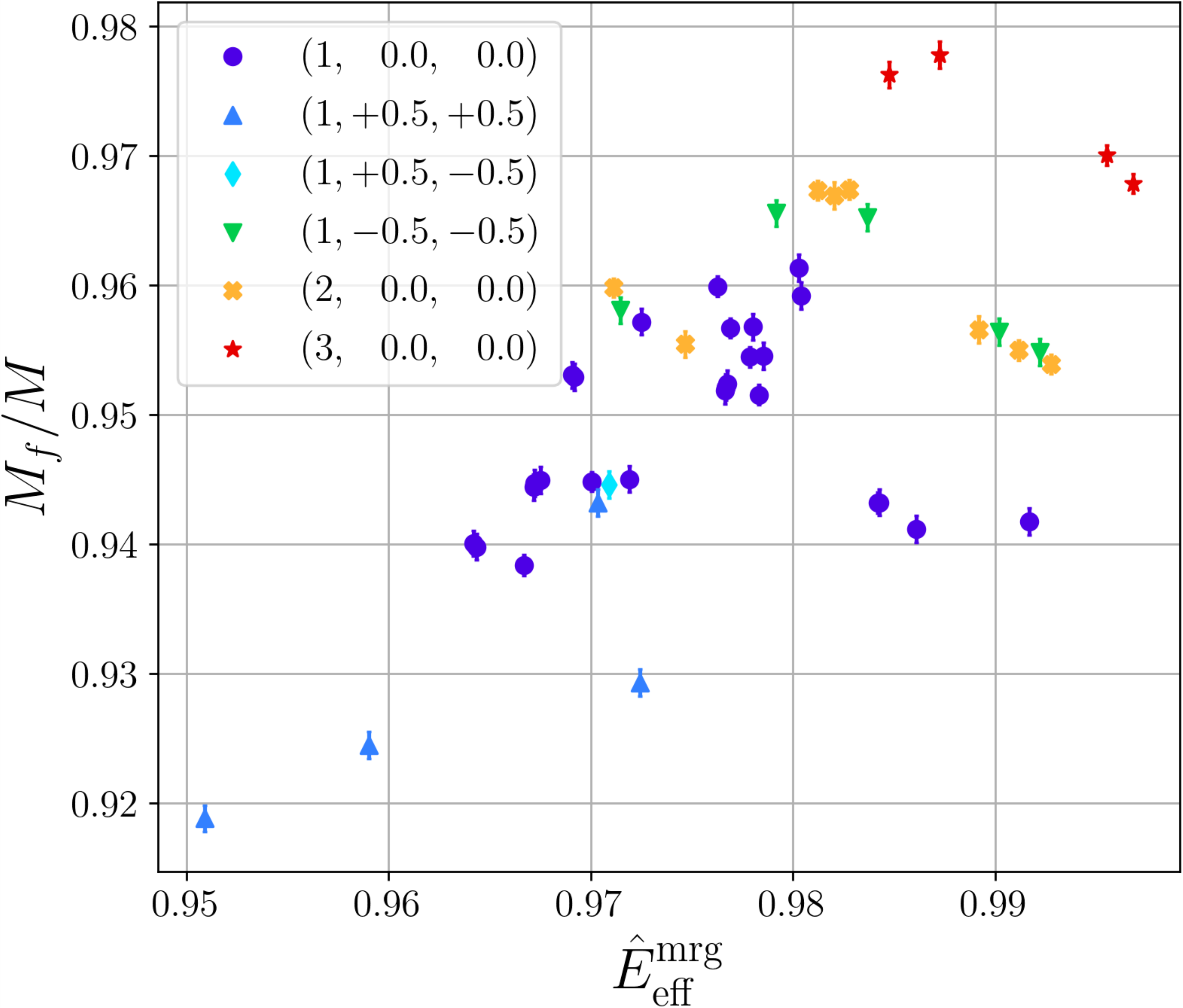}
    \includegraphics[width=0.325\textwidth]{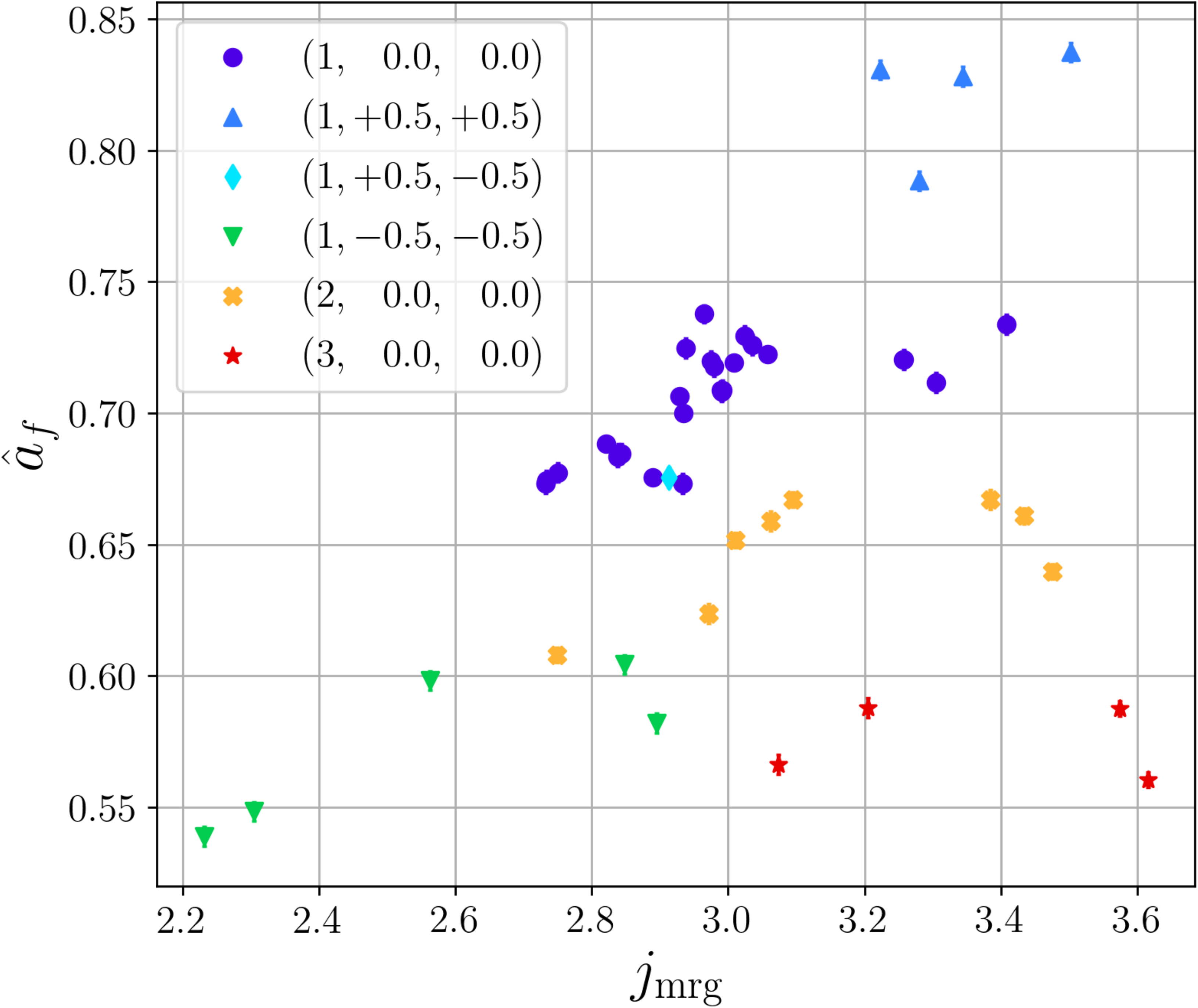}
    \caption{First panel: masses and spins of remnants generated by dynamical captures (filled markers).
    We also report the values for quasi-circular mergers as given 
    by the fits of Ref.~\cite{Jimenez-Forteza:2016oae,Nagar:2020pcj} (empty circles).
    Middle and right panels: remnant masses and spins plotted against the effective energy $\hat{E}_{\rm eff}$ and
    the $\nu$-rescaled angular momentum $j$, respectively. Both quantities on the $x$-axise are evaluated at merger.
    In the legends we report $(q,\chi_1,\chi_2)$.}
  \label{fig:remnants}
\end{figure*} 
We now discuss the properties of the remnants for coalescing configurations.
The corresponding masses and spins are reported in Tables~\ref{tab:runs_q1_nospin}
and~\ref{tab:runs_q1_spin_q23}, and shown in Fig.~\ref{fig:remnants},
where each color/marker type highlight a different $(q,\chi_1,\chi_2)$-combination. 
We also report, for reference, the quasi-circular values obtained from the post-merger fits of 
Refs.~\cite{Jimenez-Forteza:2016oae,Nagar:2020pcj} (empty circles).

Similarly to the quasi-circular case, the individual black hole spins strongly
influence the properties of the remnant.
Following Ref.~\cite{Carullo:2023kvj}, we plot the spins and
the masses against the $\nu$-rescaled angular momentum $j$ and 
the effective energy $\hat{E}_{\rm eff}$, respectively. Both quantities
are evaluated at the merger time $t_{\rm mrg}$.
The effective energy $\hat{E}_{\rm eff}$ 
is defined according to the usual effective-one-body map~\cite{Buonanno:1998gg}, 
\be
\hat{E}_{\rm eff} = 1 + \frac{\hat{E}^2-1}{2\nu}.
\ee
It is important to note that, in the test-mass limit, $\hat{E}$ reduces to the total
rest mass of the system, while $\hat{E}_{\rm eff}$ becomes the energy of the particle;
we come back to the (nonspinning) test-particle limit in Sec.~\ref{sec:testmass}.
To evaluate $j$ and $\hat{E}_{\rm eff}$ at merger, we integrate the angular momentum
and energy fluxes up to $t_{\rm mrg}$ and remove the radiated fluxes from the initial 
values $(J_0,E_0)$; then $(j_{\rm mrg}, \hat{E}_{\rm eff}^{\rm mrg})$ are
computed accordingly. We recall, in passing, that the description of the main
post-merger features for non-circular BBHs
can be achieved by considering the gauge-invariant dynamical impact parameter $\hat{b}_{\rm mrg}$, 
as discussed in Refs.~\cite{Albanesi:2023bgi,Carullo:2023kvj}. 

Binaries whose individual spins are anti-aligned with the orbital angular momentum 
produce heavier and slower-rotating remnants than binaries with nonspinning progenitors.
The opposite occurs for spin-aligned binaries, where the remnants are lighter and faster-rotating.
This scaling can be understood as follows. Progenitors with 
aligned spins ($\chi_i>0$) tend to have more stable orbits, and therefore spend
more time at small separations. They thus emit a larger amount of radiation, and ultimately 
produce less massive and faster-rotating remnants. On the contrary, 
the plunge starts at larger separations for binaries with anti-aligned spins ($\chi_i<0$), 
so that the two black holes spend 
less time close to each other. As a consequence, the system emits less gravitational radiation, 
generating a more massive and slower-rotating remnant.
A similar scaling is also observed for quasi-circular BBHs, where
nonspinning progenitors lead to $M_f = 0.952\,M$ and $\hat{a}_f= 0.686$, while
$\chi_i=0.5$ ($-0.5$) leads to $M_f = 0.933\,M$ (0.962) 
and $\hat{a}_f = 0.831$ (0.527)~\cite{Jimenez-Forteza:2016oae,Nagar:2020pcj}.

Finally, we examine the impact of the finite resolution 
and the integration procedure on the remnant properties. 
We start by considering the energetics for all the physical configurations
that have been run both at $N_M=128$ and $N_M=256$. We then compute the differences
between the remnant properties found at the two resolutions,
finding that the highest
absolute differences among equal mass binaries occur for 
the nonspinning configuration with $(\hat{E}_0, p_\varphi^0) \simeq (1.051,4.654)$, and
correspond to $(\Delta \hat{M}_f, \Delta \hat{a}_f) = (0.0013, 0.0036)$, where $\hat{M}_f=M_f/M$.
If we also consider the nonspinning unequal mass binaries, 
the maximum differences are
$\Delta M_f = 0.0014\,M$ and $\Delta \hat{a}_f = 0.0048$, and
occur for the $q=3$ configuration with
$(\hat{E}_0,\,p_\varphi^0) \simeq (1.017,4.398)$.
For all the configurations that have been run only at one resolution, 
we conservatively assign errors equal to the
maximum differences found in our dataset.  

To evaluate the error linked to the integration procedure, 
we consider a configuration that can be safely integrated both in the time
and frequency domains. We examine the case with $(\hat{E}_0,p_\varphi^0)\simeq(1.011,3.950)$,
obtaining $(\hat{M}_f^{\rm TDI}, \hat{a}_f^{\rm TDI}) = (0.9417, 0.6351)$
and $(\hat{M}_f^{\rm FFI}, \hat{a}_f^{\rm FFI}) = (0.9432, 0.6409)$, so that
the absolute differences are $(\Delta \hat{M}_f, \Delta \hat{a}_f) = (0.0015,0.0058)$. Note
than these errors are slightly larger that the maximum discrepancies associated
to the resolution.

The total error on the remnant properties is finally computed
as the square root of the quadrature sum of the resolution and integration errors;
they are shown in in Fig.~\ref{fig:remnants}, and also reported
in Tables~\ref{tab:runs_q1_nospin} and~\ref{tab:runs_q1_spin_q23}.

\section{Accuracy of EOB models for scatterings and dynamical captures} 
\label{sec:eob_nr}
\begin{figure*}[t]
  \centering 
    \includegraphics[width=0.24\textwidth]{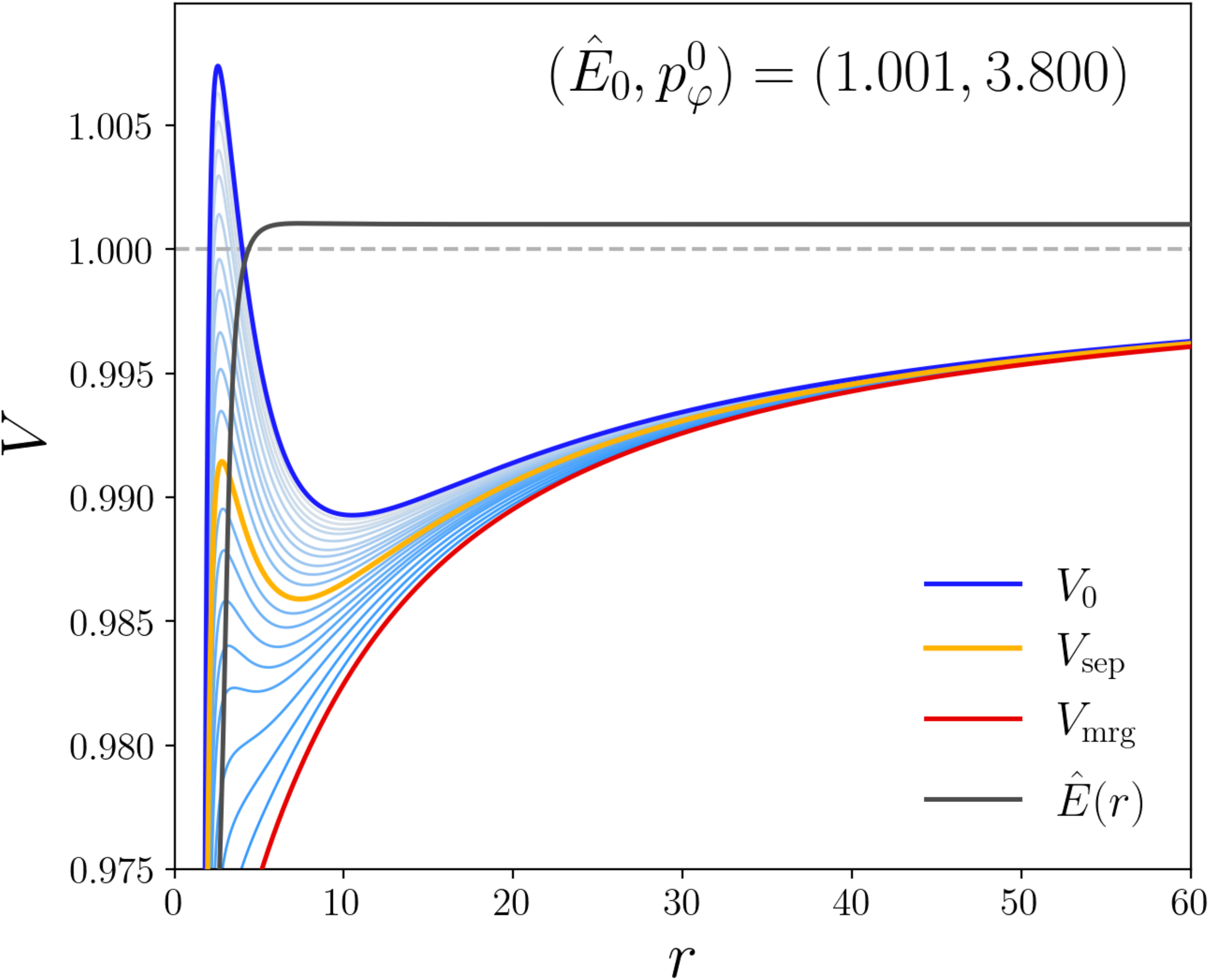}
    \includegraphics[width=0.24\textwidth]{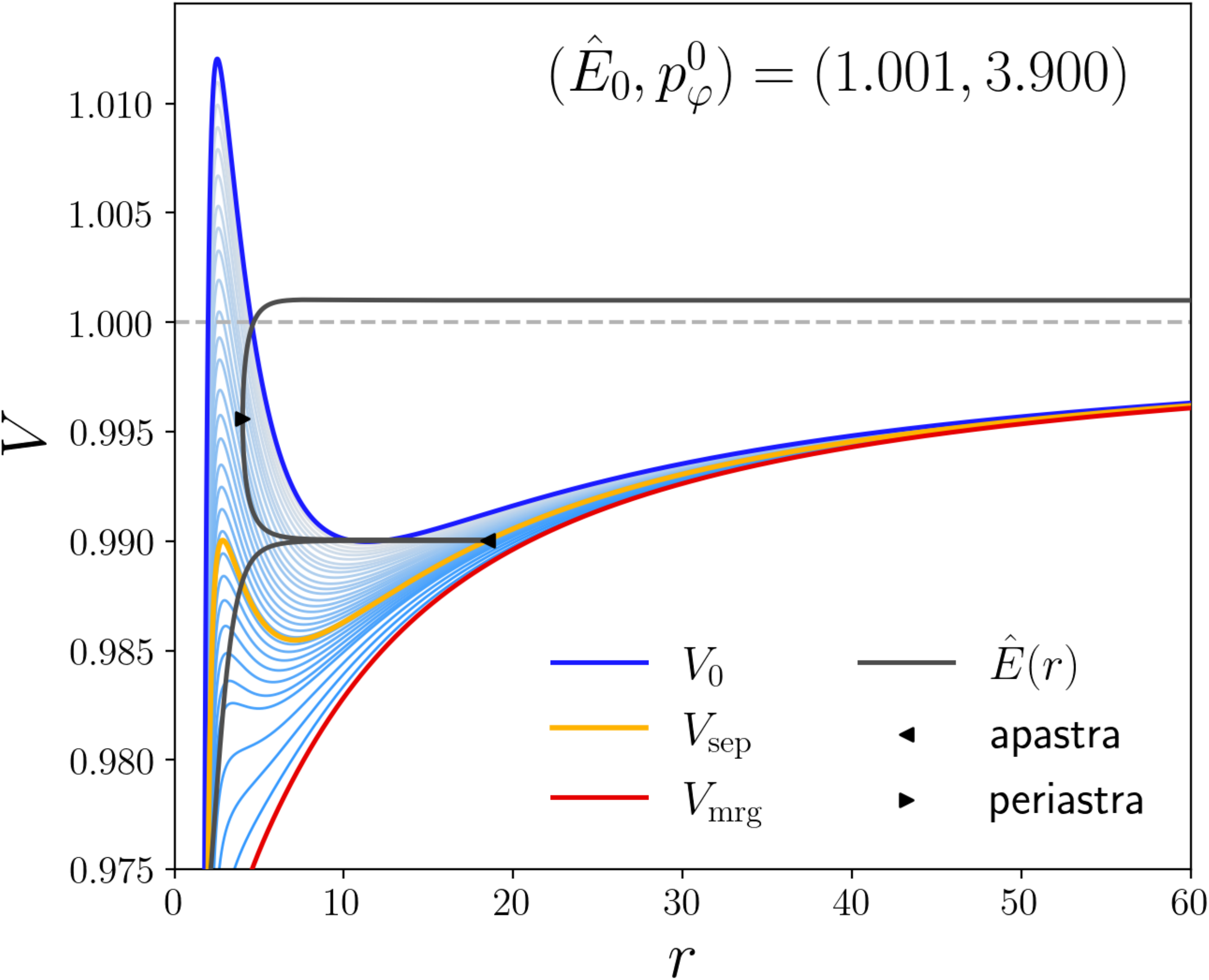}
    \includegraphics[width=0.24\textwidth]{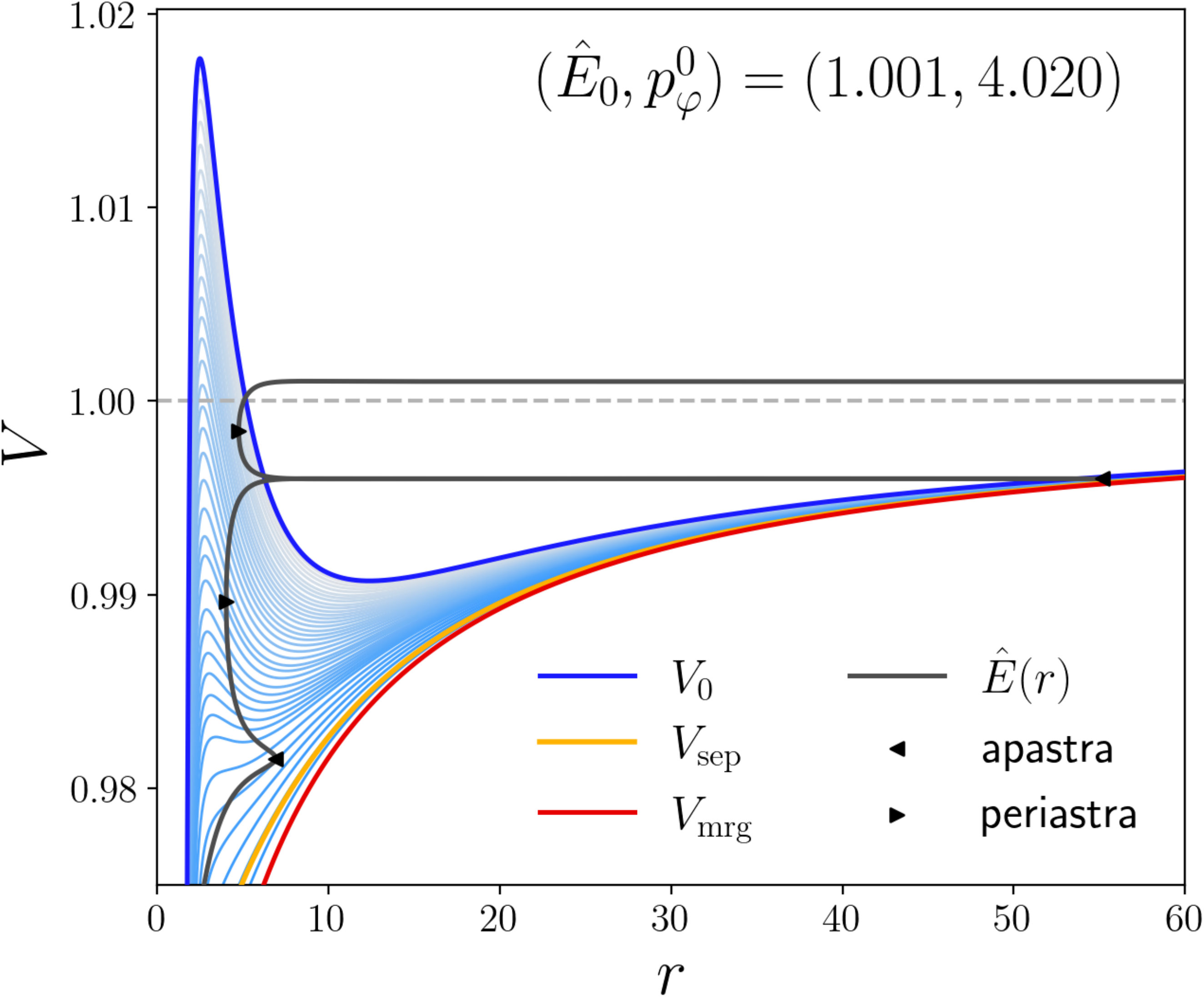}
    \includegraphics[width=0.24\textwidth]{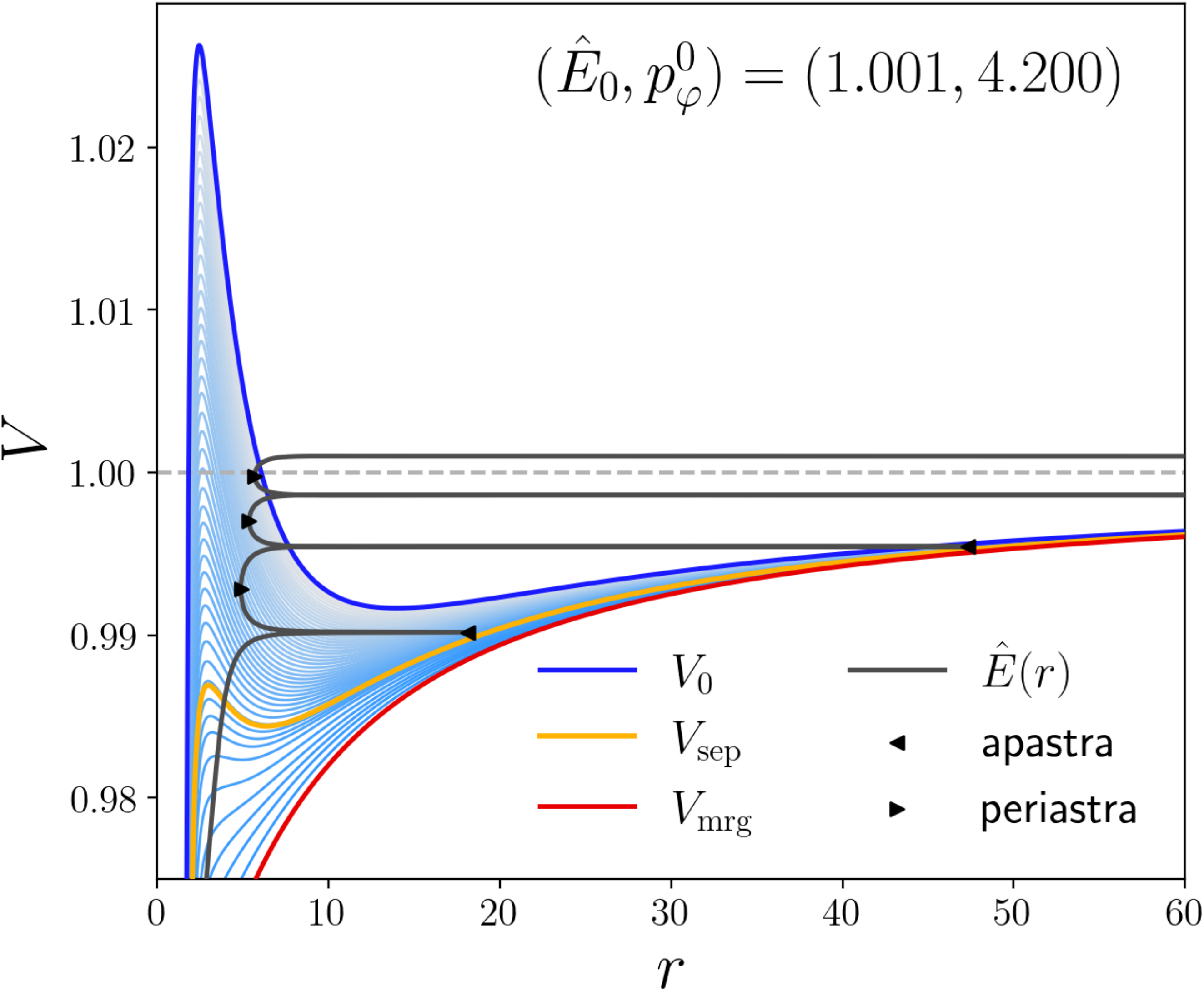}
    \caption{Time evolution of effective EOB potentials $V(r;p_\varphi)$  for different physical configurations
    with $\hat{E}_0=1.001$. From left to right, we have a direct capture, followed by a double encounter, 
    a triple encounter, and finally, a quadruple encounter.     
    The evolution of the energy $\hat{E}$ (dark gray) is overlapped to radial potentials (blue shades). 
    We highlight the initial potential (dark blue), 
    the potential at the separatrix-crossing (orange),
    and the potential at merger (red). 
    We also mark apastra and periastra, when defined.
    The first apastron of the quadruple encounter is not shown for graphical reasons, 
    since it occurs at $r \sim 172$.}
  \label{fig:eob_potentials}
\end{figure*}
Numerical relativity stands as the golden standard for generating 
accurate waveforms for compact binaries. However, exploring large parameter spaces,
as the one studied in this work, is extremely expensive
from the computational point of view. 
Herein lies the advantage of (semi-)analytical methods, 
that offer a more manageable approach, albeit at the cost of 
introducing analytical approximations. Furthermore,
these methods often incorporate calibrations performed on quasi-circular binaries. 
Therefore, to ensure the region of validity of analytical models, it is necessary
to assess their reliability through comparisons with NR simulations. 
In this paper we consider the EOB framework, and in particular we test 
the reliability of the \DALI{} model~\cite{Nagar:2023zxh,Nagar:2024dzj} in the scattering and dynamical capture
scenarios. Previous versions of this model have been already tested for these kind of
configurations in Refs.~\cite{Damour:2014afa,Gamba:2021gap,Hopper:2022rwo,Damour:2022ybd,Andrade:2023trh}, 
but here we extend this analysis by considering a larger portion of the orbital parameter space, 
and focusing on the transition from scattering to plunge. 
We also mention that series of scattering configurations at fixed energy could be used to test
PM predictions~\cite{Damour:2022ybd,Rettegno:2023ghr}, but we 
do not carry out this analysis here, since the energies for the scatterings considered in this work 
are similar to the ones analyzed in Ref.~\cite{Rettegno:2023ghr} (see Table I therein).


\subsection{Effective-one-body model}
\label{sbsec:eob}
For a complete and up-to-date description of \DALI{}, we point 
to Refs.~\cite{Chiaramello:2020ehz,Nagar:2023zxh,Nagar:2024dzj}.
Here we describe the most relevant aspects for our analysis. 
The basic idea of EOB models is to map the two-body problem into 
the motion of a single particle that moves in an effective metric. This metric 
is a $\nu$-deformation of Schwarzschild or Kerr, depending on whether 
the spins of the black holes are vanishing or not.   
The map is created by encoding the conservative PN equations of motion in an EOB 
Hamiltonian~\cite{Buonanno:1998gg,Damour:2001tu}. 
Notably, by setting the symmetric mass ratio to zero in the metric, and thus in the Hamiltonian,
we recover the motion of a test-particle in Schwarzschild or Kerr. In other words,
the comparable mass regime is smoothly connected to the test-mass limit by $\nu$. This limit 
is further discussed in Sec.~\ref{sec:testmass} for the non-spinning case.
State-of-the-art EOB model use 5PN-accurate Hamiltonians, which are completed with numerically-informed
coefficients. In particular, \DALI{} includes two free effective coefficients, $a_6$ and $c_3$,
that are informed by nonspinning and spinning NR simulations, respectively~\cite{Nagar:2023zxh,Nagar:2024dzj}.
This calibration is performed on {\it quasi-circular} waveforms of the SXS catalog~\cite{SXS:catalog,Buchman:2012dw,Chu:2009md,Hemberger:2013hsa,Scheel:2014ina,
Blackman:2015pia,Lovelace:2011nu,Lovelace:2010ne,Mroue:2013xna,Lovelace:2014twa,Kumar:2015tha,Lovelace:2016uwp,Chu:2015kft,Varma:2018mmi}.
The dissipative effects due to gravitational wave emission, that starts from 2.5PN, are
taken into account including in the Hamilton's equation a radiation reaction
force obtained through energy-balance equations~\cite{Bini:2012ji}. Note that \DALI{}
consider the quasi-circular limit of radial radiation reaction valid for generic orbits. 
The angular component of the radiation reaction is the standard quasi-circular prescription
of \TEOB{\texttt{-Giotto}}, dressed with a {\it generic} quadrupolar Newtonian prefactor~\cite{Chiaramello:2020ehz},
which generalizes ${\cal F}_\varphi$ to generic planar orbits. The reliability of these prescriptions
have been extensively tested, both in the test-mass limit and in the 
comparable mass case~\cite{Chiaramello:2020ehz,Albanesi:2021rby,Albanesi:2022ywx,Nagar:2023zxh}.
The EOB dynamics is then obtained solving the corresponding
Hamilton's equations. While analytical methods are available for early quasi-circular inspirals~\cite{Nagar:2018gnk},
these equations have to be solved numerically for quasi-circular late inspirals or elliptic/hyperbolic-like orbits.
The EOB dynamics is then used to compute the multipolar waveform at infinity. 

Each multipole of the analytical inspiral waveform includes a generic Newtonian prefactor.
We further discuss the inspiral EOB waveform used in \DALI{} in Sec.~\ref{sbsec:rwz_eob}.
The analytical wave is then completed with a ringdown model,
that describes the waveform generated after the merger of the two black holes. The smooth
match between the inspiral and ringdown waveforms is typically achieved through Next-to-Quasi-Circular (NQC)
corrections. Both the ringdown model and the NQC corrections in \DALI{} are based on quasi-circular NR simulations.
For this reason, in this work we switch-off the NQC corrections, as done in the analysis
of GW190521~\cite{LIGOScientific:2020iuh} performed in Ref.~\cite{Gamba:2021gap}. 

As anticipated, our goal is to test the reliability of \DALI{} 
for scatterings and dynamical captures, without incorporating numerical non-circular 
information into the semi-analytical model. However, 
the merger-ringdown waveform could be improved considering numerical simulations of 
highly eccentric binaries and dynamical captures~\cite{Andrade:2023trh,Carullo:2023kvj}.
The enhancement of the conservative sector via numerical information should also be studied, but
no calibration on NR simulations with generic orbits has been achieved so far.
Both generalizations are left to future works.

As typically done in the EOB literature, we consider rescaled EOB phase space 
variable $(r,p_{r_*},\varphi,p_\varphi)$, which are related to the physical ones by $r=R/M$, 
$p_{r_*}=P_{r_*}/\mu$, and $p_\varphi=P_\varphi/(\mu M)$. 
The latter variable has been already extensively used in this work. 


\subsection{Orbital parameter space}
\label{sbsec:eobnr_parspace}
\begin{figure}[t]
  \centering 
    \includegraphics[width=0.48\textwidth]{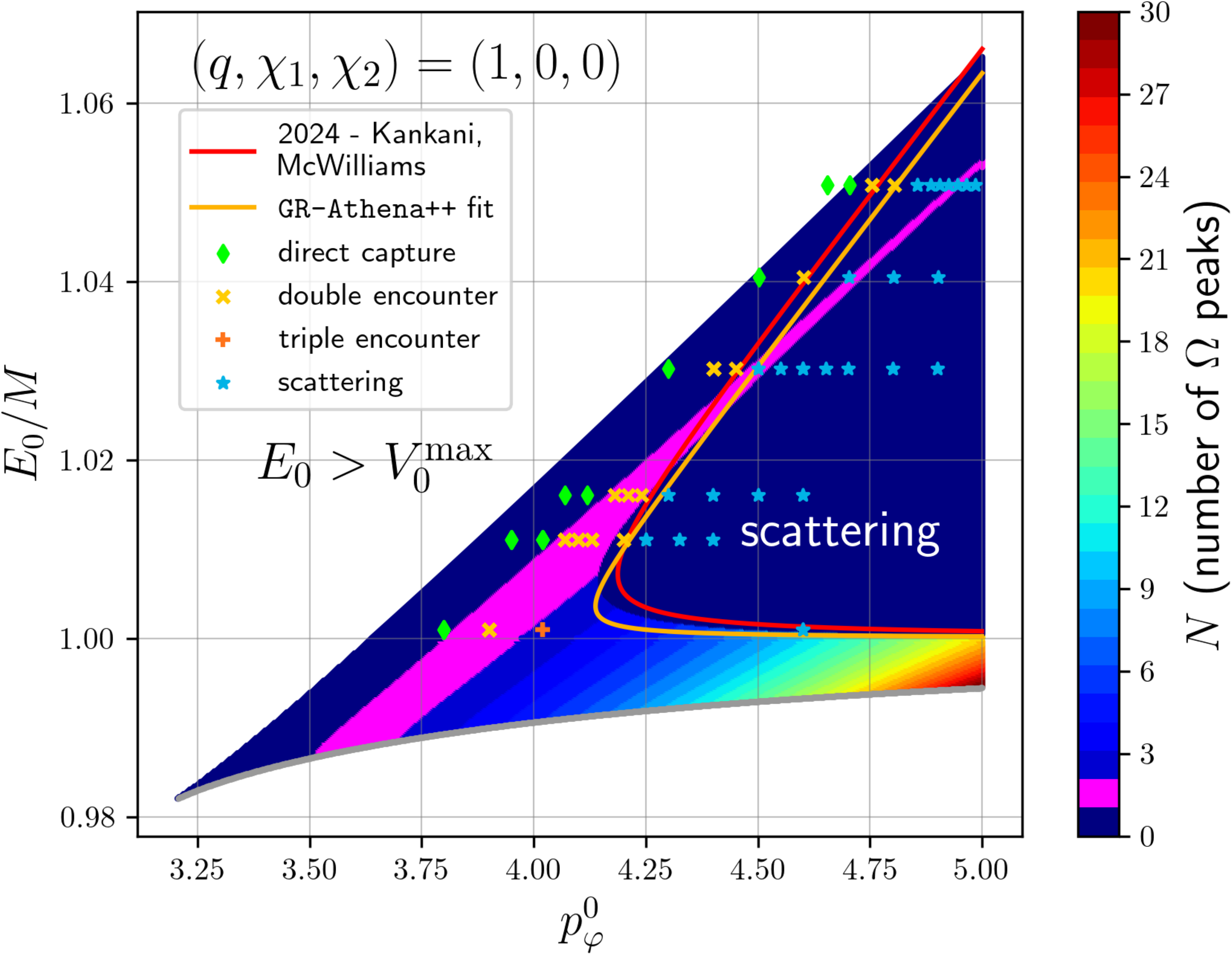}
    \caption{Phenomenologies of equal mass nonspinning systems across the parameter space.
    The colormap highlight the number of peaks of the orbital frequency $N$, 
    that is a proxy of the number of encounters, as predicted by \DALI{}.
    Quasi-circular binaries are highlighted with a gray line.
    The NR simulations considered in this work are reported with markers and colors that correspond to
    different phenomenologies. We also show the scattering-capture fits discussed in
    Sec.~\ref{sbsbsec:phenom_q1_nospin}.}
  \label{fig:parspace_TEOB_q1_nospin}
\end{figure}
\begin{figure*}
  \centering 
    \includegraphics[width=0.24\textwidth]{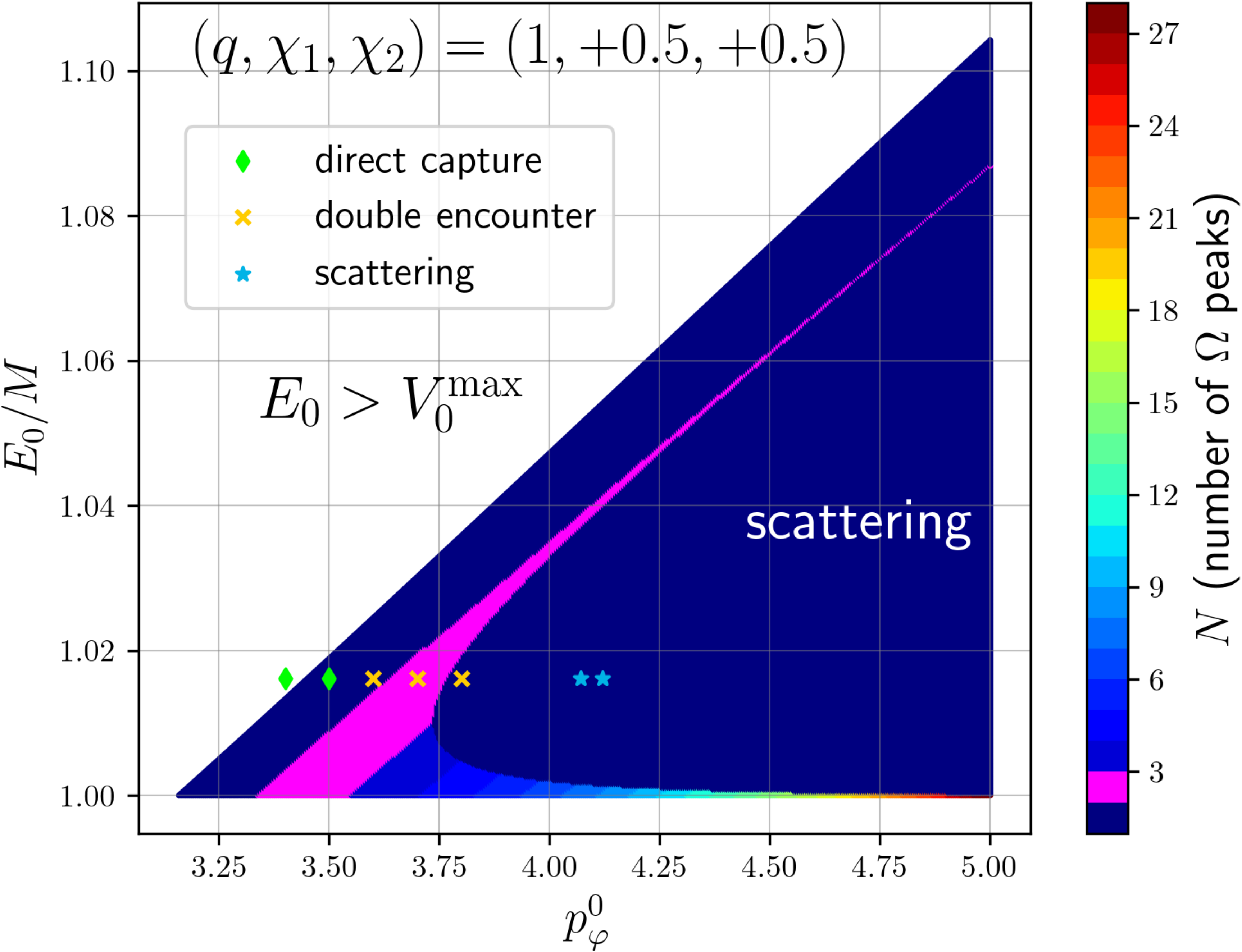}
    \includegraphics[width=0.24\textwidth]{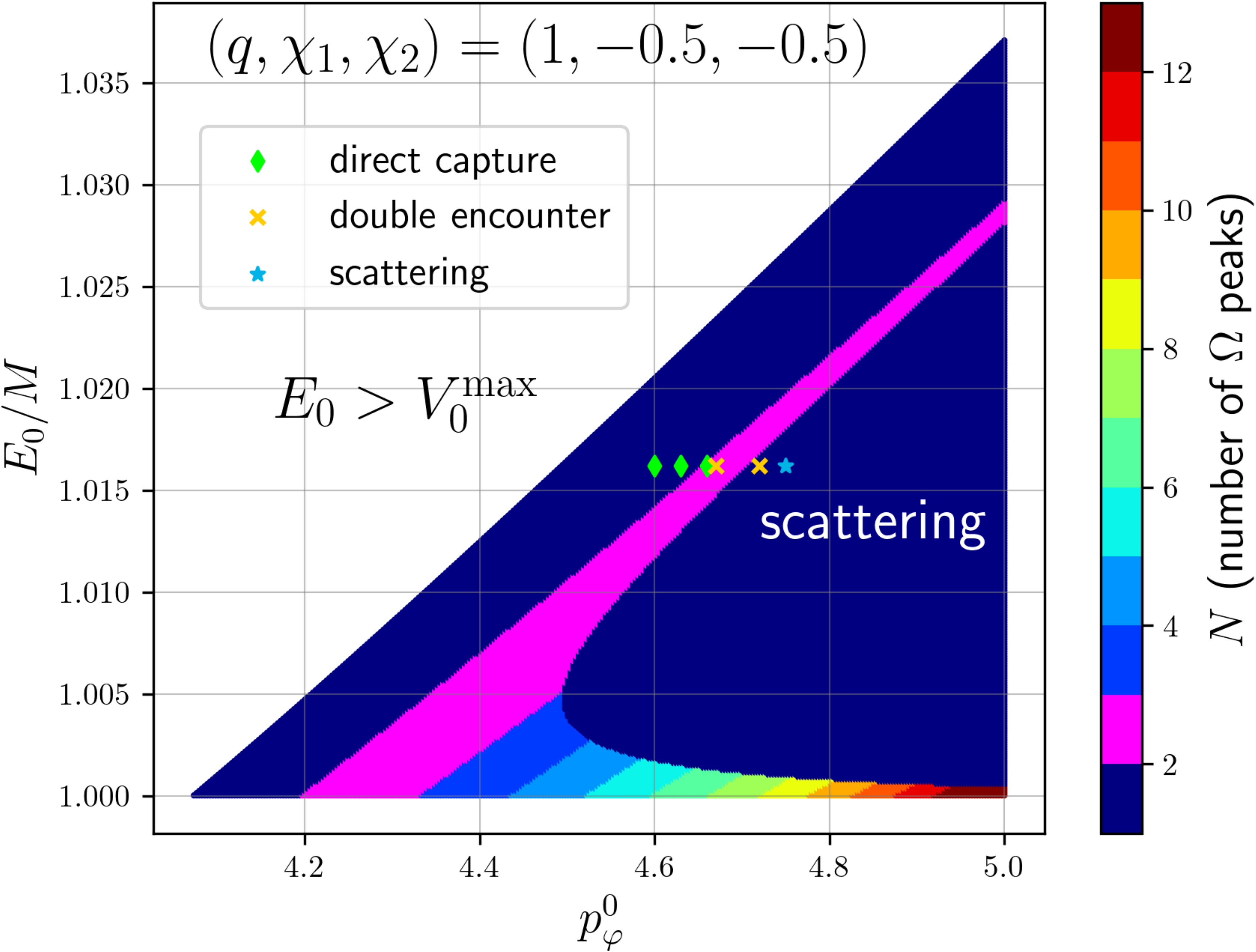}
    \includegraphics[width=0.24\textwidth]{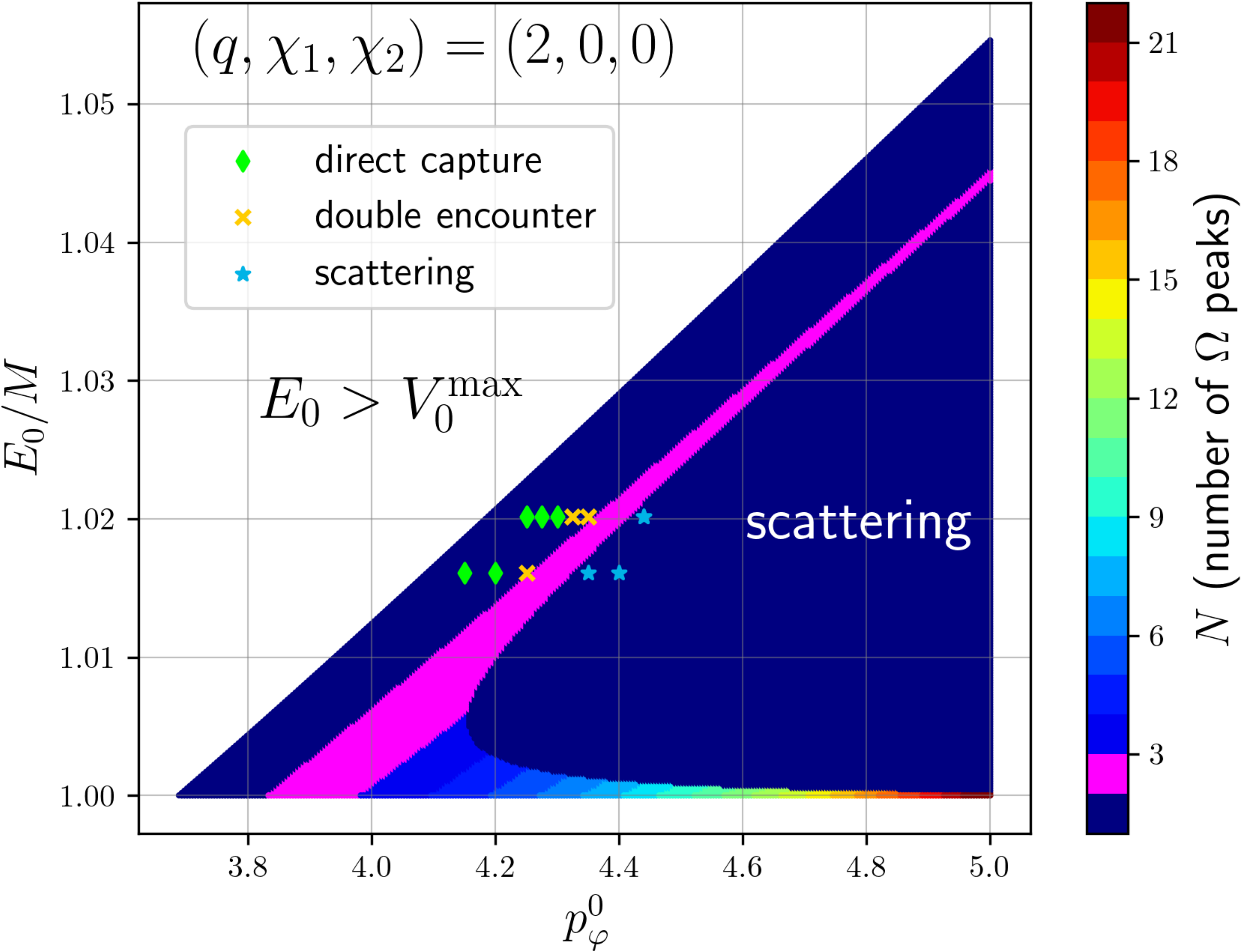}
    \includegraphics[width=0.24\textwidth]{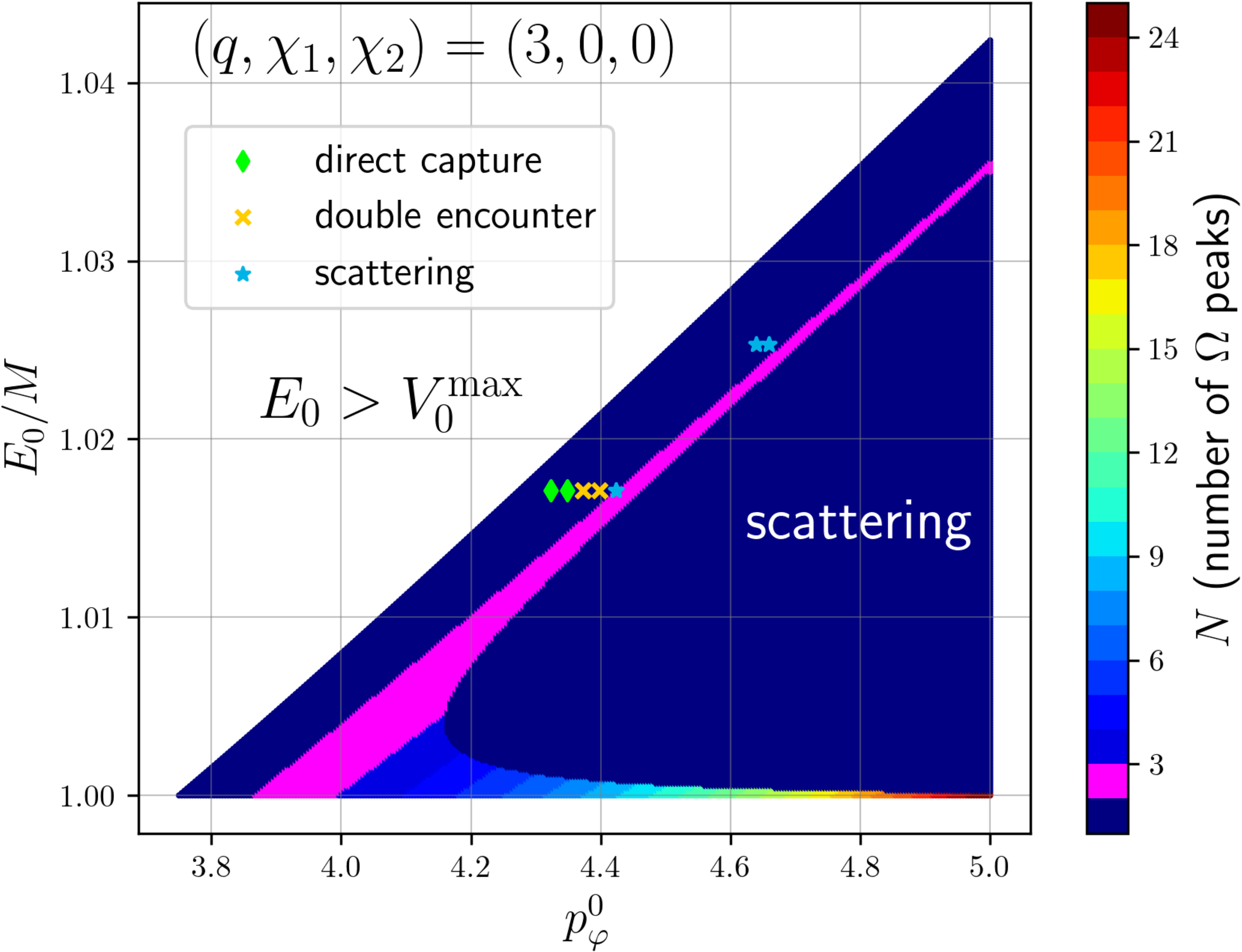}
    \caption{Phenomenologies for different $(q,\chi_1,\chi_2)$ configurations; 
    analogous to Fig.~\ref{fig:parspace_TEOB_q1_nospin}, but only for initially
    unbound orbits. The colormap highlights the number of orbital frequency peaks
    as predicted by \DALI{}, while the markers correspond to the numerical simulations.
    Note the different vertical scales.}
  \label{fig:parspace_TEOB_4}
\end{figure*}
An extensive exploration of the parameter space can be easily achieved with semi-analytical models
that are able to generate hundreds of waveforms in less than a second
on conventional consumer-grade computing machines. Such an exploration
for initially unbound configurations has been
performed in Ref.~\cite{Nagar:2020xsk} for nonspinnig systems with different
mass ratios. However, the reliability of these
predictions has to be checked using computationally expensive NR simulations. 

We consider both bound and unbound initial data, examining 
pairs of energy and angular momentum that satisfy 
$V_{\rm min}^0 \leq E_0 \leq V_{\rm max}^0$, where $V_{\rm min}^0$ ($V_{\rm max}^0$)
is the minimum (maximum) of the initial EOB radial potential $V(r;p_\varphi)$.
The latter is defined 
by setting to zero the radial momentum in the EOB Hamiltonian,
\be
V(r;p_\varphi) \equiv H_{\rm EOB}(r,p_\varphi, p_{r_*}=0). 
\ee
The condition $E_0=V_{\rm min}^0$ thus identifies stable circular orbits, that slowly evolve
under the effects of the radiation reaction;
configurations with smaller energies are not allowed.  
On the other hand, orbital configurations with $E_0>V_{\rm max}^0$ can
occur, but in this case we aprioristically know that we will have a direct capture. 
We thus focus on initial energies that lie between the initial minimum and 
maximum values of the effective potential. 

Since GWs are emitted, the system loses angular momentum, 
causing the effective potential $V(r;p_\varphi)$ to change over time.
Analyzing the evolution of this potential provides valuable insights into the dynamics of the system,
as already shown, for example, in Ref.~\cite{Albanesi:2023bgi}, where test-particles 
orbiting around Schwarzschild black holes along eccentric orbits were considered.
Here, instead, we focus on configurations with comparable masses. The evolutions of energies
and effective potentials for four initially unbound, equal mass, nonspinning systems are illustrated in 
Fig.~\ref{fig:eob_potentials}.
For all the cases, we consider an energy slightly above the parabolic limit, $\hat{E}_0 = 1.001$, 
and select the angular momenta in order to obtain a direct capture and configurations with two, three and four
encounters. Note that first three configurations have been also simulated numerically with \GRA{}.
We also highlight the potential at three specific times: at the beginning of the evolution
(blue), at the separatrix-crossing (orange), and at merger time (red). The latter
quantity is estimated from the peak of the pure orbital frequency (without
spin-orbit coupling) as $t_{\rm mrg}^{\rm EOB} =  t^{\rm peak}_{\Omega_{\rm orb}}-3$~\cite{Nagar:2020pcj}.
We recall that the separatrix is
an eccentric generalization of the last stable orbit~\cite{OShaughnessy:2002tbu,Stein:2019buj}, 
and stable orbits are no longer allowed after the crossing of this quantity; this crossing can be identified,
in terms of the potential, as $\hat{E}=V_{\rm max}$. 
In the case of the direct capture show in Fig.~\ref{fig:eob_potentials} ($p_\varphi^0=3.800$), 
enough radiation is emitted to cause the separatrix to be crossed at the first encounter, 
leading the system to plunge without experiencing any outgoing motion.
The other configurations are instead more interesting:
the systems still become bound at the first encounter, but conserving the property $\hat{E}<V_{\rm max}$. As a
consequence, after the first close encounter we have 
radial motion confined between the two radial turning points, the apastron and the periastron.
As long as the binding energy
is negative and $\hat{E}<V_{\rm max}$, 
the two radial turning points can be computed by solving $\hat{E}=V(r)$ at any time during the evolution. 
Once that the separatrix is crossed,
the periastron ceases to exist also for these configurations, leading the two black holes to merge.
We conclude the discussion on the effective potentials by mentioning that in the scattering
scenario, we simply have a rebound on the (evolving) potential barrier.

We now start to span the equal mass nonspinning parameter space more systematically.
For initially unbound configurations, we consider a very large radius at which 
the influence of the radiation reaction force is completely negligible ($r_0 = 5\cdot 10^{3}$ for 
practical purposes). For bound configurations, we start the EOB evolution at the apastron,
which is uniquely determined by $(E_0,p_\varphi^0)$. The phenomenology of the waveform 
can be tracked by looking at the number of peaks of the orbital frequency $\Omega\equiv\dot{\varphi}$,
denoted as $N$.
Each peak marks a close encounter between the two black holes, so that $N\geq2$ marks
double and multiple encounters, while $N=1$ marks either direct captures or scatterings. 
The result for the nonspinning equal mass case is reported in Fig.~\ref{fig:parspace_TEOB_q1_nospin}
up to $p_\varphi^0=5$. Quasi-circular configurations are highlighted 
with a gray line. 
The region in which we have two encounters (magenta)
shrinks as the energy increases, sharpening the transition from scatterings
to direct captures. This implies that, according to the analytical prediction, 
double encounters are more likely to occur for low energies.

On the same figure, we report the numerical results. 
We show all the equal mass nonspinning configurations considered in this
work. The markers and the colors highlight different phenomenologies:
direct captures are highlighted with green diamonds, double encounters 
with yellow crosses, and scatterings with blue stars; the only 
triple encounter of our dataset is marked with an orange cross. 
While the NR configurations are started at a smaller separation ($D\sim 100\, M$), 
than the EOB predictions ($r_0 = 5\cdot 10^{3}$), the effects of the radiation
reaction (and the junk radiation) are already negligible at the initial NR separation.

Especially for low energies, the phenomenologies predicted by \DALI{}
are in agreement with the corresponding numerical ones. 
Notably, the EOB model predicts the correct number of close passages also for the triple
encounter. However, as we further discuss below, this does not guarantee a low EOB/NR mismatch, since 
the encounters can occur at rather different times. 
The reliability of the EOB waveform is further discussed in Sec.~\ref{sbsec:eobnr_mm}.

The disagreement between the EOB prediction and the NR simulations grows as the initial energy increases. 
In particular,
for high energy the NR transition from scattering to merger occurs at lower orbital angular momenta 
than predicted by the EOB model.
This behavior is better highlighted by the fits of this transition,
that have been already discussed in Sec.~\ref{sbsbsec:phenom_q1_nospin}.
It is interesting to observe that the slope of these fits
diverges from the double encounter region predicted by the EOB model for high-energy. 
As a consequence, in the high-energy region of the parameter space, we can have NR scatterings 
that occur at energies above the maximum of the EOB potential, i.e. with $E_0>V^0_{\rm max}$.
Since the potential is computed from the EOB Hamiltonian  $H_{\rm EOB}$, this means
that even the solely conservative sector of the EOB model is no longer accurate for high energies.
This does not come as a surprise, since $H_{\rm EOB}$ is based on PN results 
that lose their reliability at high relative velocities. Moreover, the calibration 
of the effective parameters $a_6$ and $c_3$ is performed only on quasi-circular binaries.
Higher-order PN information or PM results, together with a calibration
performed on a wider region of the parameter space, could mitigate this problem.
This exploration, which is far from trivial, is left to future work. 
Moreover, while the aforementioned observation highlight the limitation
of the EOB Hamiltonian, also the radiation reaction force
is PN-approximate; more accurate fluxes would probably improve
the EOB/NR agreement. The interplay between conservative and dissipative 
sectors require in depth studied, as already highlighted in Ref.~\cite{Nagar:2024dzj}
for initially bound configurations.

We extend this analysis also to the spinning cases with $\chi_{\rm eff}=\pm 0.5$
and to the the nonspinning $q=\lbrace 2,3\rbrace$. The corresponding parameter
spaces as predicted by \DALI{}, together with the numerical results, are
reported in Fig.~\ref{fig:parspace_TEOB_4}. Even if for these values of $(q,\chi_1,\chi_2)$ we have
simulated less orbital configurations than for the equal mass nonspinning case, a few considerations
can be made.
First of all, the numerical simulations confirm that \DALI{} is able to reproduce
the phenomenologies of spinning scatterings and dynamical captures, 
at least for the energies and spins considered. However, one should also notice that
the double encounter region ($N=2$, magenta) was captured with a higher accuracy in 
the nonspinning case (cfr. with Fig.~\ref{fig:parspace_TEOB_q1_nospin}). 
Second, the accuracy of the EOB model seems to degrade for higher mass ratios, since
all the three $q=3$ scattering configurations 
are in a region where \DALI{} predicts dynamical captures. However, two of them 
are at the relatively high energy $E_0 \simeq 1.025 \,M$, and all of them 
have large scattering angles, $\chi>327^\circ$ (see Table~\ref{tab:runs_q1_spin_q23}). 
The EOB/NR agreement for the $q=2$ case is instead better, 
and the three corresponding scattering configurations
are indeed in the unbound region of \DALI{}. However, in the $q=2$ case, all the 
scattering angles are below $310^\circ$.
A more detailed analysis of spinning configurations and unequal mass systems
is postponed to future work. 

%
\begin{figure}[t]
  \centering 
    \includegraphics[width=0.48\textwidth]{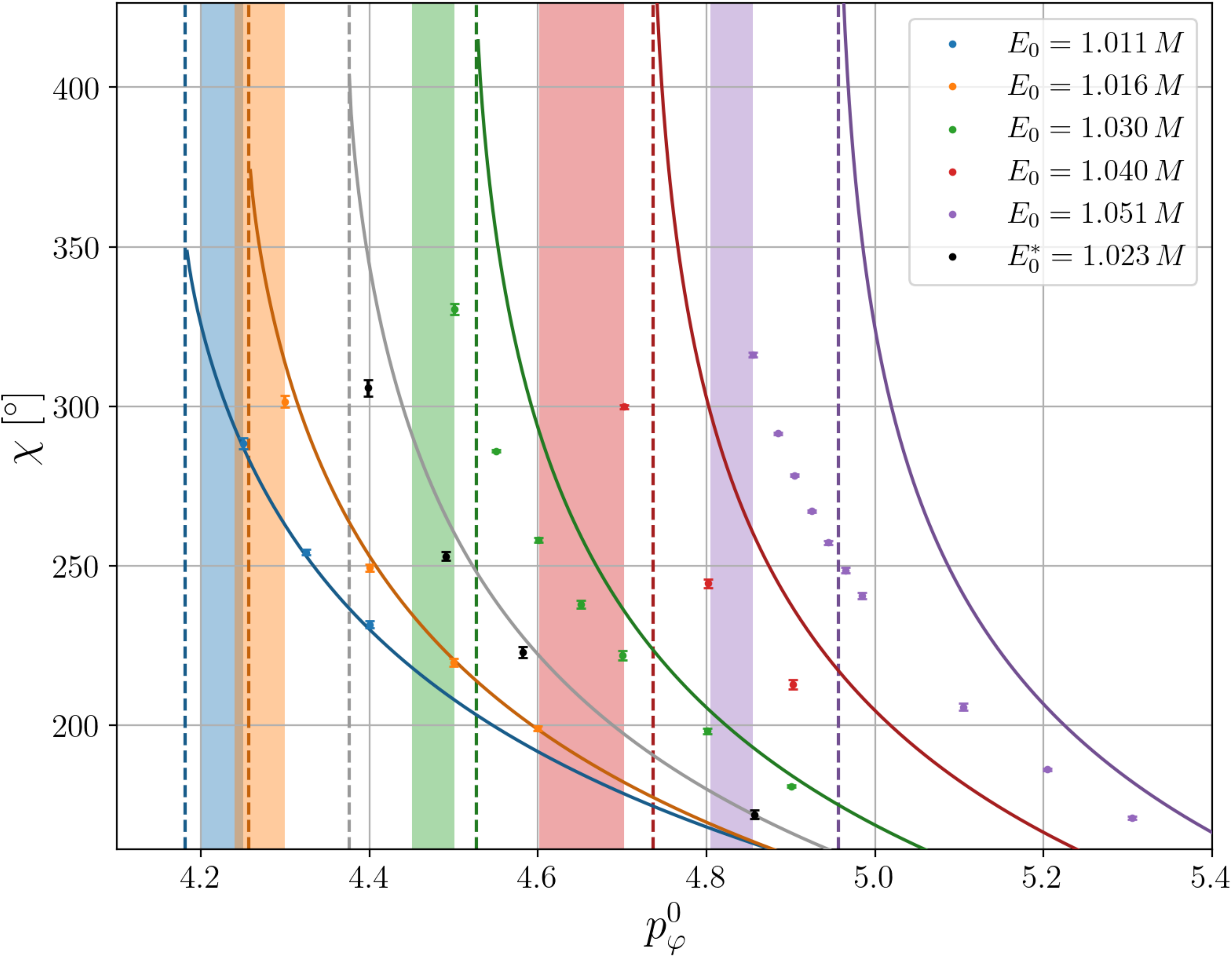}
    \caption{Scattering angles $\chi$ for equal mass nonspinning systems. 
    Different colors highlight different initial energies.
    Markers and error bars correspond to the \GRA{} results. The colored
    bands highlight the region in which the NR scattering-capture transition occurs.
    We also report the EOB prediction using lines of the same colors, showing a
    rather good agreement
    in the low energy regime $E_0\lesssim 1.02$, even for high scattering angles.
    The vertical dashed lines mark the EOB scattering-capture transition. 
    The NR data for $E_0\simeq1.023\,M$ are taken from Ref.~\cite{Damour:2014afa}.}
  \label{fig:scat_q1_nospin}
\end{figure}
\subsection{Nonspinning equal mass scatterings}
\label{sbsec:eobnr_q1_nospin_scat}
\begin{table}[t]
	\caption{\label{tab:scats_q1_nospin}
	EOB/NR scattering angles for the equal mass nonspinning configurations considered in 
	this work. We also include the configurations of Ref.~\cite{Damour:2014afa} ($E_0 \simeq 1.023\,M$).}
\begin{center}
\begin{ruledtabular}
\begin{tabular}{c c | c | c } 
$\hat{E}_0$ & $p_\varphi^0$	& $\chi_{\rm NR}\pm \Delta \chi_{\rm NR} \,[{}^{\circ}]$ & $\chi_{\rm EOB} [{}^\circ]$ \\
\hline
\hline
1.01103 & 4.24996 & $288.4 \pm 1.7 $ & $287.3 $ \\
1.01103 & 4.32496 & $254.3 \pm 0.8 $ & $253.0 $ \\
1.01103 & 4.39996 & $231.6 \pm 1.1 $ & $230.1 $ \\
\hline
1.01607 & 4.30002 & $301.5 \pm 1.8 $ & $313.4 $ \\
1.01607 & 4.40002 & $249.4 \pm 1.1 $ & $253.0 $ \\
1.01607 & 4.50002 & $219.7 \pm 1.2 $ & $220.6 $ \\
1.01607 & 4.60002 & $199.1 \pm 0.7 $ & $198.7 $ \\
\hline
1.02256 & 4.39861 & $305.8 \pm 2.6 $ & $345.8 $ \\
1.02257 & 4.49039 & $253.0 \pm 1.4 $ & $265.6 $ \\
1.02258 & 4.58209 & $222.9 \pm 1.7 $ & $227.7 $ \\
1.02259 & 4.85709 & $172.0 \pm 1.4 $ & $171.8 $ \\
1.02259 & 5.04039 & $152.0 \pm 1.3 $ & $151.3 $ \\
1.02259 & 5.49863 & $120.7 \pm 1.5 $ & $120.0 $ \\
1.02259 & 5.95687 & $101.6 \pm 1.7 $ & $101.2 $ \\
1.02259 & 6.41510 & $88.3 \pm 1.8 $ & $88.1 $ \\
1.02259 & 6.87333 & $78.4 \pm 1.8 $ & $78.3 $ \\
1.02259 & 7.33153 & $70.7 \pm 1.9 $ & $70.6 $ \\
\hline
1.03023 & 4.50070 & $330.4 \pm 1.8 $ & $\cdots$ \\
1.03023 & 4.55071 & $286.0 \pm 0.4 $ & $353.3 $ \\
1.03023 & 4.60072 & $258.1 \pm 0.8 $ & $292.7 $ \\
1.03023 & 4.65073 & $237.9 \pm 1.3 $ & $259.4 $ \\
1.03023 & 4.70073 & $222.0 \pm 1.4 $ & $236.5 $ \\
1.03023 & 4.80075 & $198.3 \pm 0.9 $ & $205.4 $ \\
1.03023 & 4.90077 & $180.9 \pm 0.3 $ & $184.3 $ \\
\hline
1.04045 & 4.70208 & $299.8 \pm 0.6 $ & $\cdots$ \\
1.04045 & 4.80212 & $244.5 \pm 1.3 $ & $300.5 $ \\
1.04045 & 4.90217 & $212.8 \pm 1.5 $ & $237.0 $ \\
\hline
1.05078 & 4.85464 & $316.2 \pm 0.8 $ & $\cdots$ \\
1.05078 & 4.88467 & $291.4 \pm 0.4 $ & $\cdots$ \\
1.05078 & 4.90468 & $278.3 \pm 0.3 $ & $\cdots$ \\
1.05078 & 4.92470 & $267.1 \pm 0.5 $ & $\cdots$ \\
1.05078 & 4.94472 & $257.3 \pm 0.6 $ & $\cdots$ \\
1.05078 & 4.96474 & $248.5 \pm 0.8 $ & $416.4 $ \\
1.05078 & 4.98476 & $240.7 \pm 1.0 $ & $351.8 $ \\
1.05079 & 5.10488 & $205.7 \pm 1.1 $ & $241.0 $ \\
1.05079 & 5.20497 & $186.1 \pm 0.5 $ & $205.3 $ \\
1.05079 & 5.30507 & $171.1 \pm 0.5 $ & $182.3 $ \\
\end{tabular}
\end{ruledtabular}
\end{center}
\end{table}
We now focus on the nonspinning equal mass scattering configurations. The numerical results from \GRA{}
are shown with colored markers in Fig.~\ref{fig:scat_q1_nospin}. 
The angles and corresponding errors are computed as discussed in Sec.~\ref{sbsbsec:postproc_scats}.
We also highlight the regions in which the scattering-capture transition occurs
with colored vertical bands. The latter are estimated
as detailed in Sec.~\ref{sbsbsec:phenom_q1_nospin}, and are equivalent to the error bars
show in Fig.~\ref{fig:transition_fit}. 
We also include the scattering systems of Ref.~\cite{Damour:2014afa}, that
have $E_0\simeq 1.023 \,M$.
The numerical results are compared with the 
corresponding EOB scattering angles and transitions
(solid and vertical dashed-lines, respectively).
For each $(E_0, p_\varphi^0)$ pair, the EOB evolution is started at a large radius ($r_0=5\cdot 10^3$). 
As a consequence, the errors of the EOB scattering angles ($\lesssim 0.1^\circ$) 
are negligible with respect to the NR ones. 
The numerical values for the EOB/NR angles are reported in Table~\ref{tab:scats_q1_nospin}. 

The EOB scattering-capture transition is fully compatible with the NR result for $E_0\simeq1.016\,M$, 
and remains quite close to the NR case also for $E_0\simeq1.011\,M$. 
For the energy considered in Ref.~\cite{Damour:2014afa}, $E_0\simeq1.023\,M$, we do not have 
a merging configuration, and therefore we cannot formally estimate the region
in which the transition occurs. However, the vicinity of the EOB transition with the highest
NR scattering angle leads us to speculate that the EOB transition is indeed quite close to the NR one.
However, for higher energies, the EOB prediction loses reliability, as already highlighted 
in Sec.~\ref{sbsec:eobnr_parspace}, and the predicted scattering-capture transition gets farther 
away from the real (numerical) one.
Finally, note that, even at high energy, the EOB/NR disagreement decreases when 
the initial angular momentum is increased, i.e. when weaker gravitational
interactions are considered.
We do not explore in detail the weak-interaction 
region of the parameter space,  
since in this work we are mostly interested in strong interactions near 
the scattering-capture transition.

\subsection{Scatterings for asymmetric systems}
\label{sbsec:eobnr_scat_asym}
\begin{table}[t]
	\caption{\label{tab:scat_asym} Scattering angles for asymmetric binaries considered in this work. The angle
	computed from the relative separation is denoted as $\chi$, while the ones computed from the single tracks
	are denoted as $\chi_{{\rm p}_1}$ and $\chi_{{\rm p}_2}$. Complementary information on 
	these runs is reported in Table~\ref{tab:runs_q1_spin_q23}. }
\begin{center}
{\footnotesize
\begin{ruledtabular}
\begin{tabular}{c | c c | c | c | c } 
($q$, $\chi_1$, $\chi_2$) & $\hat{E}_0$ & $p_\varphi^0$ & $\chi\,[{}^\circ]$ & $\chi_{{\rm p}_1}\,[{}^\circ]$ & $\chi_{{\rm p}_2}\,[{}^\circ]$  \\
\hline
\hline
(1, +0.5, -0.5) & 1.016 &  4.300 &  $  303.6\pm2.0$ & $ 306.1 \pm 2.5 $ & $   299.00 \pm 4.7$  \\
(2, \phantom{+}0.0, \phantom{-}0.0) & 1.016 &  4.350 &  $   308.6\pm1.3$  & $309.1 \pm 1.8$ &  $308.9 \pm 1.2$ \\
(2, \phantom{+}0.0, \phantom{-}0.0) & 1.016 &  4.400 &  $   274.8\pm0.3$  & $275.4 \pm 1.8$ &  $275.0 \pm 0.4$ \\
(2, \phantom{+}0.0, \phantom{-}0.0) & 1.020 &  4.440 &  $   298.7\pm0.5$  & $299.5 \pm 2.0$ &  $299.0 \pm 0.6$ \\
(3, \phantom{+}0.0, \phantom{-}0.0) & 1.017 &  4.423 &  $   379.7\pm3.0$  & $380.5 \pm 1.4$ &  $379.4 \pm 2.5$\\
(3, \phantom{+}0.0, \phantom{-}0.0) & 1.025 &  4.640 &  $   356.9\pm0.6$  & $355.4 \pm 0.8$ &  $356.7 \pm 0.5$\\
(3, \phantom{+}0.0, \phantom{-}0.0) & 1.025 &  4.660 &  $   327.8\pm0.5$  & $326.4 \pm 0.3$ &  $327.8 \pm 0.6$\\
\end{tabular}
\end{ruledtabular}
}
\end{center}
\end{table} 
While in the previous section we have discussed scattering angles for equal mass non-spinning systems,
we now briefly turn our attention to these quantities in the case of asymmetric systems.
We recall that we compute the scattering angle $\chi$ from the relative track $(x,y)=(x_1-x_2, y_1-y_2)$ converted
in polar coordinates, as detailed in Sec.~\ref{sbsbsec:postproc_scats}. If $m_1=m_2$ 
and $\chi_1=\chi_2$, this is equivalent to computing the scattering angles from the single tracks 
$(x_1,y_1)$ and $(x_2,y_2)$, modulo numerical inaccuracies linked to resolution and extrapolation. 
However, for asymmetric systems, this is not the case.
For this reason, in Table~\ref{tab:scat_asym} we also report
the scattering angles computed from the single tracks of the asymmetric systems. For the unequal mass cases,
we extrapolate the track of the heaviest black hole using a $1/r$-polynomial of the fourth order, since
higher-order polynomials lead to overfitting due to the shortness of the track. In all the other cases,
we use an SVD as discussed in Sec.~\ref{sbsbsec:postproc_scats}.
However, in all the asymmetric cases analyzed in this work, the error linked to the extrapolation
prevents us to really appreciate the difference in the scattering angles of the two black holes.
Future works may focus on reducing these errors to obtain better estimates and highlight the differences
in the scattering angles of the single black holes.

\subsection{EOB/NR unfaithfulness}
\label{sbsec:eobnr_mm}
A common evaluation metric for the goodness of waveform models is 
the so-called mismatch (or unfaithfulness). For two time domain signals 
$h_1(t)$ and $h_2(t)$, this quantity can be computed as
\begin{equation}
\label{eq:F}
    \MM = 1-\max_{\phi_0, t_0}\frac{\langle h_1, h_2 \rangle}{\sqrt{\langle h_2, h_2 \rangle \langle h_1, h_1 \rangle}}\,,
\end{equation}
where $\phi_0$, $t_0$ denote a reference phase and time 
and the inner product is defined using the Fourier transforms 
$\tilde{h}_1(f)$ and $ \tilde{h}_2(f)$ as
\begin{equation}
    {\langle h_1, h_2 \rangle} = 4 \Re \int \frac{\tilde{h}_1(f) \tilde{h}^*_2(f)}{S_n(f)} df.
\end{equation}
The function $S_n(f)$ here considered is the zero-detuned, high-power 
noise spectral density of Advanced LIGO~\cite{aLIGODesign_PSD}. 
In this work, we consider the frequency range $[11,512]\,{\rm Hz}$. The mismatches 
are computed with the function \texttt{optimized$\_$match} 
implemented in {\tt pyCBC}~\cite{Biwer:2018osg} using the (2,2) multipole. 

As argued in the previous sections, for high energies the model loses reliability,
and therefore we do not consider EOB/NR mismatches for the configurations with $E_0\gtrsim 1.025$.
We consider mismatches for all the available and complete \GRA{} waveforms with lower energies. 
The NR double encounters that have not been completed up to merger are discarded;
these configurations are marked with asterisks in Tables~\ref{tab:runs_q1_nospin}
and~\ref{tab:runs_q1_spin_q23}. 

\begin{figure}[t]
  \centering 
    \includegraphics[width=0.48\textwidth]{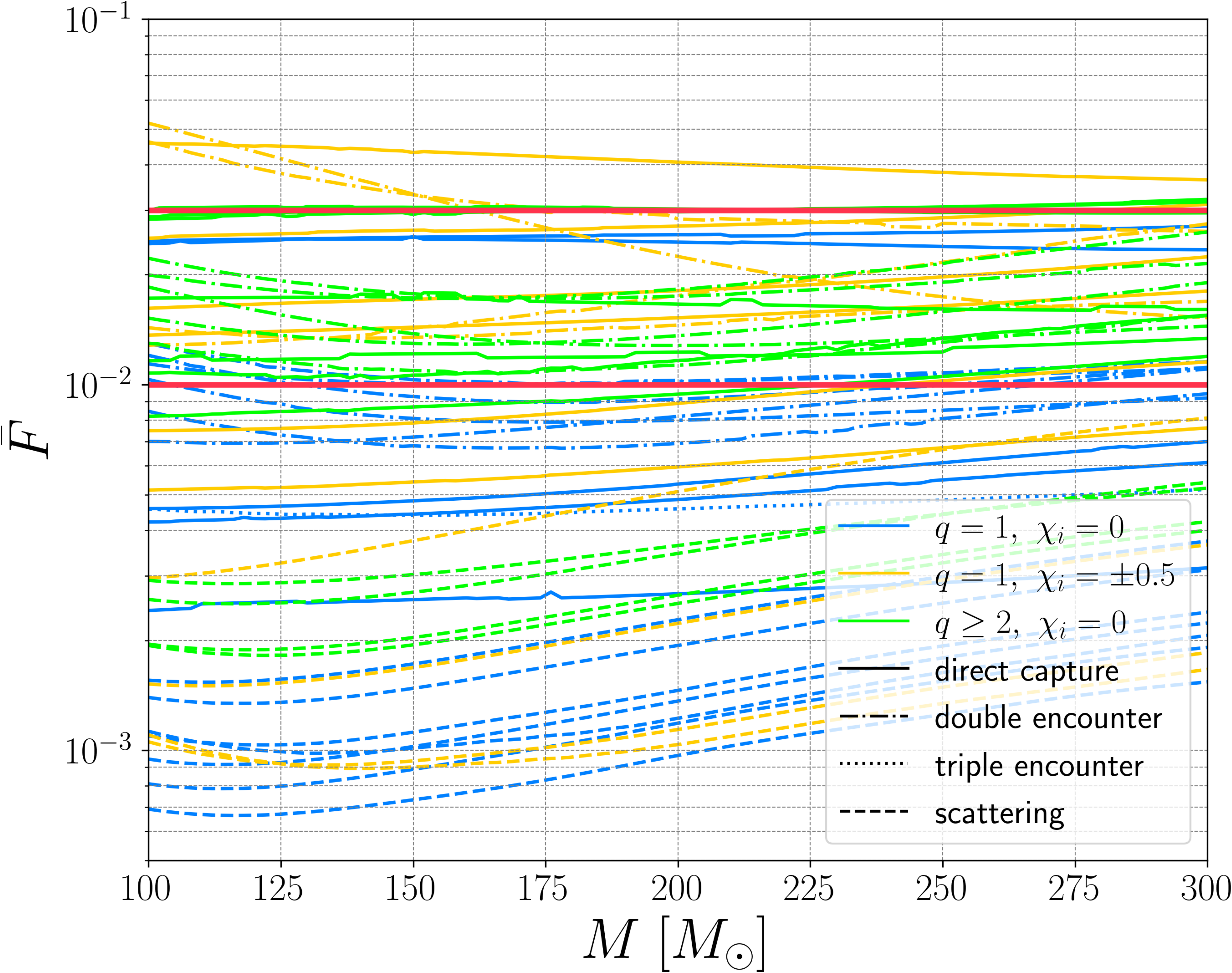}
    \caption{Optimized mismatches for equal mass nonspinning systems
    (blue), $q=\lbrace 2,3\rbrace$ nonspinning (green), and spinning equal mass (yellow).
    For the $q=1$ nonspinning case we consider only the configurations with $E\leq 1.016\,M$. 
    Scatterings are highlighted with dashed lines, double encounters with dash-dotted lines,
    direct merger with solid lines. The only triple-encounter of our
    dataset is denoted with a (blue) dotted line. 
    The two red horizontal lines mark the $1\%$ and $3\%$ thresholds.
    The corresponding values for
    the reference mass $M=200 M_\odot$ are reported in Table~\ref{tab:mm}.
    }
  \label{fig:mm_q1_nospin}
\end{figure}

Before computing the mismatches, it is useful to recall that small 
inaccuracies in the initial data can lead to drastically different phenomenologies, especially 
for configurations close to the scattering-capture transition. 
Therefore, following previous works in the literature~\cite{Gamba:2021gap,Andrade:2023trh}, we
optimize the initial energy and angular momentum used to perform the EOB evolution. 
A similar approach is followed when comparing eccentric bound 
configurations\footnote{In the bound eccentric case, the optimization is typically performed on 
the gauge-dependent eccentricity. However,
energy and angular momentum could be used to initialize initially bound evolutions too.}~\cite{Ramos-Buades:2021adz,Nagar:2024dzj}.  
An optimization of the initial data on a small interval can be justified 
for parameter estimation purposes, but one should keep in mind that this might
hide small biases on the recovered intrinsic parameters.
Table~\ref{tab:mm} shows the mismatches for the reference mass $200\,M_\odot$, both before and after the 
optimization of the initial conditions.
The dependence of the optimized mismatches against the total mass is instead shown in Fig.~\ref{fig:mm_q1_nospin}.
Different colors refer to different $(q,\chi_1,\chi_2)$-configurations: blue lines for
nonspinning equal mass, green for nonspinning unequal mass,
yellow for spinning equal mass. 
The different line styles mark different phenomenologies: dashed
lines for scatterings, dash-dotted for double encounters, and solid for direct
captures. The only triple encounter of our dataset is marked with a dotted line. 

The scattering configurations have mismatches that are at the $10^{-3}$ level, even before the
initial data optimization; the latter procedure only marginally improves the EOB/NR agreement. 
This result shows that \DALI{} is highly faithful in this energy and angular momentum regime, thus confirming the 
results discussed in Sec.~\ref{sbsec:eobnr_q1_nospin_scat}. 
On the contrary, the optimization is essential to lower the mismatches for configurations with multiple encounters.
In particular, for the equal mass nonspinning cases with two encounters we have $\bar{F}_{\rm opt}\sim 1\%$.
The triple encounter has instead an optimized mismatch that is below the $5\cdot10^{-3}$ level.
Improved initial data are also useful to lower the mismatches for some direct captures (solid lines).
Interestingly, the optimization has a marginal 
effect on the direct captures with lowest angular momenta for each energy. 
This is probably a consequence of the fact that the ringdown part dominates these signals, while the ringdown model
employed in \DALI{} is calibrated only on quasi-circular data. On the other hand, the optimization 
strongly improves $\bar{F}$ for direct captures with a quasi-circular whirl before
the plunge. This occurs, for example, for the equal mass cases with
$(\hat{E}_0,p_\varphi^0)\simeq(1.016,4.012)$,
where the mismatch drop from $31.4\%$ to $0.5\%$ after optimization. 
The equal mass nonspinning configuration with lowest energy and angular momentum,
$(\hat{E}_0,p_\varphi^0)\simeq(1.001,3.800)$, is the only direct capture without quasi-circular whirl 
whose mismatch is strongly improved by the optimization 
procedure (from $4.36\%$ to $0.27\%$).
The mismatches for the equal mass nonspinning configurations are typically around or below 
the $1\%$ threshold after optimization.
The two cases with higher final mismatches, that are direct captures 
with low angular momentum, are still within the $3\%$ threshold. 

Generally, the nonspinning unequal mass configurations have higher mismatches than the previous case, 
but they are still within, or at most around, the $3\%$ threshold.
On the other hand, for the spinning binaries we have three cases above 
the $3\%$ threshold, which however still remain below $5\%$.

We mention, in passing, that the $(q,\chi_1,\chi_2)=(1,0,0)$ cases with higher 
energies ($\hat{E}_0\geq1.030$) require an optimization in a larger region, as
can be imagined by looking at the results of Secs.~\ref{sbsec:eobnr_parspace} 
and~\ref{sbsec:eobnr_q1_nospin_scat}. 
In these higher energy cases, the mismatches for the scattering configurations are 
typically below the $2\%$, while the merger configurations have $\bar{F}_{\rm opt}\sim 7-8\%$. 
The higher mismatches, together with the fact that the optimization has to be performed on 
a larger region, are a direct consequence of the lower reliability of \DALI{} in this energy regime. 
\begin{figure*}[t]
  \centering 
    \includegraphics[width=0.32\textwidth]{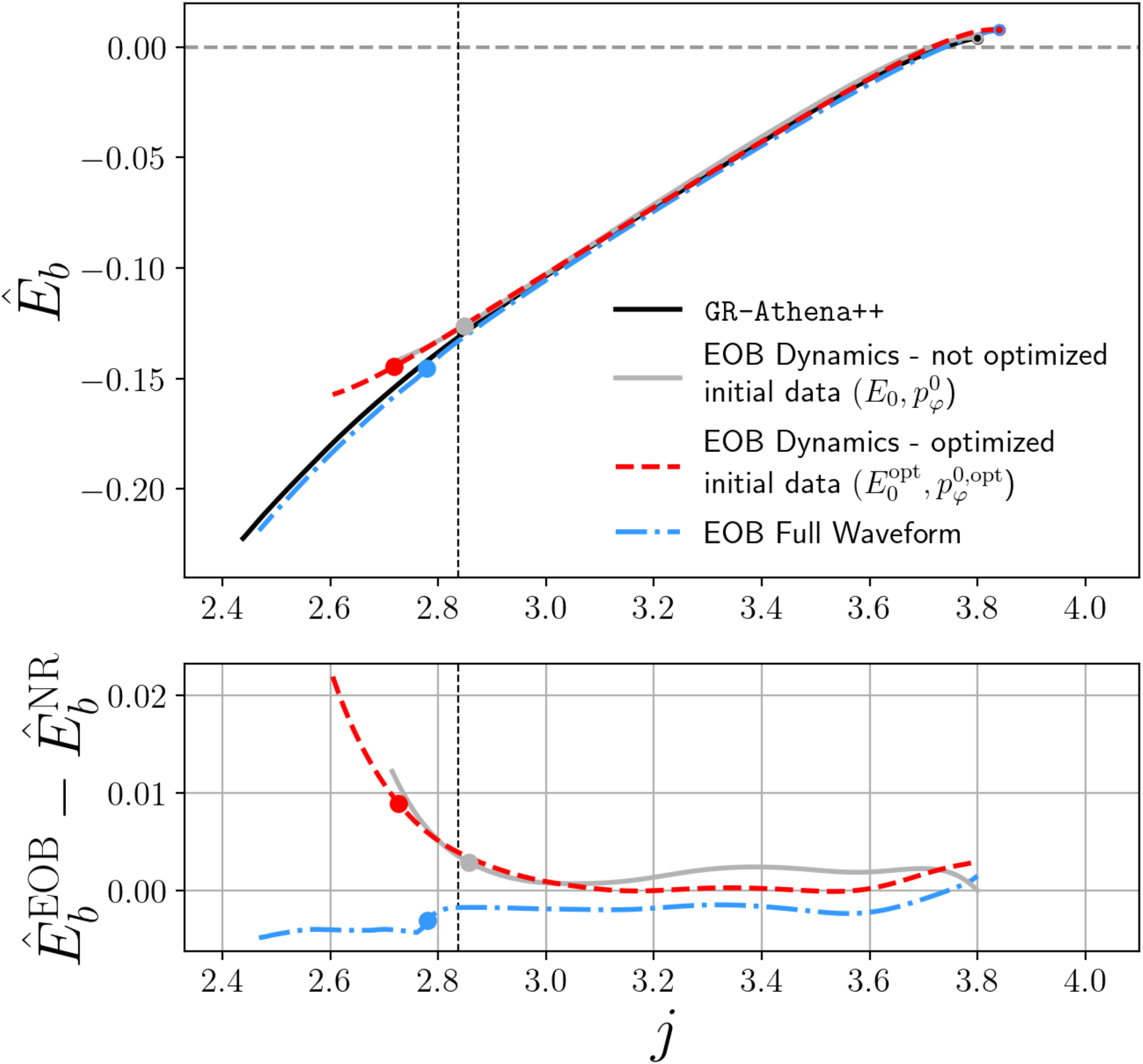}
    \includegraphics[width=0.32\textwidth]{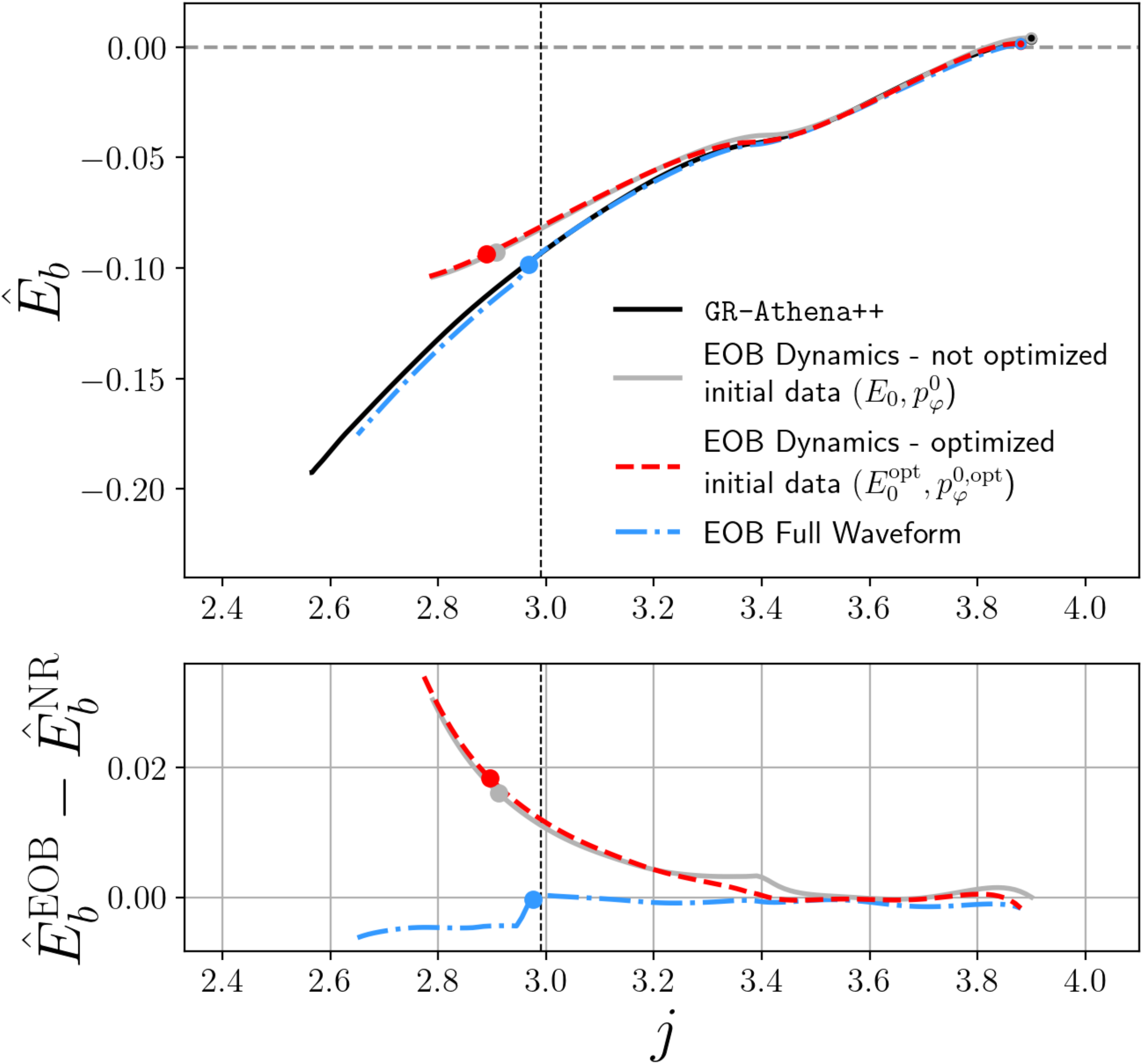}
    \includegraphics[width=0.32\textwidth]{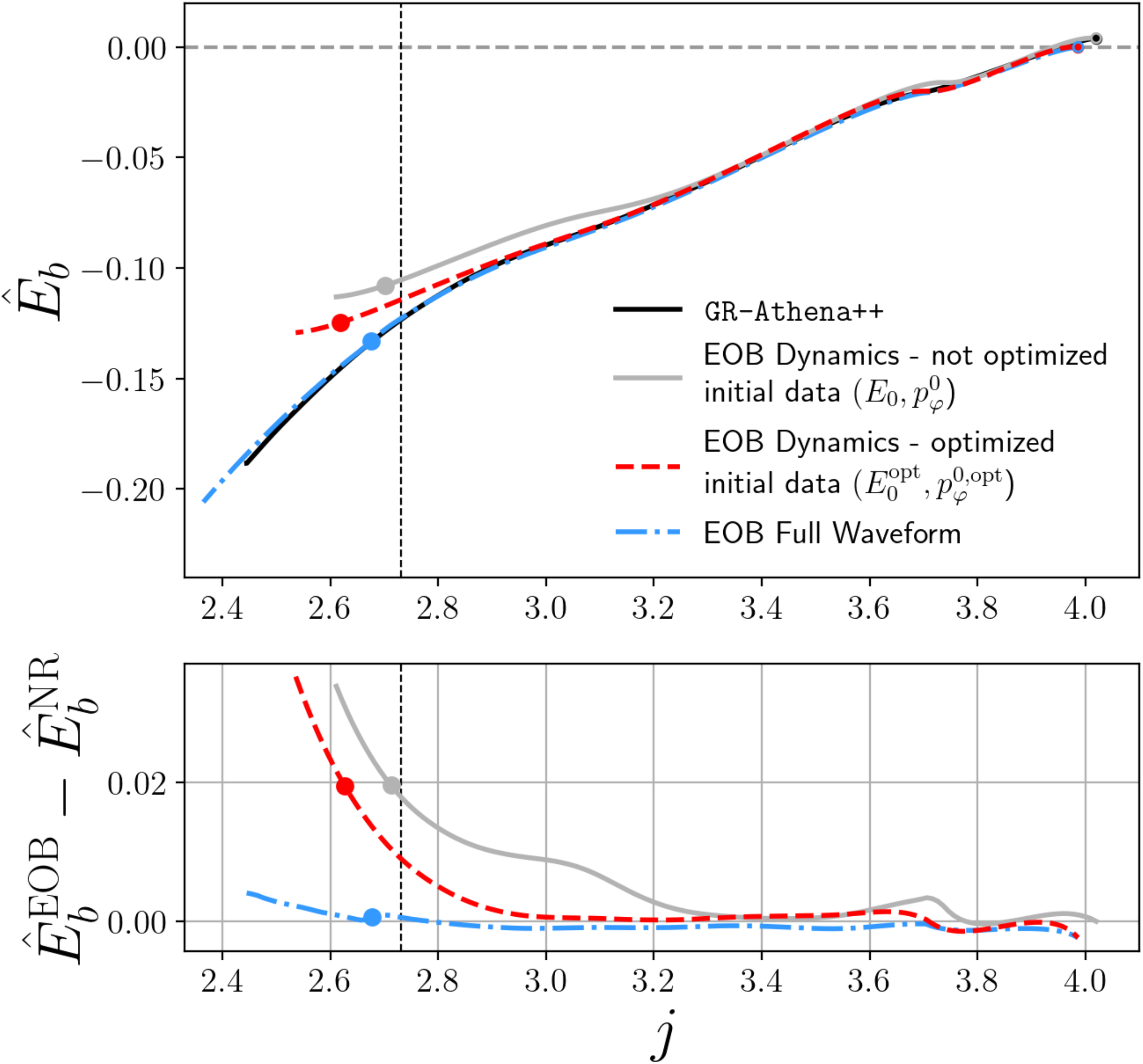}
    \caption{Energetic curves $\hat{E}_b(j)$ for the equal mass nonspinning dynamical capture,
    double encounter, and triple encounter with $\hat{E}_0 = 1.001$. 
    The NR curves (black) are computed
    considering all the multipoles up to $\l=5$, while the analytical gray and dashed red curves are obtained from the
    EOB dynamics, respectively before and after the initial data optimization discussed in Sec.~\ref{sbsec:eobnr_mm}. 
    We also show the energetics 
    computed from the EOB waveform
    (dash-dotted blue), see text for details and interpretation.
    The EOB/NR differences are reported in the bottom panels.
    The NR merger time is marked with a dashed
    vertical line, while the estimated EOB merger times are highlighted with markers. }
  \label{fig:eobnr_energetics}
\end{figure*}

We conclude this section with a remark on the mass range shown in Fig.~\ref{fig:mm_q1_nospin}. 
The reason behind the choice $M\in [100,300]M_\odot$ is twofolds. On the one hand, 
the only significant event in the O3 LVK 
catalog~\cite{LIGOScientific:2021usb,KAGRA:2021vkt} that can be interpreted as
a dynamical capture is GW190521~\cite{LIGOScientific:2020iuh, Gamba:2021gap}, which
has a total source-frame mass of approximately $150 M_\odot$.
This is also the only significant intermediate mass BBHs observed so far~\cite{LIGOScientific:2021tfm}.
Second, taking into account lower masses would mean to only consider the low-frequency (in geometrized units) 
part of the waveform, i.e. the portion associated to the precursor. However,
since we are integrating the numerical $\psi_4^{22}$ with a fixed-frequency method that removes the 
low-frequency part of the signal (see also discussion in Sec.~\ref{sbsbsec:postproc_waves}), the 
precursor of the numerical waveform is not reliable. Consequently, the EOB/NR mismatches
for initially unbound configurations 
can reach, and even overcome, the $10\%$ level in the low masses regime if FFI waveforms are employed. However,
this is only an artefact of our numerical integration, and it is not related to the 
accuracy of the EOB model itself,
that,
thanks to the inclusion of the generic Newtonian prefactor~\cite{Chiaramello:2020ehz},
is indeed able to correctly reproduce the low-frequency part of the waveform.
 
The latter statement can be confirmed by considering a dynamical capture and
computing $h_{22}$ with a TDI rather than an FFI\footnote{We recall that the 
time-domain integration is only reliable
for relatively short signals, such as direct captures, but fails for longer
signals. See discussion in Sec.~\ref{sbsbsec:postproc_waves}.}.
Time-domain methods, when practicable, are indeed able to reproduce the precursor 
of the numerical waveform. In general,
the precursor is generated before the strong gravitational interaction of the two black holes,
and therefore we expect to have the same precursor
for all the configurations with same initial energy. 
We thus proceed by considering the
nonspinning equal mass case with ($\hat{E}_0,p_\varphi^0)\simeq(1.011,3.950)$.
We compute the numerical waveform with a time-domain integration, and then we calculate
the EOB/NR mismatch.
We find $\bar{F}_{\rm opt}\sim2.5\%$ for $M\in [50,300]M_\odot$,
and lower mismatches for lower masses. In particular,
we find $\bar{F}_{\rm opt}=0.7\%$ for $10\,M_\odot$ .

\subsection{EOB/NR energetics}
\label{sbsec:eobnr_energetics}

While we have already presented some numerical energetic curves 
in Sec.~\ref{sec:numrel_simulations} (and in particular in Figs.~\ref{fig:series_E1.011},~\ref{fig:updown_spin} and~\ref{fig:series_q123}),
in this section we discuss the comparisons with the energetics obtained from \DALI{}.
This test is indeed a well established gauge-invariant diagnostic to the study the accuracy of the 
EOB dynamics~\cite{Nagar:2015xqa,Damour:2011fu}.
We recall that the numerical curves are obtained by subtracting from the initial values $(E_0,p_\varphi^0)$
the integrated energy and angular momentum fluxes,
which we compute using all the multipoles up to $\l=5$.
The EOB counterparts are evaluated along the EOB dynamics, using in particular
the time evolution of the Hamiltonian $H_{\rm EOB}$ and the canonical angular momentum
$p_\varphi$ found by solving the equations of motion.
We report in Fig.~\ref{fig:eobnr_energetics} the results for the series of simulations with
the lowest initial energy considered in this work ($\hat{E}_0 = 1.001$). 
We select this series because it has a diverse range of coalescing phenomenologies:
direct capture, double encounter, and triple encounter. 
For the proper EOB energetics computed from the dynamics,
we report two different cases: i) the ones computed from the not-optimized 
NR initial data $(E_0, p_\varphi^0)$ (gray), ii) the curves
obtained considering the initial data that optimize the mismatches as discussed in 
Sec.~\ref{sbsec:eobnr_mm} (dashed red); the enhanced initial data are also reported
in Table~\ref{tab:mm}.
While the effect of the optimization is marginal  
(but not completely negligible) 
for the direct capture and the double counter, the improved initial data for the triple encounter
provide a much more accurate curve. However, the optimization is
also useful to improve the EOB/NR agreement for other direct captures and
double encounters considered in this work.
In general, we observe that the agreement between analytical and numerical results 
decreases toward the merger.
While in the NR cases the merger time is identified by the last peak of the (2,2) amplitude 
(see also Appendix~\ref{appendix:tmrg}),
we recall that for the EOB case we estimate it from the peak of the pure orbital frequency computed without
spin-orbit coupling as $t_{\rm mrg}^{\rm EOB} =  t^{\rm peak}_{\Omega_{\rm orb}}-3$~\cite{Nagar:2020pcj}.
In particular, the EOB energetics obtained from the dynamics systematically end at higher energies
and smaller angular momenta with respect to the NR case. It should be however considered that these analytical
energetics are computed from the dynamics, and therefore they do not contain 
any contribution from the quasi-circular ringdown model of \DALI{}.
In other words, the most significant comparison between the analytical and numerical
results is for the pre-merger portion of the evolution, where the agreement
is indeed better. 

To confirm the latter statement, we also compute the integrated analytical fluxes from the complete
optimized EOB waveform, and subtract them to the initial data. The corresponding energetics are
shown with dash-dotted blue lines in Fig.~\ref{fig:eobnr_energetics}. Note that these EOB curves are less meaningful 
than the ones previously discussed, since the energetics
are mainly a diagnostic for the dynamics. However, they are still useful to highlight that the inclusion
of a ringdown model in the EOB waveform makes the analytical results closer to the numerical curves also after 
the merger. 
The small discrepancies between the energetics computed from the EOB dynamics and the EOB waveform
are mainly linked to the fact that each multipole of the waveform includes a noncircular 
Newtonian correction~\cite{Chiaramello:2020ehz,Albanesi:2021rby}
(see also discussion in Sec.~\ref{sbsec:rwz_eob}),
while the angular radiation reaction ${\cal F}_\varphi$ only include the (2,2) Newtonian noncircular correction.
Moreover, when computing the EOB fluxes, we sum up to $\l=5$ for consistency with the NR curves, 
while the EOB dynamics incorporates multipoles up to $\l=8$. However, we checked that 
these additional modes are highly subdominant during the inspiral,
and are not a primary source of discrepancy.
Note that in the EOB/NR difference 
computed considering the energetics from the EOB waveform, there is a small jump 
near the EOB merger time. 
This discontinuity arises because the NQC corrections 
were not included in the computation of the EOB waveform.

We conclude this discussion by recalling that the \DALI{} dynamics is only numerically informed through 
the $a_6$ and $c_3$ coefficients, that are tuned on quasi-circular NR simulations.
As already mentioned when analyzing other diagnostics,
the inclusion of non-circularized simulations in this procedure, together with
more non-circular analytical information, would probably 
further improve the model, and thus the comparisons of the energetics.

\section{Test-mass limit} 
\label{sec:testmass}
\begin{figure*}[t]
  \centering 
    \raisebox{.14\height}{
    \includegraphics[width=0.32\textwidth]{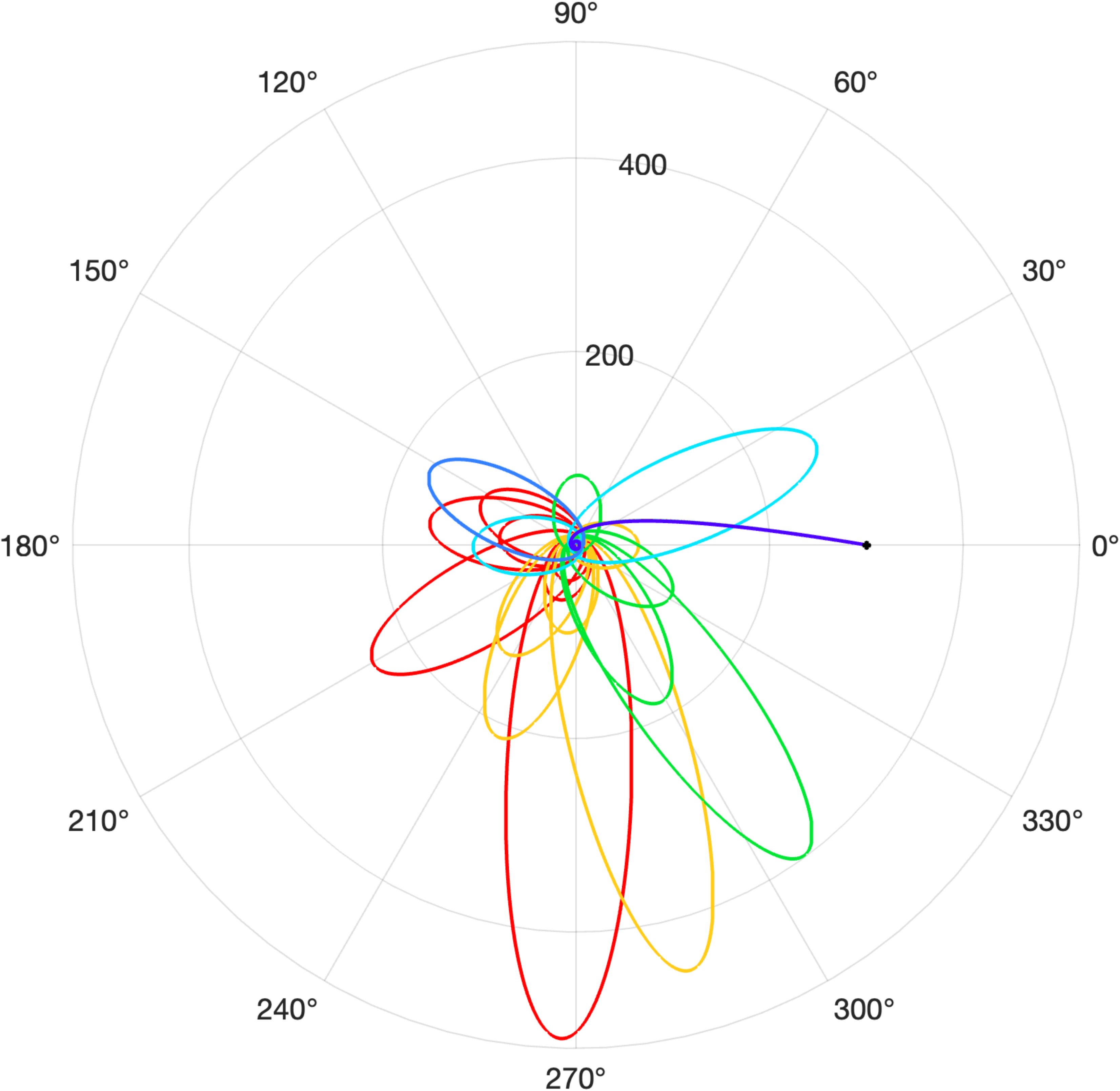}
    }
    \hspace{0.2cm}
    \includegraphics[width=0.64\textwidth]{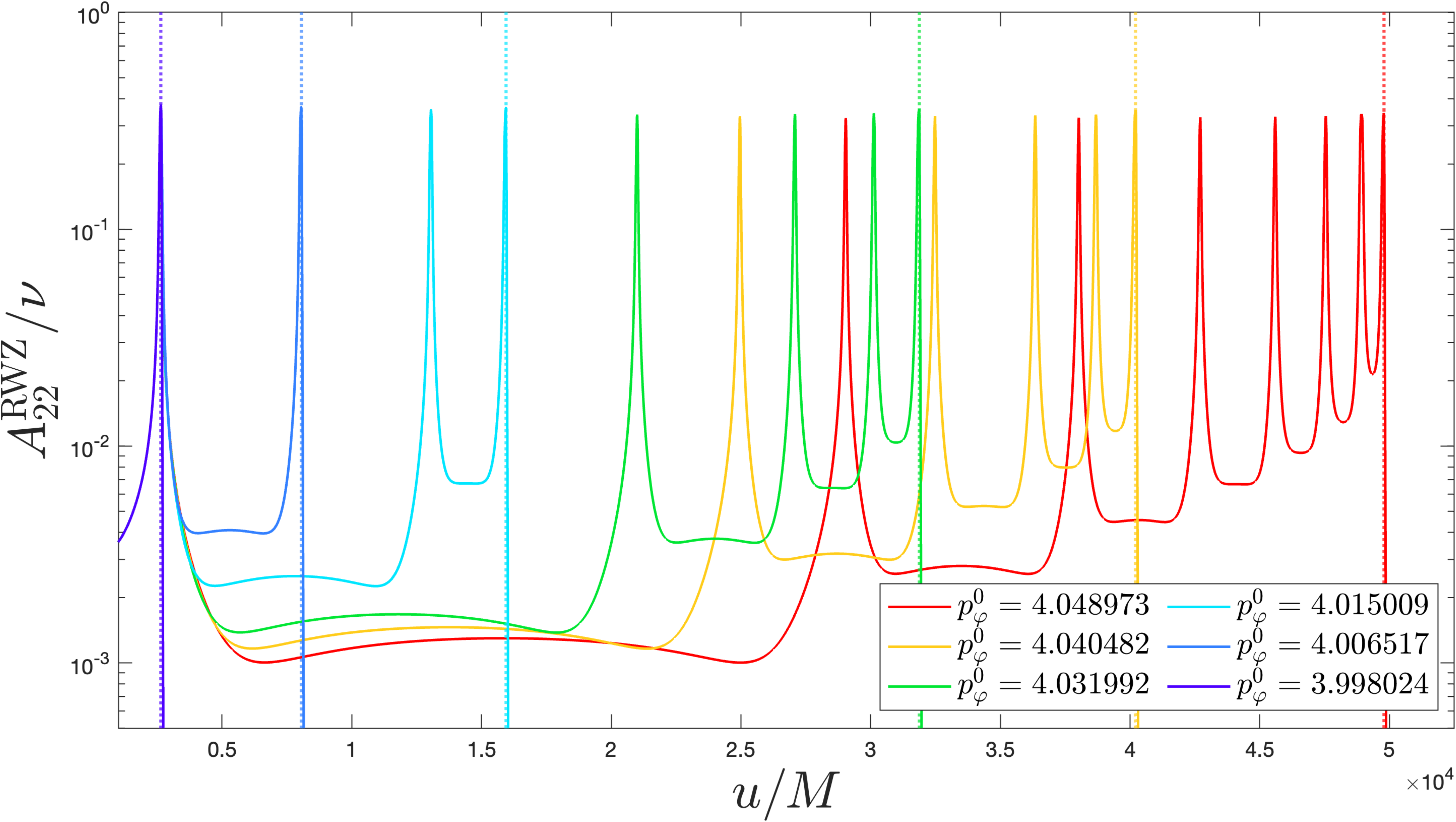}
    \caption{Left panel: trajectories for test-mass dynamical captures with $\nu=10^{-2}$, $r_0=300$, and 
     $\hat{E}_0 = 1.000001$. The black dot marks the initial position, that is the same for all configurations. 
     Right panel: corresponding (2,2) Zerilli $\nu$-rescaled amplitudes computed with the time-domain code
     \RWZ{}; vertical logarithmic scale.  
     The initial angular momenta are reported in the legend.
     The vertical dotted lines mark the amplitude maxima.}
  \label{fig:dyn_captures_rwz}
\end{figure*}
While extreme mass-ratio inspirals (EMRIs) might be relevant for future
space-based detectors, such as LISA~\cite{Babak:2017tow,Berry:2019wgg}, 
the test-particle limit is also an useful controlled
laboratory to test and validate analytical prescriptions
for binaries with generic mass 
ratio~\cite{Nagar:2006xv,Damour:2007xr,Bernuzzi:2010xj,Albanesi:2021rby,Placidi:2021rkh,Albanesi:2023bgi}. In particular,
the test-mass limit is naturally included in the EOB approach,
since the effective EOB metric is a $\nu$-deformation of the Schwarzschild metric 
(or the Kerr one, if spinning black holes are considered). 
For these reasons, we now shift our focus to test particles captured by 
Schwarzschild black holes. Corresponding waveforms at linear order in 
the mass ratio can be derived as solutions of the RWZ
equations with a test particle source term~\cite{Regge:1957td,Zerilli:1970se,Nagar:2005ea,Martel:2005ir}.
These equations are solved numerically
with the time-domain code \RWZ{}~\cite{Bernuzzi:2010ty,Bernuzzi:2011aj,Bernuzzi:2012ku}.
We then compare the numerical results with 
the corresponding fully analytical EOB waveforms, adopting different prescriptions for the non-circular corrections. 


\subsection{Test-particle dynamics}
\label{sbsec:testmass_dynamics}
Time-like geodesics in Schwarzschild spacetime can be conveniently parametrized by the Schwarzschild coordinate
time, so that the test-particle Hamiltonian depends only on spatial coordinate and momenta.
More specifically, the $\mu$-rescaled Hamiltonian reads
\be
\label{eq:Hschw}
\Hschw = \sqrt{A(r)\left(1 + \frac{p_\varphi^2}{r^2}\right) + p_{r_*}^2},
\ee 
where $p_{r_*}$ is the conjugate momentum of the tortoise coordinate $r_*$, defined
as $p_{r_*}^2 = A/B \,p_r^2$. The Schwarzschild metric potentials $A$ and $B$ are defined as
$A=1-2/r$ and $B=A^{-1}$. It is useful to explicitly keep $B$ in the 
formal definition of $p_{r_*}$, so that it remains valid in the Kerr and EOB cases
with opportune generalizations of these two metric potentials~\cite{Damour:2014sva}. 
Dissipative effects linked to the GW emission can be included in the dynamics by adding a 
radiation reaction force ${\cal F}$. 
The equations of motion thus read
\begin{align}
\dot{r} =& \frac{A}{\Hschw} p_{r_{*}}  , \\ 
\dot{\varphi} =&\frac{A}{\Hschw}\frac{p_\varphi}{r^2}, \\
\dot{p}_{r_*} =& A \hat{\F}_r - \frac{A}{r^2\Hschw}\left[ p_\varphi^2\left( \frac{3}{r^2} 
                - \frac{1}{r} \right) +1 \right]  , \\
\dot{p}_\varphi =& \hat{\F}_{\varphi},
\end{align}
where $\hat{\F}_{r,\varphi} \equiv {\cal F}_{r,\varphi}/\nu$ are the radial and angular components 
of the radiation reaction force. For this term we use the resummed PN 
expressions discussed in Ref.~\cite{Chiaramello:2020ehz} and extensively
tested for bound dynamics in Ref.~\cite{Albanesi:2021rby}. 
We consider systems with $\nu=10^{-2}$. This choice only affects
the relevance of the radiation reactions during the inspiral, since no $\nu$-corrections
are included in the conservative dynamics, that is fully described by Eq.~\eqref{eq:Hschw}. 
However, for the description
of astrophysical EMRIs, one should consider an EOB Hamiltonian informed 
with gravitational self-force (GSF) results, as detailed in 
Refs.~\cite{Nagar:2022fep}. Fluxes with higher-order PN corrections
has been also proven to be essential to consistently reproduce
GSF calculations~\cite{Albertini:2022rfe,Albertini:2022dmc,Albertini:2023aol}.
However, as mentioned earlier, in this analysis we are interested in 
using the waveforms generated by a test-mass dynamical capture to assess 
the analytical EOB prescription for the waveform, and also to gain insights into
properties of the numerical waveform.
For these reasons, the prescription of the radiation
reaction discussed in Refs.~\cite{Chiaramello:2020ehz,Albanesi:2021rby} is more than sufficient for our goals. 

The trajectories of six dynamical captures with initial separation $r_0=300$, initial energy 
$\hat{E}_0 = 1.000001$, and different initial angular momenta
are shown in the left panel of Fig.~\ref{fig:dyn_captures_rwz}.
The angular momenta, whose values are reported in the legend of the right panel,
are chosen by requiring to have multiple encounters in a reasonable time window.  
Since for high-mass ratios the relevance of the radiation reaction is inhibited, 
we chose an initial energy just above the parabolic limit, in order to obtain 
captures with multiple encounters.
This result also points out that the initially unbound portion of the
parameter space for large mass ratio binaries is largely dominated by scattering and 
direct captures~\cite{Nagar:2020xsk}. 
Depending on formation scenarios, comparable masses BBHs might be more likely to be
sources of dynamical captures with multiple close encounters.

\begin{figure}[t]
  \centering 
    \includegraphics[width=0.48\textwidth]{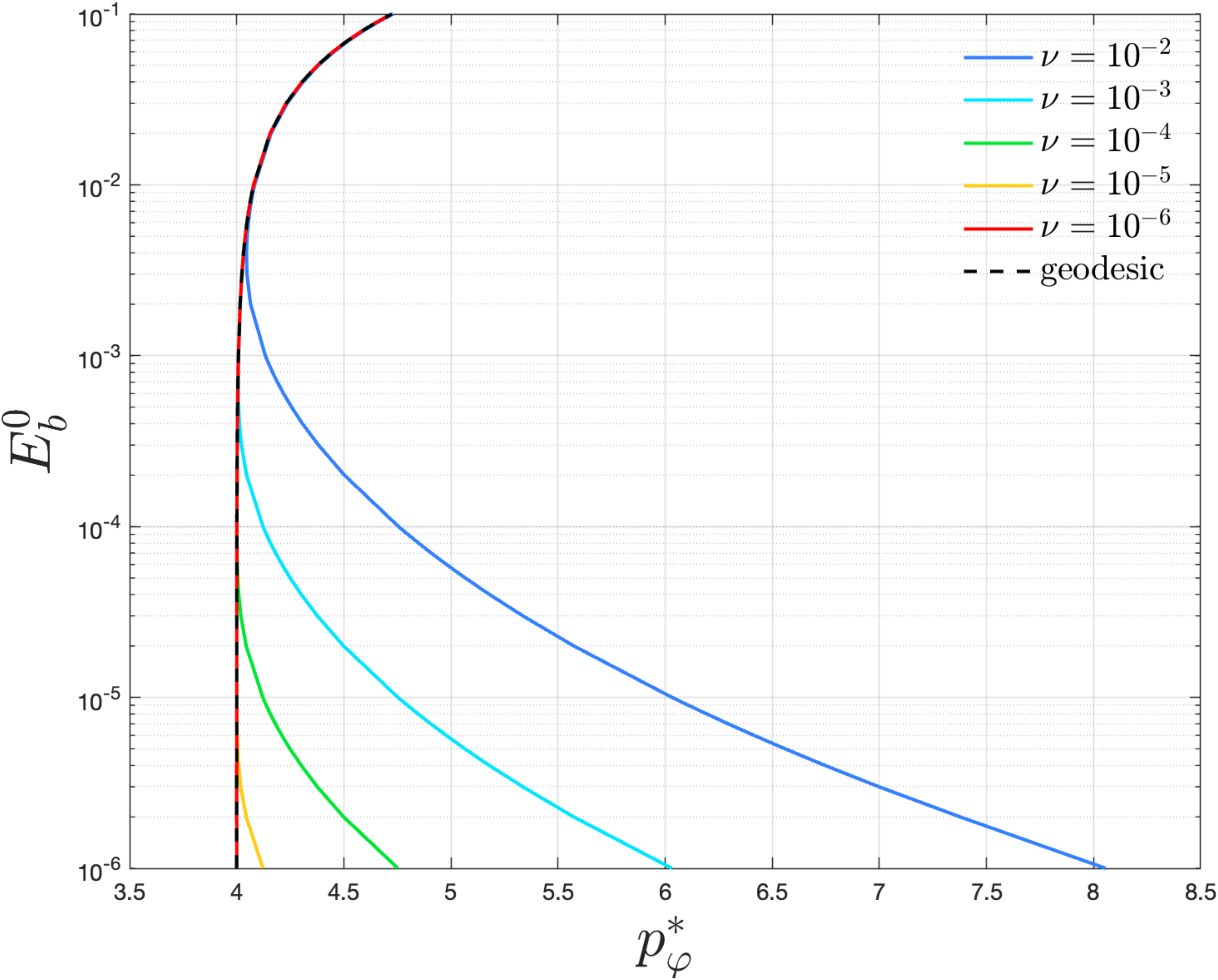}
    \caption{Impact of the radiation reaction on the the critical initial angular momentum $p_\varphi^*$ 
    at which the scattering-capture transition occurs in the test-particle limit. 
    On the $y$-axis we show the initial (not $\nu$-reduced) binding energy 
    $E_b^0\equiv\hat{E}_0-1$ in logarithmic scale. Solid lines are obtained by
    considering EOB evolutions with different symmetric mass ratios, whose effects
    are solely included in the dissipative sector. In all the cases, we start the evolution from $r_0=300$.
    The black dashed line marks the geodesic case ($\nu=0$), 
    computed fully analytically from Eq.~\eqref{eq:pph_transition_schw}.}
  \label{fig:schw_transition}
\end{figure}
The relevance of the radiation reaction
is also visualized in Fig.~\ref{fig:schw_transition},
where we show the critical initial angular momentum $p_\varphi^*$ at which the scattering-capture transition
occurs for different mass ratios and energies.
While for $\nu\neq0$ we have a dissipative contribution to the dynamics and 
we need to use a numerical ODE solver for the Hamilton's equations, 
for the geodesic case the critical angular momentum can by computed as~\cite{Damour:2022ybd}
\be
\label{eq:pph_transition_schw}
p_\varphi^{*,{\rm Schw}} = \sqrt{\frac{1}{u_0^2} \left(\frac{\hat{E}^2}{1-2 u_0} - 1\right)},
\ee
where
\be
u_0 = \frac{4-3 \hat{E}^2 + \hat{E} \sqrt{9 \hat{E}^2-8}}{8}.
\ee
As can be appreciated from Fig.~\ref{fig:schw_transition}, the region of the parameter space
in which dynamical captures with multiple encounters are allowed strongly shrinks with the mass ratio,
and, for the energies considered, practically vanishes for $\nu=10^{-6}$. 
Finally, while the geodesic case leads to a logarithmic divergence~\cite{Damour:2022ybd} in the scattering angle
at $p^*_\varphi (\hat{E})$, the presence of a radiation reaction limits these angles to 
finite values. 

\subsection{Numerical solution of RWZ equations}
\label{sbsec:rwz}
Given the test-particle dynamics discussed in the previous section, we can compute 
the corresponding waveform at linear order in perturbation theory by solving the
inhomogeneous RWZ 
equations with test-particle source term~\cite{Regge:1957td,Zerilli:1970se,Nagar:2005ea,Martel:2005ir},
\be
\label{eq:rwz}
\p_t^2 \Psi_\lm^\oe - \p^2_{r_*} \Psi_\lm^\oe + V_\l^\oe \Psi_\lm^\oe = S_\lm^\oe,
\ee
where the master function $\Psi_\lm^\oe$ is linked to 
the waveform multipoles by a simple normalization 
$\Psi_\lm^\oe = h_\lm / \sqrt{(\l+2)(\l+1)\l(\l-1)}$.
We solve these equations with the time-domain code \RWZ{}~\cite{Bernuzzi:2010ty,Bernuzzi:2011aj,Bernuzzi:2012ku},
where the Dirac $\delta$ in the source term $S_\lm^\oe$ is approximated by a narrow Gaussian.
Since the motion of the particles can extend to large separations ($r \sim 500$), 
a large computational domain is needed.
We thus use as time and radial coordinates 
$(\tau,\rho)\in \mathbb{R}\times [R_*^-,S^+]_{R_*^+} = \mathbb{R}\times [-100,700]_{600}$,
that coincide with the Schwarzschild time and tortoise radial coordinates
in the compact region $[R_*^-,R_*^+)$, where the motion of the particle is confined.
An hyperboloidal layer~\cite{Zenginoglu:2010cq,Bernuzzi:2011aj} is then matched to these coordinates
at $\rho=r_*=R_*^+$.
The compactification variable $\rho(r_*)$ maps $[R_*^+,\infty)$ in $[R_*^+,S]$.
The time coordinate $\tau$ is determined requiring that the time Killing vector
is the same in the two coordinate systems, and that the outgoing light-rays are left invariant. 
Within these choices, the surface $\rho=S$ corresponds to future null infinity, $S^+$,
where the waveform is extracted,
avoiding in this way errors linked to finite-distance extraction. 
The grid is uniform with radial step $\Delta \rho = 0.02$, and the time evolution
is performed with a 4th-order Runge-Kutta scheme.

The amplitudes of the (2,2) Zerilli waveforms that corresponds to the dynamics discussed in
the previous section are shown in the right panel of Fig.~\ref{fig:dyn_captures_rwz}. An interesting feature
is that the waveform generated near the apastron is non-vanishing, and its relevance increase
as the particle loses energy; this is particularly evident for the configurations
with many encounters before merger. 
Since this portion of the waveform corresponds to a low orbital (and thus waveform)
frequency, this feature is lost when integrating with a frequency cut-off, as in the FFI methods
that we use to obtain $h_\lm$ from $\psi_4^\lm$ in the comparable mass case. 
In Appendix~\ref{appendix:rwz_integration},
we use these test-mass data to gain insight on the numerical integration of $\psi_4^\lm$
performed in the comparable mass case.

\subsection{Noncircular EOB waveforms in the test-mass limit}
\label{sbsec:rwz_eob}
\begin{figure}[t]
  \centering 
    \includegraphics[width=0.48\textwidth]{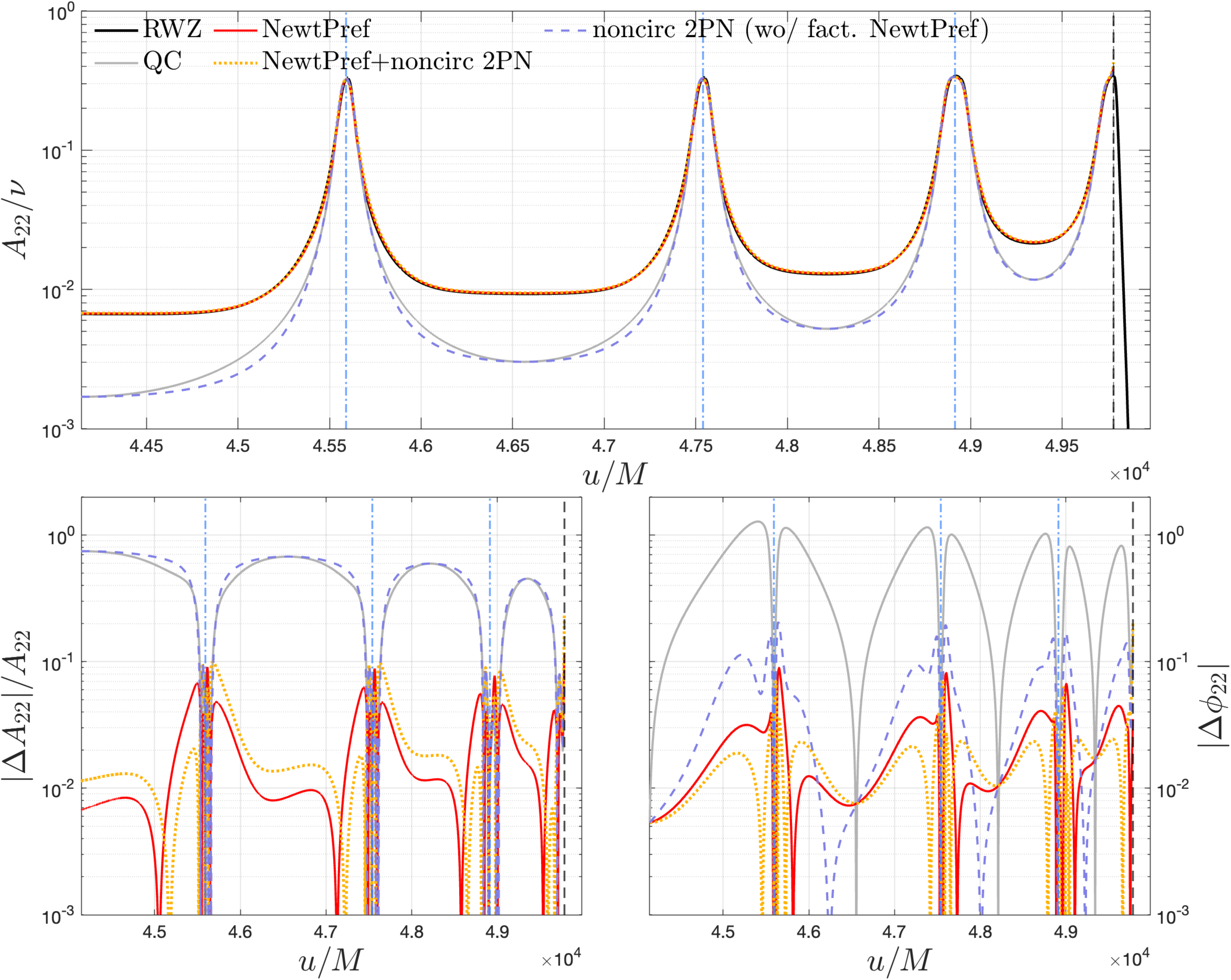}
    \caption{Zerilli (black) and EOB (2,2) waveforms evaluated along 
    the last part of the test-particle dynamics with $p_\varphi^0=4.04897$ 
    (red lines in Fig.~\ref{fig:dyn_captures_rwz}).
    Periastra are marked
    with dash-dotted vertical lines, while the dashed vertical line
    marks the peak of $A_{22}^{\rm RWZ}$.
    We consider different EOB noncircular corrections:
    i) none (solid gray), 
    ii) generic Newtonian prefactor
    (solid red), iii) 2PN with generic NP factorization 
    (dotted orange),
    iv) 2PN without generic NP factorization 
    (dashed blue). In the bottom panels we report the phase
    and relative amplitude differences for each analytical prescription.}
  \label{fig:rwz_eob}
\end{figure}
Capturing the correct waveform behavior in highly eccentric configurations is a delicate matter,
and not all the prescriptions with noncircular information at the same PN order provide
comparable results. We will thus use the Zerilli waveforms discussed in the previous section to test 
the analytical EOB prescriptions for the inspiral waveform
and establish which one is the most accurate.

We start by recalling the analytical structure of the EOB inspiral waveform.
Each multipole is factorized as~\cite{Damour:2008gu,Placidi:2021rkh}
\be
\label{eq:hlm_factorization}
h_\lm = h_\lm^{(N,\epsilon)_{\rm c}} \hat{h}_\lm^{(N,\epsilon)_{\rm nc}}  
       \hat{h}_\lm^{(\epsilon)_{\rm c}} \hat{h}_\lm^{(\epsilon)_{\rm nc}},
\ee
where $\epsilon$ marks the parity of the modes; we have $\epsilon=0$ for even modes ($\l+m$ even),
1 otherwise. The factors in Eq.~\eqref{eq:hlm_factorization} are briefly discussed in the following. 
\begin{itemize}
\item $h_\lm^{(N,\epsilon)_{\rm c}}$ is the Newtonian circular term that, aside from a normalization,
it is given by $\propto \left(r \Omega\right)^{\l+\epsilon} e^{-i m \varphi}$. 
\item $ \hat{h}_\lm^{(N,\epsilon)_{\rm nc}}$ is the noncircular Newtonian correction, written in terms of 
explicit time-derivatives of the radius and frequency~\cite{Chiaramello:2020ehz}. For the (2,2) mode, this correction 
reads 
\be
\label{eq:NewtPref22_wave}
\hat{h}_{22}^{(N, 0)_{\rm nc}}  = 1 - \frac{\ddot{r}}{2 r \Omega^2} - \frac{\dot{r}^2}{2 r^2 \Omega^2} + \frac{2 \mathrm{i} \dot{r}}{r \Omega} + \frac{\mathrm{i} \dot{\Omega}}{2 \Omega^2}.
\ee
This contribution, together with $h_\lm^{(N,\epsilon)_{\rm c}}$, constitutes the {\it generic} Newtonian prefactor. 
\item $\hat{h}_\lm^{(\epsilon)_{\rm c}}$ is the PN circular correction, that is further factorized in 
effective source term,
tail contribution, and residual phase/amplitude series~\cite{Damour:2008gu}. The resummation 
of these series is a crucial aspect of the model, especially for the amplitude corrections known as $\rho_\lm$~\cite{Messina:2018ghh,Nagar:2019wrt}.
\item $\hat{h}_\lm^{(\epsilon)_{\rm nc}}$ is the noncircular PN
correction~\cite{Khalil:2021txt,Placidi:2021rkh,Albanesi:2022xge,Placidi:2023ofj},
where the instantaneous part
is written as discussed in Ref.~\cite{Albanesi:2022xge}, and the eccentric tail
is taken from Ref.~\cite{Placidi:2021rkh}; for the latter,
we consider the expression written in terms of
$\left(r, p_{r_*}, \dot{p}_{r_*} \right)$.
We remark that the eccentric tail is obtained from a low-eccentricity expansion,
and that we consider the expressions up to 2PN. This factor is not included yet in 
\DALI{}, but has been tested in previous works~\cite{Placidi:2021rkh,Albanesi:2022xge}.
\end{itemize}
Due to the finite PN accuracy, the choice of the factorization 
and subsequent resummation is a crucial in order to obtain accurate EOB waveforms. 
In particular, it has been shown in previous works that the factorization of the noncircular Newtonian 
correction $ \hat{h}_\lm^{(N,\epsilon)_{\rm nc}}$ strongly improves the analytical/numerical 
agreement for noncircular planar dynamics~\cite{Placidi:2021rkh,Albanesi:2022ywx,Albanesi:2022xge}. 
We thus perform a similar test also in the dynamical capture scenario considered in this work. 

Among the ones shown in Fig.~\ref{fig:dyn_captures_rwz},  we analyze in detail the configuration
with highest number of close encounters.
Then, in Fig.~\ref{fig:rwz_eob}, we compare the numerical waveform with different analytical prescriptions.
Each analytical waveform is computed up to the amplitude peak of the Zerilli waveform.
We do not complete the analytical waveforms with a ringdown model, nor we apply NQC corrections.
The first analytical prescription that we consider is the quasi-circular one (solid gray), that, not surprisingly,
is not able to correctly reproduce the amplitude and the phase of the Zerilli waveform.
Even the solely inclusion of the Newtonian noncircular correction produces a waveform 
that is far more accurate (solid red), especially at apastron, when the particle reaches the highest
orbital separation from the Schwarzschild black hole. The strong improvement
of the analytical waveform near the apastron was also noted in Ref.~\cite{Albanesi:2022xge}
for eccentric configurations (see Fig.~2 therein). The addition of 2PN noncircular terms
to the Newtonian-factorized prescription 
provides a small correction to the waveform (dotted orange), marginally improving the phase. 
However, the inclusion of 2PN corrections {\it without} factorizing the generic Newtonian prefactor
yields a waveform (dashed light-blue) whose amplitude is not reliable, especially at apastron. Only the phase 
is improved with respect to the quasi-circular waveform, but still performs worse than the prescriptions
with the factorization of the generic Newtonian prefactor. 
A similar result was found in Ref.~\cite{Placidi:2021rkh}, where 
fully PN-expanded prescriptions, such as the one proposed in Ref.~\cite{Khalil:2021txt},
and waveforms with the generic Newtonian prefactor factorized, such as
the one considered in \DALI{}, were tested against numerical data.  
While the phase was accurate for the whole inspiral for both typologies,
the amplitude of the fully PN-expanded waveform performed poorly at apastron,
especially for highly-eccentric binaries (see Fig.~15 and~18 of Ref.~\cite{Placidi:2021rkh}). 
The findings depicted in Fig.~\ref{fig:rwz_eob} 
further support these observations, indicating the necessity to refrain from 
PN-expanding the time-derivatives, at least at Newtonian level, to accurately describe 
the amplitude at apastron.
In this way, these derivatives are evaluated using the 
complete equations of motion rather than the PN-expanded ones.
When computing the waveform
from the dynamics, these derivatives can be easily obtained by numerical differentiation.
However, when higher-order time derivatives are included in the radiation reaction,
an iterative procedure has to be used in order to 
solve the Hamilton's equations with a standard ODE integrator~\cite{Damour:2012ky,Chiaramello:2020ehz}.

\section{Conclusions} 
\label{sec:conc}
Several aspects of this work involved investigating initially unbound comparable mass two-black-hole systems, 
which can result in either scatterings or dynamical captures, depending on the impact of
the back-reaction of the GW emission.
As a part of this study, we have performed NR simulations for 84 physical configurations, 
including nonspinning systems up to mass ratio $q=3$
and equal mass cases, both with and without spin.
Largely guided by the predictions of the state-of-the-art EOB model \DALI{}~\cite{Nagar:2023zxh,Nagar:2024dzj}, 
we have chosen NR initial data with the goal of reproducing a large set of phenomenologies.
Varying the initial energy and angular momentum,
we have then performed
the time evolution of different physical systems 
with the code \GRA{}~\cite{Daszuta:2021ecf,Cook:2023bag}. 
Afterward, we have discussed the post-processing applied to
the output of the NR code 
explaining, in particular, the procedures followed to compute strains, scattering angles, 
and energetics.
Different physical configurations have been further examinated in Sec.~\ref{sbsec:phenom},
focusing on the phenomenology of waveforms and dynamics; in the case of 
coalescing systems, we also studied the properties of the remnants.
Of the different $(q,\chi_1,\chi_2)$-configurations considered in this work, 
the equal mass nonspinning case is the one that has been studied in 
greater detail. 
Regressions of the scattering-capture transition in the latter case have 
been also presented in Sec.~\ref{sbsbsec:phenom_q1_nospin}, 
and subsequently compared with the recent results of Ref.~\cite{kankani:2024may},
finding consistency between the two estimates. 

\texttt{Dal\'i}, the avatar of \TEOB{} considered in this work, 
has been also compared with the aforementioned numerical results in Sec.~\ref{sec:eob_nr}.
After having discussed the prediction of the analytical model across the parameter space,
we have examined its agreement with the NR simulations.
Lending focus on the scattering-capture transition, we found a remarkable 
EOB/NR agreement for initial energies up to $E_0\sim 1.02\,M$;
the EOB/NR agreement also encompass scattering angles, as discussed in Sec.~\ref{sbsec:eobnr_q1_nospin_scat}.
Inevitably, the EOB model loses reliability for higher energies, especially near the transition region,
since the EOB Hamiltonian and the radiation reaction are PN-approximate, and thus not
reliable for high velocities; however, the analytical/numerical agreement is progressively 
restored for large angular momenta, even at the highest energy considered in this work ($E_0\sim 1.05\,M$).

We then shifted our attention to the mismatches between EOB and NR (2,2) waveforms, showing that
the unfaithfulness computed with the LIGO noise is typically below the $1\%$ for the 
configurations with $E_0 \lesssim 1.02 \,M$, 
with a few cases that reach the $3\%$. 
Improving the ringdown model adopted in \DALI{} is necessary
to further enhance the agreement in the direct capture scenario. 
Note that the current post-merger model implemented in our EOB model is indeed informed
only on quasi-circular data, but an extension to generic planar orbits could be achieved 
following the path traced in Ref.~\cite{Carullo:2023kvj}. 
Scatterings show the lowest mismatches among the configurations
examined, while the unfaithfulness for the multiple encounters typically lays 
in between these unbound configurations and direct captures.

In recent years PM computations for the dynamics of two-body systems 
have been conducted using a variety of methods, 
some of which adapted from the field of particle physics, e.g.~\cite{Guevara:2018wpp,Bern:2019nnu,
Bern:2021dqo,Bern:2021yeh,Manohar:2022dea,
Saketh:2021sri,DiVecchia:2021bdo,DiVecchia:2022nna,
Mogull:2020sak,Riva:2021vnj,Jakobsen:2021smu,Jakobsen:2022psy,
Dlapa:2021npj,Dlapa:2021vgp,Dlapa:2022lmu,Dlapa:2024cje,Bini:2021gat,Bini:2022wrq,Bini:2022enm}.
The inclusion of these recent PM results in the EOB dynamics is not a trivial 
task~\cite{Bini:2018ywr,Antonelli:2019ytb,Buonanno:2024vkx}, 
and even the PM predictions for scattering angles need a proper EOB-inspired 
resummation to provide results in agreement with NR simulations, especially for low impact 
parameters~\cite{Damour:2022ybd,Rettegno:2023ghr}.
However, the inclusion of PM results in the EOB dynamics should be further studied to improve EOB 
models in the high velocity regime, where PN results naturally lose their reliability.  

Finally, the test-mass limit studied in Sec.~\ref{sec:testmass} 
reaffirms its utility as a controlled laboratory setting for gaining deeper 
insights into properties that are also applicable to comparable mass binaries.
In particular, the numerical waveforms found by solving the Zerilli equation have been used to: 
i) corroborate the higher
accuracy of the analytical noncircular corrections proposed 
in Refs.~\cite{Chiaramello:2020ehz,Placidi:2021rkh,Albanesi:2022xge}
compared to noncircular corrections fully expanded with the PN equations of motion, ii) gain
insights on the numerical integration procedure carried out in the NR case. 

On the numerical side, future explorations will involve 
different aspects. First of all, producing more accurate numerical waveforms for scattering
and captures is of primary interest, since they are the benchmark to validate and calibrate semi-analytical models.
While the standard extraction of the Weyl scalar at finite distance can still be used to obtain 
the strain by numerical integration, CCE~\cite{Bishop:1998uk,Babiuc:2010ze} or metric perturbations extraction~\cite{Abrahams:1995gn,Pazos:2006kz,Bishop:2016lgv}
would probably provide more reliable results. Second, while we started to explore also unequal mass configurations
and spin effects, a more refined exploration is needed in order to extend the calibration,
and thus the reliability, of analytical models. 

Ultimately, the outcomes of this research underline the synergies between numerical and analytical methods. 
While numerical simulations are crucial for validating and informing semi-analytical models, 
predictions from robust analytical techniques can offer a preliminary exploration of the parameter 
space and indicate the regions where NR simulations should be focused.

\begin{acknowledgments}
  S.A. and A.N. would like to thank T. Damour, C. Lousto, and J. Healy for useful
  discussions. 
  S.A. also expresses gratitude for the hospitality and stimulating environment of the IHES, 
  where part of this work was conducted. This visit was supported by the
  {\it “2021 Balzan Prize for Gravitation: Physical and Astrophysical Aspects”}, 
  awarded to T. Damour.
  S.A. acknowledges support from the Deutsche Forschungsgemeinschaft (DFG) project ``GROOVHY'' 
  (BE 6301/5-1 Projektnummer: 523180871).
  A.R. and D.R. acknowledge support from NASA under award No. 80NSSC21K1720.
  D.R. acknowledges U.S. Department of Energy, Office of Science, Division of
  Nuclear Physics under Award Number(s) DE-SC0021177.
  R.G. acknowledges support by the Deutsche Forschungsgemeinschaft (DFG) under Grant No.
  406116891 within the Research Training Group RTG 2522/1 and from NSF Grant PHY-2020275
  (Network for Neutrinos, Nuclear Astrophysics, and Symmetries (N3AS)).
  S.B. and B.D. knowledges support by the EU Horizon under ERC Consolidator Grant,
  no. InspiReM-101043372. 
  S.B., B.D. and F.Z. acknowledge support by the EU H2020 under ERC Starting
  Grant, no.~BinGraSp-714626.
  D.R. acknowledges support from the Sloan Foundation.
  
  The numerical simulations were performed on the national HPE Apollo Hawk 
  at the High Performance Computing Center Stuttgart (HLRS). 
  The authors acknowledge HLRS for funding this project by providing access 
  to the supercomputer HPE Apollo Hawk under the grant numbers INTRHYGUE/44215
  and MAGNETIST/44288. 
  Simulations were also performed on TACC's Frontera (NSF LRAC
  allocation PHY23001) and on Perlmutter. This research used resources
  of the National Energy Research Scientific Computing Center, a DOE
  Office of Science User Facility supported by the Office of Science of
  the U.S.~Department of Energy under Contract No.~DE-AC02-05CH11231.
  Simulations were also performed on SuperMUC\_NG at the
  Leibniz-Rechenzentrum (LRZ) Munich.  
  The authors acknowledge the Gauss Centre for Supercomputing
  e.V. (\url{www.gauss-centre.eu}) for funding this project by providing
  computing time on the GCS Supercomputer SuperMUC-NG at LRZ
  (allocations {\tt pn68wi} and {\tt pn36jo}).
  Postprocessing and development run were performed on the ARA cluster
  at Friedrich Schiller University Jena. 
  The ARA cluster is funded in part by DFG grants INST
  275/334-1 FUGG and INST 275/363-1 FUGG, and ERC Starting Grant, grant
  agreement no. BinGraSp-714626.
  The authors are indebted to Beppe Starnazza and Sergio Conforti.
  
  \TEOB{} is publicly available at  
  {\footnotesize \url{https://bitbucket.org/teobresums/teobresums/src/Dali/}}.
   The avatar of \DALI{} used in this work is marked with the tag
  \href{https://bitbucket.org/teobresums/teobresums/commits/tag/2405.20398}{2405.20398}.

\end{acknowledgments}

\appendix 

%
\begin{figure*}[t]
  \centering 
    \includegraphics[width=0.32\textwidth]{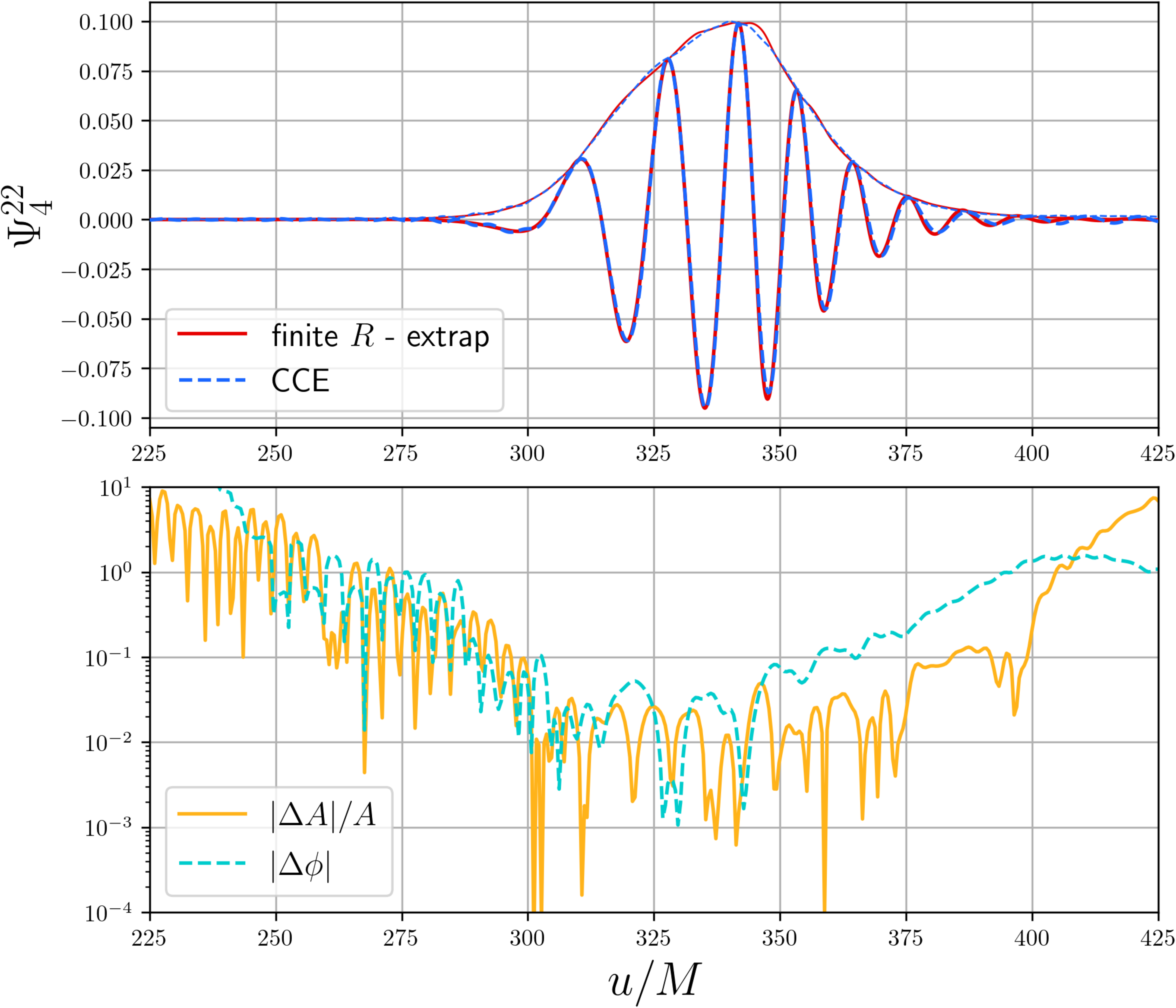}
    \includegraphics[width=0.32\textwidth]{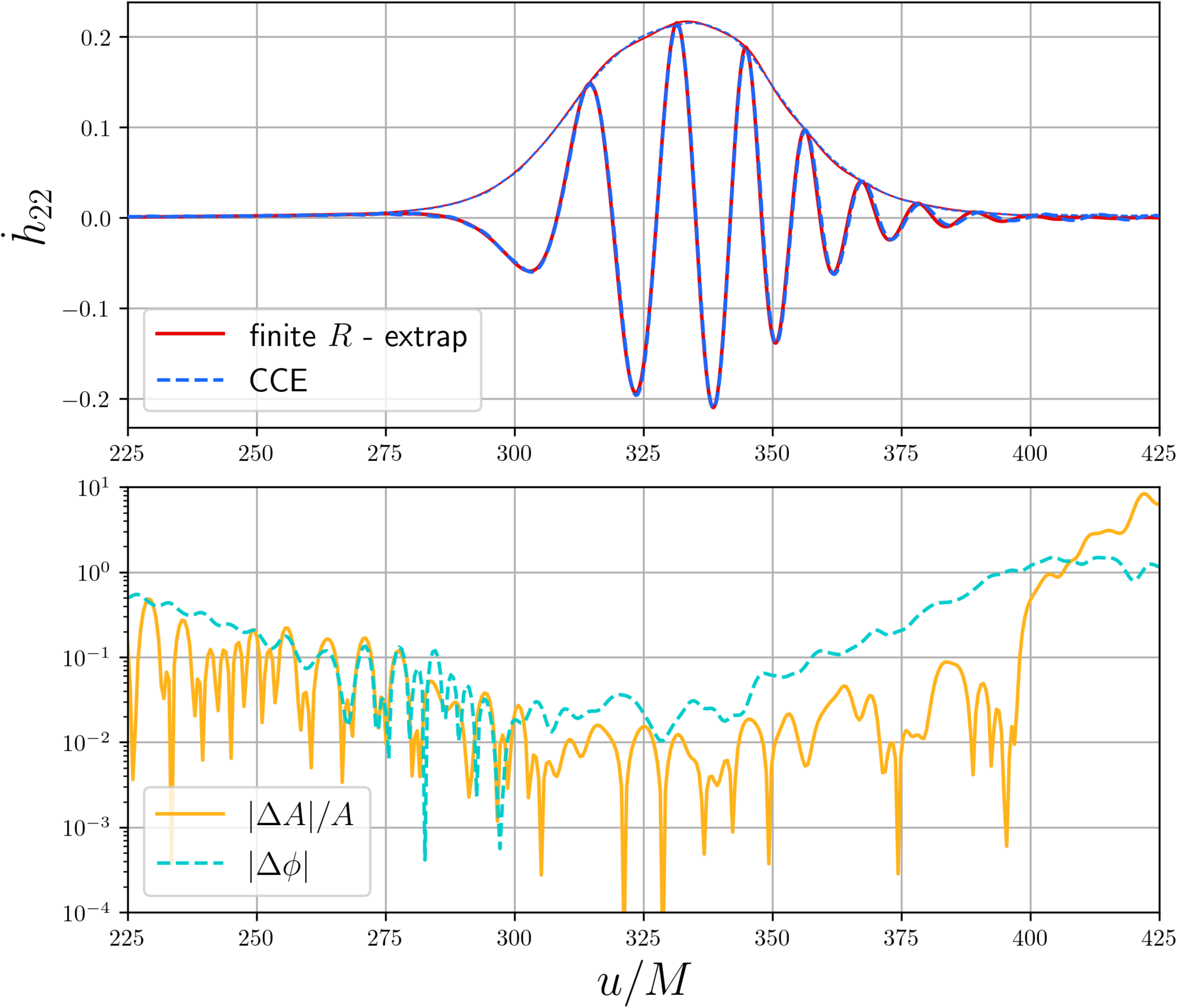}
    \includegraphics[width=0.32\textwidth]{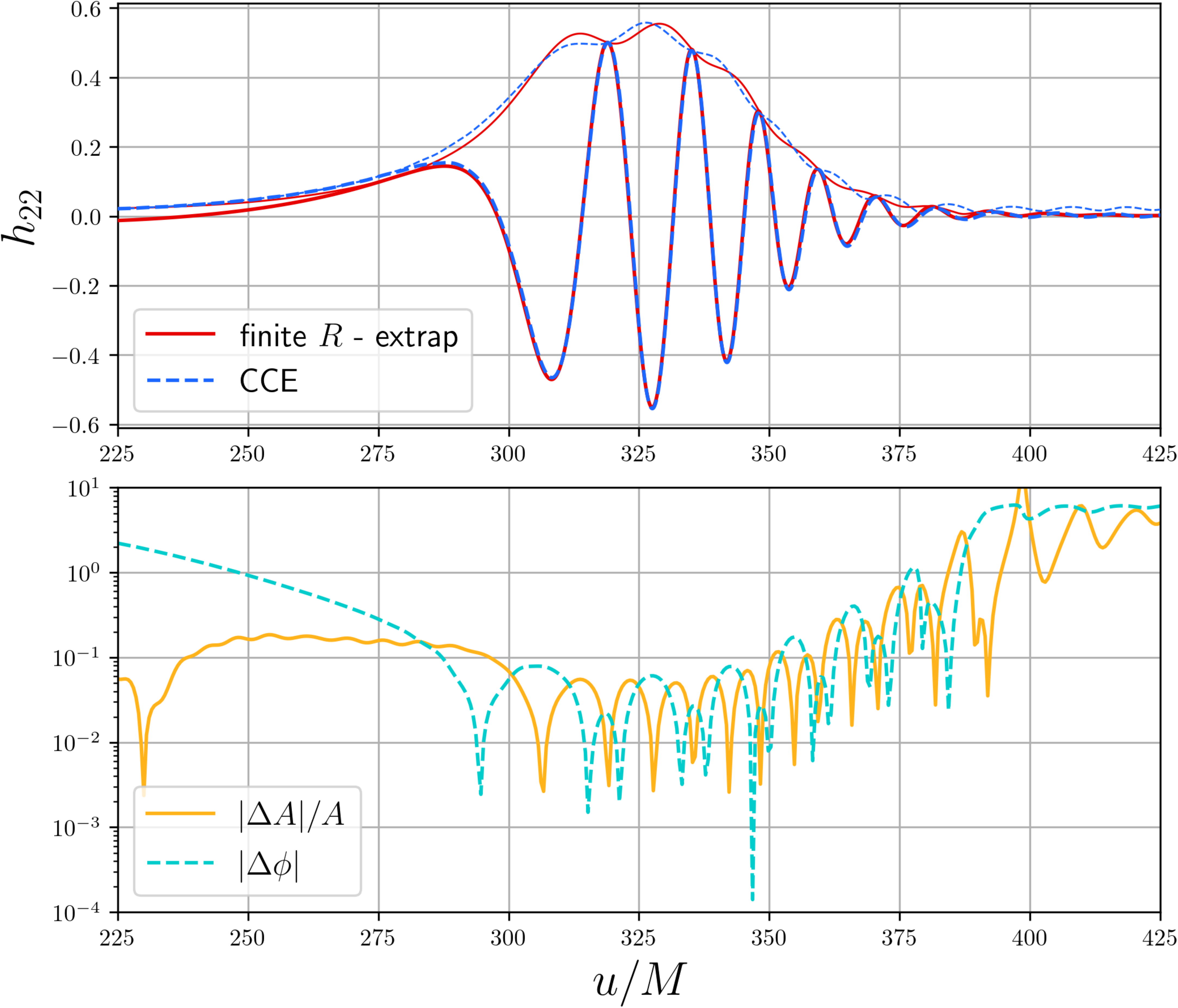}
    \caption{
    Comparisons between extrapolated finite-distance waveforms (solid red) and 
    CCE waveforms (dashed blue) for the equal mass nonspinning configuration
    with $(\hat{E}_0,p_\varphi^0)\simeq(1.011, 3.950)$. We show, from left to right,  
    $\psi_4^{22}$, $\dot{h}_{22}$, and $h_{22}$. In the CCE case, we extract the News at
    infinity and we obtain the Weyl scalar and the strain by numerical
    derivation and integration, respectively. Finite-distance waveforms
    are obtained in post-processing from $\psi_4$ as detailed
    in Sec.~\ref{sbsbsec:postproc_waves}. 
    In the bottom panels we show 
    the phase differences in radians (dashed cyan) 
    and the relative amplitude differences (solid orange).}
  \label{fig:cce_test}
\end{figure*}
\section{CCE waveforms}
\label{app:cce_waves}
As mentioned in Sec.~\ref{sbsbsec:postproc_waves},
we also computed CCE waveforms
using the public code \texttt{PITTNull}~\cite{Bishop:1998uk,Babiuc:2010ze}
for some configurations. 
The metric fields are expanded~\cite{Babiuc:2010ze} and 
the coefficients are stored during 
the \GRA{} evolution at every $10$th time step of evolution and for eight different radial 
intervals, ranging from $[76,84]\,M$ to $[296,304]\,M$ (see midpoints below).
These expansion coefficients are used to construct the initial data in 
the \texttt{PITTNull} code. Additionally,
we set the radius of the initial data world tube at the middle of each
extraction interval, so that we have $R_w = \lbrace 80,90,100,120,140,160,200,300\rbrace\,M$. 

Unfortunately, the expansion coefficients include numerical noises~\cite{Babiuc:2010ze}; 
leaving these numerical
noises untreated and carrying them into the initial data of the \texttt{PITTNull} code,
introduce even larger noises in the final solution of nonlinear characteristic
equations~\cite{Babiuc:2010ze}. 
As such, we remove the high-frequency noises using the built-it filter \texttt{fftwfilter} 
with a max frequency of $\omega_{\rm max} = 1.5\, M^{-1}$.

CCE data for a significant physical configuration are reported
in Fig.~\ref{fig:cce_test} (dashed blue).
More precisely, we consider the News given in output
by \texttt{PITTNull} for $R_w=200\,M$. 
The Weyl scalar is then obtained by numerical differentiation with a 4th-order centered stencil scheme;
we checked that this result is equivalent to the Weyl scalar given directly by \texttt{PITTNull}.
The CCE strain is instead obtained by applying an FFI to the News with cut-off frequency $f_0 = 0.007\, M^{-1}$.
Despite the need of just one integration step, time-domain methods seem to not yield 
reliable strains; this issue may be linked to high-frequency noise still present in our CCE data. 
We select a large radius because the initial portion of the News for small world tubes 
exhibits an unphysical bump, which tends to diminish only for larger radii. 
This is likely associated with the proximity to the punctures at the beginning of the simulation.
Since the CCE waveform with $R_w = 300 \,M$ is visibly affected
by numerical dissipation, we consider the previous radius, i.e., $R_w = 200 \,M$.

In Fig.~\ref{fig:cce_test} we also report the extrapolated finite-distance waveform,
computed by integrating the extrapolated Weyl scalar, that is obtained from Eq.~\eqref{eq:extrapolation}
using $\psi_4^{22}$ extracted at $R=100\,M$.
The integration steps are performed in a similar manner to the CCE case, employing an FFI with 
$f_0 = 0.007\, M^{-1}$, which is also the standard procedure used throughout 
this paper for equal mass binaries, as detailed in Sec.~\ref{sbsbsec:postproc_waves}.

The relative amplitude differences (solid orange) and the phase differences (dashed light-blue) 
between the two different methods are reported in the bottom panels of Fig.~\ref{fig:cce_test}.
As can be seen, these difference are quite large for the precursor and the ringdown.
This is however expected, since in these cases the signal is weaker, 
and thus more difficult to resolve and contaminated by numerical noise. We thus mainly focus
our discussion on the GWs generated by the close encounter and the subsequent merger.

The comparison reveals that the amplitude difference for $\psi_4^{22}$ is not completely negligible, 
even in the aforementioned time window, where it oscillates between $\sim 0.2\%$ and $\sim 4\%$. 
The phase difference reaches at most the 0.05 radians before merger, but rapidly increases afterwards. 
Similar discrepancies are observed for $\dot{h}_{22}$. 
The differences for the strain are instead higher, reaching $|\Delta A|/A\sim 10\%$ and
$|\Delta \phi|\sim 0.1$ radians even during the close encounter/merger.
The amplitudes of the strain reported in the right
upper panel show a priori unexpected oscillations, both in the CCE
and extrapolated finite-distance waveforms. This is likely to be an artefact
of the FFI, since these oscillations increase for lower cut-off, 
and decrease for higher ones. 

The test performed in this appendix indicates that obtaining accurate CCE waveforms
for dynamical capture is not straightforward and poses several challenges.
However, the presence of less integration steps with respect
to the extrapolation of finite distance waveforms, 
together with the direct computation at future null infinity,
lead us to think that, with improved 
CCE data, it should be possible to obtain more accurate results. 
CCE waveforms should be revisited in future works, considering for example 
improved high-frequency filters.
We postpone a more thoughtful analysis to the future. 

\section{Integration tests in the extreme mass-ratio limit}
\label{appendix:rwz_integration}
\begin{figure*}[t]
  \centering 
    \includegraphics[width=0.49\textwidth]{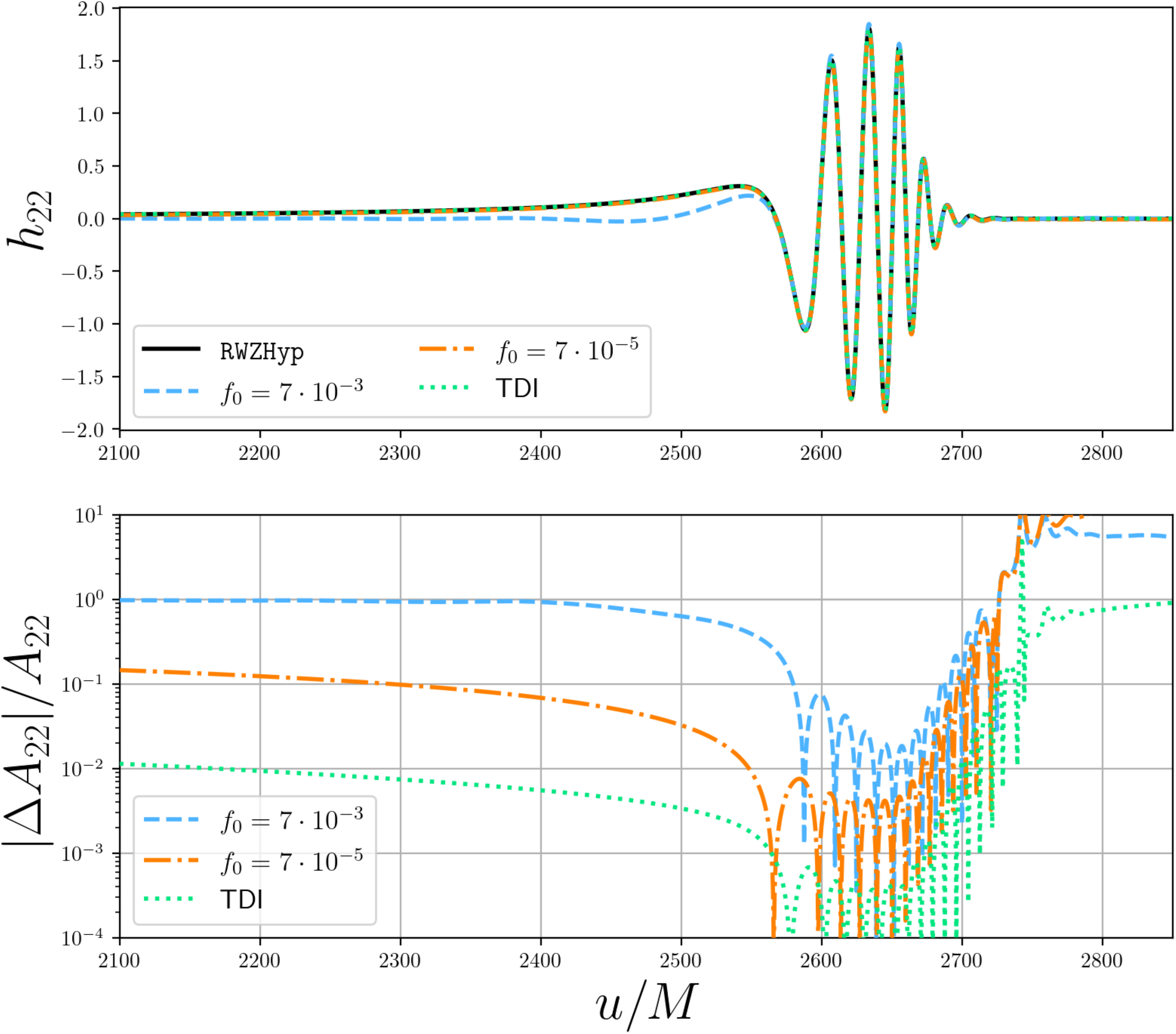}
    \includegraphics[width=0.49\textwidth]{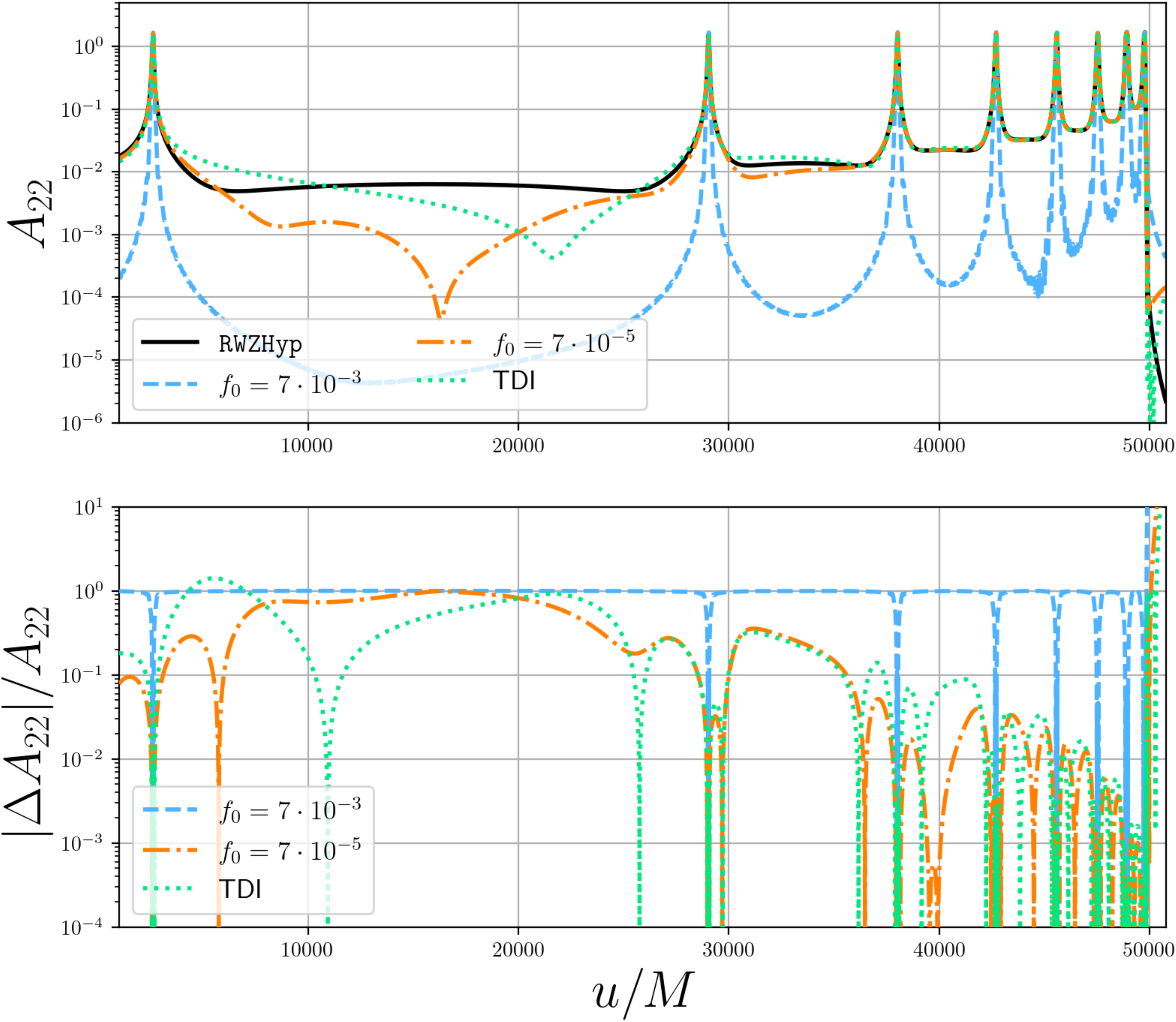}
    \caption{Left panels: Real part of the (2,2) Zerilli waveform obtained 
    by solving numerically Eq.~\eqref{eq:rwz} with \RWZ{} 
    along the a dynamics with $\nu=10^{-2}$, $r_0=300$, $\hat{E}_0=1.000001$, and $p_\varphi^0=3.988024$. 
    This waveform (black) 
    is compared with different integration tests, obtained by double-integrating the second time derivative
    of the solution. We consider FFI with two different 
    frequency cut-offs, $f_0=7\cdot 10^{-3}$ (dashed blue) and $f_0=7\cdot 10^{-5}$ (dot-dashed orange), 
    and a time-domain integration where the numerical drift is removed by subtracting a fitted constant
    at each integration step (dotted green). The relative amplitude difference with the original 
    Zerilli waveform are shown in the bottom panel with the same color scheme.
    Right panels: similar analysis for the configuration with $p_\varphi^0=4.048973$; in the upper panel
    we show the amplitude in logarithmic scale rather than the real part.}
  \label{fig:rwz_integration}
\end{figure*}
Waveforms generated by test-particle dynamics can be obtained by numerically integrating
the RWZ equations~\eqref{eq:rwz}. To accomplish this task,
we use the time-domain code \RWZ{}, that extracts the waveform at future null infinity.
On the contrary, the NR code \GRA{} produces the Weyl scalar $\psi_4^\lm=\ddot{h}_\lm$ at finite 
distance (if no CCE is used), so that in order to obtain the strain at future null infinity
we need three integration steps: one to extrapolate $\psi_4^\lm$ using Eq.~\eqref{eq:extrapolation}, and
two additional steps to obtain $h_\lm$. As argued in Sec.~\ref{sbsec:postproc}, through this paper
we have used an FFI with cut-off at $f_0=0.007\,M^{-1}$. This precise value is not
strictly physically motivated, but cut-off at higher frequencies would remove a large portion of the signal, 
while too low cut-off frequencies would generate evident drifts in the NR waveform. 

We can thus use the RWZ waveforms to test the TD and FD integrations. 
After having obtained $h_{22}$ by numerically solving the Zerili equation with \RWZ{}, we compute 
$\psi_4^{22}$ by computing the second-time derivative with a 4th-order centered
stencil scheme. The Weyl scalar is then integrated back with different methods.
We start by considering, as a show-case example, the test-mass dynamical capture with lowest angular momentum,
that is indeed a direct capture.
We start by considering an FFI with cut-off frequency $f_0=0.007 \,M^{-1}$, similarly to 
the NR case; the result is shown in the left panels of Fig.~\ref{fig:rwz_integration} with a dashed blue line.
Comparing this waveform with the one obtained directly from \RWZ{}
(black line), it is immediately clear the precursor of the FFI waveform is not reliable. This
fact is made even more explicit by the relative amplitude difference shown in the bottom panel. 
The FFI wave becomes more accurate for the merger-ringdown signal, but the discrepancies with
the \RWZ{} amplitude are still between the $1-8\%$. 
Since the test-mass waveforms produced by \RWZ{} are not contaminated by noise as much as
the full NR ones, we can consider a lower cut-off frequency.
The result for $f_0=7\cdot 10^{-5}$ is shown with a dot-dashed orange line. This
waveform is more reliable during the whole evolution with respect to the previous case, but especially for the
low-frequency precursor. However, the amplitude of the precursor is still off by a
$10\%$, while the error on the amplitude during the merger
drops below the $1\%$. Smaller cut-off frequencies do not improve this result, and eventually lead to 
greater discrepancies. An even more accurate waveform can be obtained with a time-domain integration,
where at each step we subtract an integration constant determined by a fit on the last portion of 
the post-merger waveform (after the quasi-normal-mode ringing). The latter result
is shown with a dotted green line. The relative amplitude difference with
the precursor of the \RWZ{} waveform is at the $1\%$ level, and the difference drops below
the $10^{-3}$ threshold during the merger/early ringdown. 
Similar results can be obtained also for the other test-mass configurations with more encounters, 
as shown by the case reported in the right panels of Fig.~\ref{fig:rwz_integration}.
Note, however, that for this very long signal, none of the integration methods is 
able to correctly reproduce the waveform at the first apastron passage. The accuracy
of all the integration methods increases with the proximity to the merger. 

This results show that, when applicable, time-domain methods should be preferred to FFI.
However, in the NR case the former method works only for a few cases, typically for signals with a short 
duration, such as direct captures. Longer waveforms, like those generated by double encounters,
require an FFI. While the test-mass results seem to suggest
that the cut-off frequency used in the NR case is rather large, we recall that the FFI procedure
adopted in this paper
yields reliable remanent properties, that are computed from the energetics $\hat{E}_b(j)$, and thus from 
the waveform through Eqs.~\eqref{eq:fluxes}. 
This was explicitly checked in Sec.~\ref{sbsec:remnants}, where the differences
between remnant properties obtained with FFI and time-domain integrations were
$(\Delta \hat{M}, \, \Delta \hat{a}_f) = (0.0015,\,0.0058)$
for the nonspinning equal mass configuration with 
$(\hat{E}_0,\,p_\varphi^0) \simeq (1.011, \,3.950)$.
Similarly, for this configuration the EOB/NR mismatch
was not significantly affected by the integration method
in the mass range $M\in[100,300] M_\odot$, as discussed at the end
of Sec.~\ref{sbsec:eobnr_mm}.

\section{Merger time for generic orbits}
\label{appendix:tmrg}
For spin-aligned, equal mass, quasi-circular BBHs, the merger time $t_{\rm mrg}$
is typically defined as the global peak of the (2,2) amplitude.
However, in highly eccentric binaries, the highest peak might not be the one
associated with the merger of the two black holes. Indeed, it can happen that bursts generated by previous
close encounters have higher peaks,
as already noted in Ref.~\cite{Gold:2012tk,Carullo:2023kvj}.
For this reason, we propone to define the merger time $t_{\rm mrg}$
as the time that corresponds to the {\it last} peak of the (2,2) amplitude.
To avoid considering local peaks associated to mode-mixing, rather than to the coalescence
of the two black holes, we also require that the peak is above the threshold
$\epsilon_A = 0.2 A^{\max}_{22}$, where $A^{\max}_{22} \equiv \max(A_{22})$ is
the global maximum of the amplitude. 
\begin{figure*}[t]
  \centering
    \includegraphics[width=0.99\textwidth]{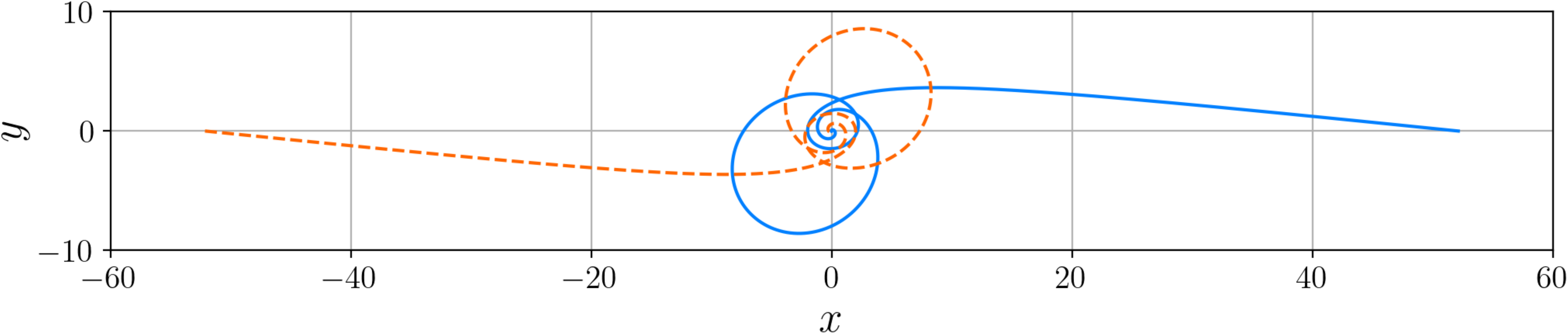}\\
    \vspace{0.3cm}
    \includegraphics[width=0.96\textwidth]{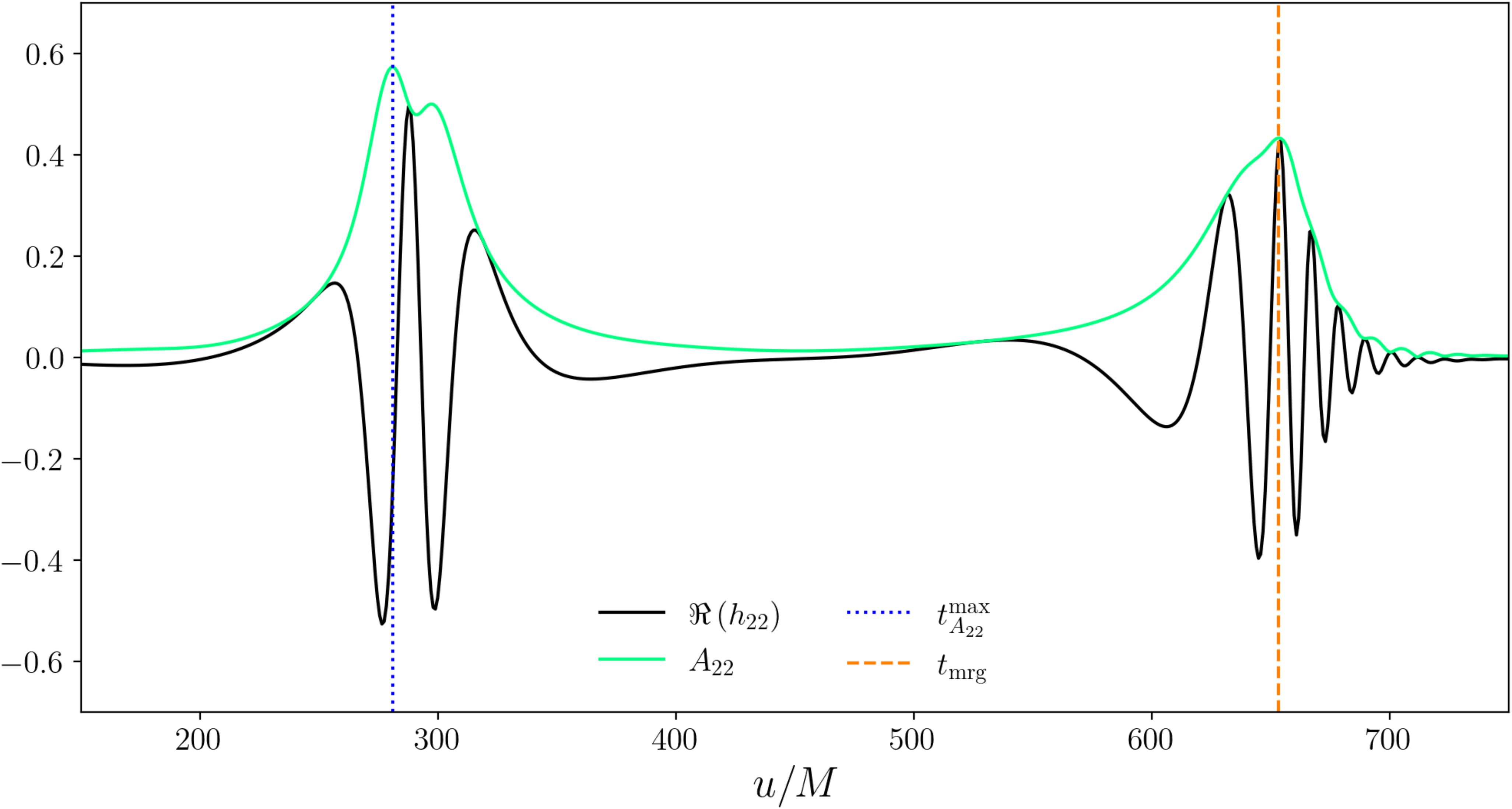}
    \caption{Upper panel: tracks of the punctures for the nonspinning equal mass configuration
    with $(\hat{E}_0,p_\varphi^0)\simeq(1.016, 4.180)$. Lower panel:
    real part and amplitude of the corresponding (2,2) waveform (black and green, respectively). 
    The dashed orange
    line marks the merger time, defined as the last $A_{22}$ peak, while the blue dotted
    line highlight the global maximum of the amplitude. The visible artefacts in the amplitudes are
    linked to integration and are also discussed in Appendix~\ref{appendix:tmrg}. }
  \label{fig:tmrg}
\end{figure*}

We report in Fig.~\ref{fig:tmrg} an example that clarifies the necessity of this definition. We consider
an equal mass nonspinning double encounter with initial energy $E_0 \simeq 1.016\,M$ and angular momentum
$p_\varphi^0 \simeq 4.180$. The global maximum of the invariant amplitude occurs at $u \simeq 280\,M$, in correspondence
of the first close encounter, where $A_{22}^{\rm max}\sim 0.57$. On the other hand, the last maximum
occurs at $u\simeq 650\,M$, shortly before the beginning of the ringdown phase,
where the amplitude only reaches $A_{22}^{\rm mrg}\sim 0.43$, so that 
the ratio with the global maximum is $A_{22}^{\rm mrg}/A_{22}^{\rm max}\sim 0.75$.
Higher initial energies can even decrease this proportion;
for the equal mass nonspinning configuration with $(E_0, p_\varphi^0)\simeq(1.051, 4.755)$, we get
$A_{22}^{\rm mrg}/A_{22}^{\rm max} \sim 0.52$. For higher energies than the ones considered in this work, 
the ratio might be even
more unbalanced, but we recall that the width of the double encounter region of the parameter space
shrinks for increasing energies,
making these configurations less likely to occur (see, e.g., Fig.~\ref{fig:parspace_TEOB_q1_nospin}
and~\ref{fig:parspace_TEOB_4}). 

We also make a remark on the effects of the integration procedure. As can be clearly seen
from the amplitude reported in Fig.~\ref{fig:tmrg}, $A_{22}$ exhibits a double peak
in correspondence of the first encounter. However, this is only an artefact linked to the
cut-off frequency used in the FFI. Indeed, higher cut-offs yield only one peak,
but also remove more physical
frequencies from the signal. As a consequence, fluxes computed with greater
cut-offs are underestimated. 
This further suggests that the post-processing of $\psi_4$ for dynamical captures and scattering is indeed 
a delicate matter, and alternative methods to extract the strain should be considered. 

We conclude this discussion by emphasizing that in cases where higher modes significantly contribute to the strain, 
such as BBHs with higher mass ratios or precessing dynamics, 
it is more appropriate to consider the invariant amplitude $A_{\rm inv}$
instead of the (2,2) waveform amplitude. This quantity is defined as~\cite{Schmidt:2017btt}
\be
\label{eq:Ainv}
A_{\rm inv} = \sqrt{\sum_\lm |h_\lm|^2},
\ee
where $h_\lm$ are the multipoles of the waveform in the co-precessing frame. 
In this work we study spin-aligned systems,
so that the difference between $A_{\rm inv}/\sqrt{2}$ and $A_{22}$ is in practice
negligible, especially for the determination of the merger time. Nevertheless, the two quantities
differ for precessing binaries or high mass ratios, so that $A_{\rm inv}$ is in general preferred.

\begin{table*}[t]
	\caption{\label{tab:mm}
	Optimized and not optimized EOB/NR (2,2) mismatches for the configuration shown in Fig.~\ref{fig:mm_q1_nospin}.
	The reference mass here considered is $M = 200\,M_{\odot}$.}
\begin{center}
\begin{ruledtabular}
\begin{tabular}{c c c | c c | c c | c c | c | c} 
$q$ & $\chi_1$ & $\chi_2$ & 
$\hat{E}_0$ & $p_\varphi^0$	& $\hat{E}_0^{\rm opt}$ & $p_\varphi^{0, {\rm opt}}$ &
$\Delta E_0/E_0$ & $\Delta p_\varphi^0/p_\varphi^0$ & 
$\bar{F}$ &  $\bar{F}_{\rm opt}$ \\
\hline
\hline
1 & $\phantom{+}0.0$ & $\phantom{+}0.0$ & 1.00100 & 3.79998 & 1.00194 & 3.84037 & $ \phantom{+}9.33\times 10^{-4}$ & $ \phantom{+}1.06\times 10^{-2}$ & $ 4.356\times 10^{-2}$ & $ 2.681\times 10^{-3}$  \\
1 & $\phantom{+}0.0$ & $\phantom{+}0.0$ & 1.00100 & 3.89998 & 1.00035 & 3.88043 & $ -6.47\times 10^{-4}$ & $ -5.01\times 10^{-3}$ & $ 1.872\times 10^{-1}$ & $ 8.187\times 10^{-3}$  \\
1 & $\phantom{+}0.0$ & $\phantom{+}0.0$ & 1.00100 & 4.01998 & 1.00000 & 3.98654 & $ -1.00\times 10^{-3}$ & $ -8.32\times 10^{-3}$ & $ 3.964\times 10^{-1}$ & $ 4.577\times 10^{-3}$  \\
\hline
1 & $\phantom{+}0.0$ & $\phantom{+}0.0$ & 1.01103 & 3.94996 & 1.01177 & 4.02312 & $ \phantom{+}7.32\times 10^{-4}$ & $ \phantom{+}1.85\times 10^{-2}$ & $ 3.068\times 10^{-2}$ & $ 2.572\times 10^{-2}$  \\
1 & $\phantom{+}0.0$ & $\phantom{+}0.0$ & 1.01103 & 4.01996 & 1.01138 & 4.06664 & $ \phantom{+}3.40\times 10^{-4}$ & $ \phantom{+}1.16\times 10^{-2}$ & $ 2.325\times 10^{-1}$ & $ 4.851\times 10^{-3}$  \\
1 & $\phantom{+}0.0$ & $\phantom{+}0.0$ & 1.01103 & 4.06996 & 1.01002 & 4.06742 & $ -1.00\times 10^{-3}$ & $ -6.24\times 10^{-4}$ & $ 4.890\times 10^{-1}$ & $ 6.911\times 10^{-3}$  \\
1 & $\phantom{+}0.0$ & $\phantom{+}0.0$ & 1.01103 & 4.09996 & 1.01075 & 4.10158 & $ -2.78\times 10^{-4}$ & $ \phantom{+}3.95\times 10^{-4}$ & $ 4.519\times 10^{-1}$ & $ 7.993\times 10^{-3}$  \\
1 & $\phantom{+}0.0$ & $\phantom{+}0.0$ & 1.01103 & 4.12996 & 1.01016 & 4.11037 & $ -8.60\times 10^{-4}$ & $ -4.74\times 10^{-3}$ & $ 3.553\times 10^{-1}$ & $ 1.028\times 10^{-2}$  \\
1 & $\phantom{+}0.0$ & $\phantom{+}0.0$ & 1.01103 & 4.24996 & 1.01205 & 4.27829 & $ \phantom{+}1.00\times 10^{-3}$ & $ \phantom{+}6.66\times 10^{-3}$ & $ 1.962\times 10^{-3}$ & $ 1.364\times 10^{-3}$  \\
1 & $\phantom{+}0.0$ & $\phantom{+}0.0$ & 1.01103 & 4.32496 & 1.01205 & 4.34848 & $ \phantom{+}1.00\times 10^{-3}$ & $ \phantom{+}5.44\times 10^{-3}$ & $ 2.062\times 10^{-3}$ & $ 1.187\times 10^{-3}$  \\
1 & $\phantom{+}0.0$ & $\phantom{+}0.0$ & 1.01103 & 4.39996 & 1.01205 & 4.41653 & $ \phantom{+}1.00\times 10^{-3}$ & $ \phantom{+}3.77\times 10^{-3}$ & $ 2.118\times 10^{-3}$ & $ 9.685\times 10^{-4}$  \\
\hline
1 & $\phantom{+}0.0$ & $\phantom{+}0.0$ & 1.01607 & 4.07001 & 1.01644 & 4.14859 & $ \phantom{+}3.65\times 10^{-4}$ & $ \phantom{+}1.93\times 10^{-2}$ & $ 5.190\times 10^{-2}$ & $ 2.454\times 10^{-2}$  \\
1 & $\phantom{+}0.0$ & $\phantom{+}0.0$ & 1.01607 & 4.12001 & 1.01708 & 4.19632 & $ \phantom{+}1.00\times 10^{-3}$ & $ \phantom{+}1.85\times 10^{-2}$ & $ 3.146\times 10^{-1}$ & $ 5.350\times 10^{-3}$  \\
1 & $\phantom{+}0.0$ & $\phantom{+}0.0$ & 1.01607 & 4.18001 & 1.01543 & 4.19203 & $ -6.30\times 10^{-4}$ & $ \phantom{+}2.87\times 10^{-3}$ & $ 5.055\times 10^{-1}$ & $ 9.132\times 10^{-3}$  \\
1 & $\phantom{+}0.0$ & $\phantom{+}0.0$ & 1.01607 & 4.21002 & 1.01519 & 4.20508 & $ -8.66\times 10^{-4}$ & $ -1.17\times 10^{-3}$ & $ 4.817\times 10^{-1}$ & $ 1.021\times 10^{-2}$  \\
1 & $\phantom{+}0.0$ & $\phantom{+}0.0$ & 1.01607 & 4.30002 & 1.01708 & 4.33832 & $ \phantom{+}1.00\times 10^{-3}$ & $ \phantom{+}8.91\times 10^{-3}$ & $ 2.470\times 10^{-3}$ & $ 2.277\times 10^{-3}$  \\
1 & $\phantom{+}0.0$ & $\phantom{+}0.0$ & 1.01607 & 4.40002 & 1.01708 & 4.42930 & $ \phantom{+}1.00\times 10^{-3}$ & $ \phantom{+}6.66\times 10^{-3}$ & $ 2.442\times 10^{-3}$ & $ 1.936\times 10^{-3}$  \\
1 & $\phantom{+}0.0$ & $\phantom{+}0.0$ & 1.01607 & 4.50002 & 1.01708 & 4.51600 & $ \phantom{+}1.00\times 10^{-3}$ & $ \phantom{+}3.55\times 10^{-3}$ & $ 2.503\times 10^{-3}$ & $ 1.459\times 10^{-3}$  \\
1 & $\phantom{+}0.0$ & $\phantom{+}0.0$ & 1.01607 & 4.60002 & 1.01708 & 4.59917 & $ \phantom{+}1.00\times 10^{-3}$ & $ -1.85\times 10^{-4}$ & $ 2.853\times 10^{-3}$ & $ 1.224\times 10^{-3}$  \\
\hline
1 & $-0.5$ & $-0.5$ & 1.01619 & 4.60002 & 1.01709 & 4.63381 & $ \phantom{+}8.88\times 10^{-4}$ & $ \phantom{+}7.35\times 10^{-3}$ & $ 2.901\times 10^{-2}$ & $ 2.714\times 10^{-2}$  \\
1 & $-0.5$ & $-0.5$ & 1.01619 & 4.63002 & 1.01720 & 4.66663 & $ \phantom{+}1.00\times 10^{-3}$ & $ \phantom{+}7.91\times 10^{-3}$ & $ 2.249\times 10^{-2}$ & $ 1.808\times 10^{-2}$  \\
1 & $-0.5$ & $-0.5$ & 1.01619 & 4.66002 & 1.01720 & 4.69164 & $ \phantom{+}1.00\times 10^{-3}$ & $ \phantom{+}6.79\times 10^{-3}$ & $ 1.484\times 10^{-1}$ & $ 8.890\times 10^{-3}$  \\
1 & $-0.5$ & $-0.5$ & 1.01619 & 4.67002 & 1.01628 & 4.67213 & $ \phantom{+}8.85\times 10^{-5}$ & $ \phantom{+}4.52\times 10^{-4}$ & $ 3.726\times 10^{-1}$ & $ 1.456\times 10^{-2}$  \\
1 & $-0.5$ & $-0.5$ & 1.01619 & 4.72002 & 1.01624 & 4.69562 & $ \phantom{+}5.23\times 10^{-5}$ & $ -5.17\times 10^{-3}$ & $ 3.914\times 10^{-1}$ & $ 1.753\times 10^{-2}$  \\
1 & $-0.5$ & $-0.5$ & 1.01619 & 4.75002 & 1.01720 & 4.74797 & $ \phantom{+}1.00\times 10^{-3}$ & $ -4.31\times 10^{-4}$ & $ 1.411\times 10^{-2}$ & $ 5.101\times 10^{-3}$  \\
\hline
1 & $\phantom{+}0.5$ & $-0.5$ & 1.01619 & 4.12001 & 1.01668 & 4.18427 & $ \phantom{+}4.79\times 10^{-4}$ & $ \phantom{+}1.56\times 10^{-2}$ & $ 2.931\times 10^{-1}$ & $ 5.958\times 10^{-3}$  \\
1 & $\phantom{+}0.5$ & $-0.5$ & 1.01619 & 4.30002 & 1.01720 & 4.33893 & $ \phantom{+}1.00\times 10^{-3}$ & $ \phantom{+}9.05\times 10^{-3}$ & $ 2.424\times 10^{-3}$ & $ 2.225\times 10^{-3}$  \\
\hline
1 & $\phantom{+}0.5$ & $\phantom{+}0.5$ & 1.01619 & 3.40001 & 1.01527 & 3.53132 & $ -9.05\times 10^{-4}$ & $ \phantom{+}3.86\times 10^{-2}$ & $ 4.713\times 10^{-2}$ & $ 4.061\times 10^{-2}$  \\
1 & $\phantom{+}0.5$ & $\phantom{+}0.5$ & 1.01619 & 3.50001 & 1.01560 & 3.63578 & $ -5.81\times 10^{-4}$ & $ \phantom{+}3.88\times 10^{-2}$ & $ 3.389\times 10^{-1}$ & $ 1.524\times 10^{-2}$  \\
1 & $\phantom{+}0.5$ & $\phantom{+}0.5$ & 1.01619 & 3.60001 & 1.01517 & 3.65271 & $ -1.00\times 10^{-3}$ & $ \phantom{+}1.46\times 10^{-2}$ & $ 3.757\times 10^{-1}$ & $ 2.242\times 10^{-2}$  \\
1 & $\phantom{+}0.5$ & $\phantom{+}0.5$ & 1.01619 & 3.70001 & 1.01517 & 3.69726 & $ -1.00\times 10^{-3}$ & $ -7.45\times 10^{-4}$ & $ 4.680\times 10^{-1}$ & $ 2.838\times 10^{-2}$  \\
1 & $\phantom{+}0.5$ & $\phantom{+}0.5$ & 1.01619 & 4.07001 & 1.01721 & 4.12375 & $ \phantom{+}1.00\times 10^{-3}$ & $ \phantom{+}1.32\times 10^{-2}$ & $ 1.056\times 10^{-3}$ & $ 1.140\times 10^{-3}$  \\
1 & $\phantom{+}0.5$ & $\phantom{+}0.5$ & 1.01619 & 4.12001 & 1.01721 & 4.16602 & $ \phantom{+}1.00\times 10^{-3}$ & $ \phantom{+}1.12\times 10^{-2}$ & $ 9.787\times 10^{-4}$ & $ 1.037\times 10^{-3}$  \\
\hline
2 & $\phantom{+}0.0$ & $\phantom{+}0.0$ & 1.01609 & 4.15000 & 1.01616 & 4.22411 & $ \phantom{+}7.02\times 10^{-5}$ & $ \phantom{+}1.79\times 10^{-2}$ & $ 4.211\times 10^{-2}$ & $ 2.970\times 10^{-2}$  \\
2 & $\phantom{+}0.0$ & $\phantom{+}0.0$ & 1.01609 & 4.20000 & 1.01619 & 4.26516 & $ \phantom{+}9.95\times 10^{-5}$ & $ \phantom{+}1.55\times 10^{-2}$ & $ 2.781\times 10^{-1}$ & $ 9.421\times 10^{-3}$  \\
2 & $\phantom{+}0.0$ & $\phantom{+}0.0$ & 1.01609 & 4.25001 & 1.01517 & 4.26705 & $ -9.05\times 10^{-4}$ & $ \phantom{+}4.01\times 10^{-3}$ & $ 4.232\times 10^{-1}$ & $ 1.131\times 10^{-2}$  \\
2 & $\phantom{+}0.0$ & $\phantom{+}0.0$ & 1.01609 & 4.35001 & 1.01711 & 4.40510 & $ \phantom{+}1.00\times 10^{-3}$ & $ \phantom{+}1.27\times 10^{-2}$ & $ 4.037\times 10^{-3}$ & $ 2.656\times 10^{-3}$  \\
2 & $\phantom{+}0.0$ & $\phantom{+}0.0$ & 1.01609 & 4.40001 & 1.01711 & 4.44775 & $ \phantom{+}1.00\times 10^{-3}$ & $ \phantom{+}1.09\times 10^{-2}$ & $ 2.640\times 10^{-3}$ & $ 2.522\times 10^{-3}$  \\
\hline
2 & $\phantom{+}0.0$ & $\phantom{+}0.0$ & 1.02014 & 4.25010 & 1.02066 & 4.34935 & $ \phantom{+}5.14\times 10^{-4}$ & $ \phantom{+}2.34\times 10^{-2}$ & $ 5.728\times 10^{-2}$ & $ 3.010\times 10^{-2}$  \\
2 & $\phantom{+}0.0$ & $\phantom{+}0.0$ & 1.02014 & 4.25010 & 1.02108 & 4.36016 & $ \phantom{+}9.18\times 10^{-4}$ & $ \phantom{+}2.59\times 10^{-2}$ & $ 5.772\times 10^{-2}$ & $ 3.023\times 10^{-2}$  \\
2 & $\phantom{+}0.0$ & $\phantom{+}0.0$ & 1.02014 & 4.27510 & 1.02091 & 4.37650 & $ \phantom{+}7.51\times 10^{-4}$ & $ \phantom{+}2.37\times 10^{-2}$ & $ 1.353\times 10^{-1}$ & $ 1.661\times 10^{-2}$  \\
2 & $\phantom{+}0.0$ & $\phantom{+}0.0$ & 1.02014 & 4.30010 & 1.01975 & 4.36402 & $ -3.81\times 10^{-4}$ & $ \phantom{+}1.49\times 10^{-2}$ & $ 3.470\times 10^{-1}$ & $ 1.166\times 10^{-2}$  \\
2 & $\phantom{+}0.0$ & $\phantom{+}0.0$ & 1.02014 & 4.32510 & 1.02010 & 4.38444 & $ -3.41\times 10^{-5}$ & $ \phantom{+}1.37\times 10^{-2}$ & $ 3.622\times 10^{-1}$ & $ 1.280\times 10^{-2}$  \\
2 & $\phantom{+}0.0$ & $\phantom{+}0.0$ & 1.02014 & 4.35010 & 1.01926 & 4.37649 & $ -8.64\times 10^{-4}$ & $ \phantom{+}6.07\times 10^{-3}$ & $ 4.040\times 10^{-1}$ & $ 1.362\times 10^{-2}$  \\
2 & $\phantom{+}0.0$ & $\phantom{+}0.0$ & 1.02014 & 4.44010 & 1.02116 & 4.50468 & $ \phantom{+}1.00\times 10^{-3}$ & $ \phantom{+}1.45\times 10^{-2}$ & $ 7.008\times 10^{-3}$ & $ 3.456\times 10^{-3}$  \\
\hline
3 & $\phantom{+}0.0$ & $\phantom{+}0.0$ & 1.01714 & 4.32295 & 1.01722 & 4.40139 & $ \phantom{+}7.63\times 10^{-5}$ & $ \phantom{+}1.81\times 10^{-2}$ & $ 5.603\times 10^{-2}$ & $ 3.009\times 10^{-2}$  \\
3 & $\phantom{+}0.0$ & $\phantom{+}0.0$ & 1.01714 & 4.34794 & 1.01721 & 4.42803 & $ \phantom{+}7.16\times 10^{-5}$ & $ \phantom{+}1.84\times 10^{-2}$ & $ 1.773\times 10^{-1}$ & $ 1.227\times 10^{-2}$  \\
3 & $\phantom{+}0.0$ & $\phantom{+}0.0$ & 1.01714 & 4.37293 & 1.01677 & 4.42946 & $ -3.58\times 10^{-4}$ & $ \phantom{+}1.29\times 10^{-2}$ & $ 3.345\times 10^{-1}$ & $ 1.745\times 10^{-2}$  \\
3 & $\phantom{+}0.0$ & $\phantom{+}0.0$ & 1.01714 & 4.39792 & 1.01710 & 4.45250 & $ -3.28\times 10^{-5}$ & $ \phantom{+}1.24\times 10^{-2}$ & $ 3.874\times 10^{-1}$ & $ 1.826\times 10^{-2}$  \\
3 & $\phantom{+}0.0$ & $\phantom{+}0.0$ & 1.01714 & 4.42291 & 1.01816 & 4.51373 & $ \phantom{+}1.00\times 10^{-3}$ & $ \phantom{+}2.05\times 10^{-2}$ & $ 2.676\times 10^{-1}$ & $ 3.633\times 10^{-3}$  \\
\end{tabular}
\end{ruledtabular}
\end{center}
\end{table*}

\begin{table*}[t]
	\caption{\label{tab:runs_q1_nospin} Equal mass nonspinning configurations considered in this work. The asterisk marks simulations that have not been evolved up to the merger.
	The quantities $N_L$, $N_M$, and $\delta x_p$ are related to the grid used and are discussed in Sec.~\ref{sbsec:gra}. }
\begin{center}
\begin{ruledtabular}
\begin{tabular}{c c c | c c c | c c c |  c | c | c | c } 
$q$ & $\chi_1$ & $\chi_2$ & $\hat{E}_0$ & $p_\varphi^0$ &  $D/M$ & $N_L$ & $N_M^{\rm max}$ & $\delta x_p$ & phenom & $M_f\,[M]$ & $\hat{a}_f$ & $\chi\,[{}^\circ]$ \\
\hline
\hline

1.00 &  $\phantom{+}0.0$ &  $\phantom{+}0.0$ &  1.00100 &  3.79998 &    99.89 &  12 & 192 & $6.429\cdot10^{-3}$ & direct capt.       & $  0.9444 \pm   0.0021$ & $  0.6834 \pm   0.0075$ &    $\cdots$      \\
1.00 &  $\phantom{+}0.0$ &  $\phantom{+}0.0$ &  1.00100 &  3.89998 &    99.88 &  12 & 192 & $7.975\cdot10^{-3}$ & double enc.\phantom{*} & $  0.9519 \pm   0.0021$ & $  0.7081 \pm   0.0075$ &    $\cdots$      \\
1.00 &  $\phantom{+}0.0$ &  $\phantom{+}0.0$ &  1.00100 &  4.01998 &    99.88 &  12 & 192 & $1.213\cdot10^{-2}$ & triple enc.        & $  0.9529 \pm   0.0021$ & $  0.6732 \pm   0.0075$ &    $\cdots$      \\
1.00 &  $\phantom{+}0.0$ &  $\phantom{+}0.0$ &  1.00100 &  4.59998 &    99.87 &  12 & 192 & $7.812\cdot10^{-3}$ & unknown            & $\cdots$ & $\cdots$ &$   237.2\pm7.9$  \\
\hline
1.00 &  $\phantom{+}0.0$ &  $\phantom{+}0.0$ &  1.01103 &  3.94996 &   102.83 &  11 & 256 & $1.172\cdot10^{-2}$ & direct capt.       & $  0.9432 \pm   0.0016$ & $  0.7204 \pm   0.0060$ &    $\cdots$      \\
1.00 &  $\phantom{+}0.0$ &  $\phantom{+}0.0$ &  1.01103 &  4.01996 &   102.82 &  11 & 256 & $1.172\cdot10^{-2}$ & direct capt.       & $  0.9448 \pm   0.0015$ & $  0.6755 \pm   0.0058$ &    $\cdots$      \\
1.00 &  $\phantom{+}0.0$ &  $\phantom{+}0.0$ &  1.01103 &  4.06996 &   102.82 &  11 & 256 & $1.172\cdot10^{-2}$ & double enc.\phantom{*} & $  0.9567 \pm   0.0015$ & $  0.7063 \pm   0.0058$ &    $\cdots$      \\
1.00 &  $\phantom{+}0.0$ &  $\phantom{+}0.0$ &  1.01103 &  4.09996 &   102.82 &  11 & 256 & $1.172\cdot10^{-2}$ & double enc.\phantom{*} & $  0.9545 \pm   0.0015$ & $  0.7193 \pm   0.0058$ &    $\cdots$      \\
1.00 &  $\phantom{+}0.0$ &  $\phantom{+}0.0$ &  1.01103 &  4.12996 &   102.82 &  11 & 256 & $1.172\cdot10^{-2}$ & double enc.\phantom{*} & $  0.9516 \pm   0.0015$ & $  0.7224 \pm   0.0059$ &    $\cdots$      \\
1.00 &  $\phantom{+}0.0$ &  $\phantom{+}0.0$ &  1.01103 &  4.19996 &   102.82 &  11 & 128 & $2.344\cdot10^{-2}$ & double enc.*       & $\cdots$ & $\cdots$ &    $\cdots$      \\
1.00 &  $\phantom{+}0.0$ &  $\phantom{+}0.0$ &  1.01103 &  4.24996 &   102.82 &  11 & 256 & $1.172\cdot10^{-2}$ & scattering         & $\cdots$ & $\cdots$ &$   288.4\pm1.7$  \\
1.00 &  $\phantom{+}0.0$ &  $\phantom{+}0.0$ &  1.01103 &  4.32496 &   102.82 &  11 & 128 & $2.344\cdot10^{-2}$ & scattering         & $\cdots$ & $\cdots$ &$   254.3\pm0.8$  \\
1.00 &  $\phantom{+}0.0$ &  $\phantom{+}0.0$ &  1.01103 &  4.39996 &   102.82 &  11 & 128 & $2.344\cdot10^{-2}$ & scattering         & $\cdots$ & $\cdots$ &$   231.6\pm1.1$  \\
\hline
1.00 &  $\phantom{+}0.0$ &  $\phantom{+}0.0$ &  1.01607 &  4.07001 &   104.27 &  11 & 256 & $1.172\cdot10^{-2}$ & direct capt.       & $  0.9412 \pm   0.0021$ & $  0.7117 \pm   0.0075$ &    $\cdots$      \\
1.00 &  $\phantom{+}0.0$ &  $\phantom{+}0.0$ &  1.01607 &  4.12001 &   104.27 &  11 & 256 & $1.172\cdot10^{-2}$ & direct capt.       & $  0.9450 \pm   0.0021$ & $  0.6733 \pm   0.0075$ &    $\cdots$      \\
1.00 &  $\phantom{+}0.0$ &  $\phantom{+}0.0$ &  1.01607 &  4.18001 &   104.27 &  11 & 256 & $1.172\cdot10^{-2}$ & double enc.\phantom{*} & $  0.9568 \pm   0.0021$ & $  0.7177 \pm   0.0075$ &    $\cdots$      \\
1.00 &  $\phantom{+}0.0$ &  $\phantom{+}0.0$ &  1.01607 &  4.21002 &   104.27 &  11 & 256 & $1.172\cdot10^{-2}$ & double enc.\phantom{*} & $  0.9545 \pm   0.0021$ & $  0.7259 \pm   0.0075$ &    $\cdots$      \\
1.00 &  $\phantom{+}0.0$ &  $\phantom{+}0.0$ &  1.01607 &  4.24002 &   104.27 &  11 & 256 & $1.172\cdot10^{-2}$ & double enc.*       & $\cdots$ & $\cdots$ &    $\cdots$      \\
1.00 &  $\phantom{+}0.0$ &  $\phantom{+}0.0$ &  1.01607 &  4.30002 &   104.27 &  11 & 256 & $1.172\cdot10^{-2}$ & scattering         & $\cdots$ & $\cdots$ &$   301.5\pm1.8$  \\
1.00 &  $\phantom{+}0.0$ &  $\phantom{+}0.0$ &  1.01607 &  4.40002 &   104.27 &  11 & 128 & $2.344\cdot10^{-2}$ & scattering         & $\cdots$ & $\cdots$ &$   249.4\pm1.1$  \\
1.00 &  $\phantom{+}0.0$ &  $\phantom{+}0.0$ &  1.01607 &  4.50002 &   104.26 &  11 & 128 & $2.344\cdot10^{-2}$ & scattering         & $\cdots$ & $\cdots$ &$   219.7\pm1.2$  \\
1.00 &  $\phantom{+}0.0$ &  $\phantom{+}0.0$ &  1.01607 &  4.60002 &   104.26 &  11 & 128 & $2.344\cdot10^{-2}$ & scattering         & $\cdots$ & $\cdots$ &$   199.1\pm0.7$  \\
\hline
1.00 &  $\phantom{+}0.0$ &  $\phantom{+}0.0$ &  1.03023 &  4.30067 &   108.21 &  11 & 256 & $1.172\cdot10^{-2}$ & direct capt.       & $  0.9418 \pm   0.0021$ & $  0.7338 \pm   0.0075$ &    $\cdots$      \\
1.00 &  $\phantom{+}0.0$ &  $\phantom{+}0.0$ &  1.03023 &  4.40069 &   108.21 &  11 & 384 & $7.812\cdot10^{-3}$ & double enc.\phantom{*} & $  0.9572 \pm   0.0021$ & $  0.6774 \pm   0.0075$ &    $\cdots$      \\
1.00 &  $\phantom{+}0.0$ &  $\phantom{+}0.0$ &  1.03023 &  4.45069 &   108.21 &  11 & 256 & $1.172\cdot10^{-2}$ & double enc.\phantom{*} & $  0.9592 \pm   0.0021$ & $  0.7293 \pm   0.0075$ &    $\cdots$      \\
1.00 &  $\phantom{+}0.0$ &  $\phantom{+}0.0$ &  1.03023 &  4.50070 &   108.21 &  11 & 256 & $1.172\cdot10^{-2}$ & scattering         & $\cdots$ & $\cdots$ &$   330.4\pm1.8$  \\
1.00 &  $\phantom{+}0.0$ &  $\phantom{+}0.0$ &  1.03023 &  4.55071 &   108.21 &  11 & 128 & $2.344\cdot10^{-2}$ & scattering         & $\cdots$ & $\cdots$ &$   286.0\pm0.4$  \\
1.00 &  $\phantom{+}0.0$ &  $\phantom{+}0.0$ &  1.03023 &  4.60072 &   108.21 &  11 & 256 & $1.172\cdot10^{-2}$ & scattering         & $\cdots$ & $\cdots$ &$   258.1\pm0.8$  \\
1.00 &  $\phantom{+}0.0$ &  $\phantom{+}0.0$ &  1.03023 &  4.65073 &   108.20 &  11 & 128 & $2.344\cdot10^{-2}$ & scattering         & $\cdots$ & $\cdots$ &$   237.9\pm1.3$  \\
1.00 &  $\phantom{+}0.0$ &  $\phantom{+}0.0$ &  1.03023 &  4.70073 &   108.20 &  11 & 128 & $2.344\cdot10^{-2}$ & scattering         & $\cdots$ & $\cdots$ &$   222.0\pm1.4$  \\
1.00 &  $\phantom{+}0.0$ &  $\phantom{+}0.0$ &  1.03023 &  4.80075 &   108.20 &  11 & 128 & $2.344\cdot10^{-2}$ & scattering         & $\cdots$ & $\cdots$ &$   198.3\pm0.9$  \\
1.00 &  $\phantom{+}0.0$ &  $\phantom{+}0.0$ &  1.03023 &  4.90077 &   108.20 &  11 & 128 & $2.344\cdot10^{-2}$ & scattering         & $\cdots$ & $\cdots$ &$   180.9\pm0.3$  \\
\hline
1.00 &  $\phantom{+}0.0$ &  $\phantom{+}0.0$ &  1.04045 &  4.50199 &   110.89 &  11 & 128 & $2.344\cdot10^{-2}$ & direct capt.       & $  0.9398 \pm   0.0021$ & $  0.7248 \pm   0.0075$ &    $\cdots$      \\
1.00 &  $\phantom{+}0.0$ &  $\phantom{+}0.0$ &  1.04045 &  4.60203 &   110.88 &  11 & 128 & $2.344\cdot10^{-2}$ & double enc.\phantom{*} & $  0.9614 \pm   0.0021$ & $  0.7199 \pm   0.0075$ &    $\cdots$      \\
1.00 &  $\phantom{+}0.0$ &  $\phantom{+}0.0$ &  1.04045 &  4.70208 &   110.88 &  11 & 128 & $2.344\cdot10^{-2}$ & scattering         & $\cdots$ & $\cdots$ &$   299.8\pm0.6$  \\
1.00 &  $\phantom{+}0.0$ &  $\phantom{+}0.0$ &  1.04045 &  4.80212 &   110.88 &  11 & 128 & $2.344\cdot10^{-2}$ & scattering         & $\cdots$ & $\cdots$ &$   244.5\pm1.3$  \\
1.00 &  $\phantom{+}0.0$ &  $\phantom{+}0.0$ &  1.04045 &  4.90217 &   110.88 &  11 & 128 & $2.344\cdot10^{-2}$ & scattering         & $\cdots$ & $\cdots$ &$   212.8\pm1.5$  \\
\hline
1.00 &  $\phantom{+}0.0$ &  $\phantom{+}0.0$ &  1.05078 &  4.65445 &   113.41 &  11 & 256 & $1.172\cdot10^{-2}$ & direct capt.       & $  0.9401 \pm   0.0020$ & $  0.7377 \pm   0.0068$ &    $\cdots$      \\
1.00 &  $\phantom{+}0.0$ &  $\phantom{+}0.0$ &  1.05078 &  4.70449 &   113.41 &  11 & 256 & $1.172\cdot10^{-2}$ & direct capt.       & $  0.9384 \pm   0.0016$ & $  0.7001 \pm   0.0062$ &    $\cdots$      \\
1.00 &  $\phantom{+}0.0$ &  $\phantom{+}0.0$ &  1.05078 &  4.75454 &   113.41 &  11 & 256 & $1.172\cdot10^{-2}$ & double enc.\phantom{*} & $  0.9599 \pm   0.0015$ & $  0.6884 \pm   0.0059$ &    $\cdots$      \\
1.00 &  $\phantom{+}0.0$ &  $\phantom{+}0.0$ &  1.05078 &  4.80459 &   113.41 &  11 & 128 & $2.344\cdot10^{-2}$ & double enc.*       & $\cdots$ & $\cdots$ &    $\cdots$      \\
1.00 &  $\phantom{+}0.0$ &  $\phantom{+}0.0$ &  1.05078 &  4.85464 &   113.41 &  11 & 256 & $1.172\cdot10^{-2}$ & scattering         & $\cdots$ & $\cdots$ &$   316.2\pm0.8$  \\
1.00 &  $\phantom{+}0.0$ &  $\phantom{+}0.0$ &  1.05078 &  4.88467 &   113.41 &  11 & 128 & $2.344\cdot10^{-2}$ & scattering         & $\cdots$ & $\cdots$ &$   291.4\pm0.4$  \\
1.00 &  $\phantom{+}0.0$ &  $\phantom{+}0.0$ &  1.05078 &  4.90468 &   113.41 &  11 & 128 & $2.344\cdot10^{-2}$ & scattering         & $\cdots$ & $\cdots$ &$   278.3\pm0.3$  \\
1.00 &  $\phantom{+}0.0$ &  $\phantom{+}0.0$ &  1.05078 &  4.92470 &   113.41 &  11 & 128 & $2.344\cdot10^{-2}$ & scattering         & $\cdots$ & $\cdots$ &$   267.1\pm0.5$  \\
1.00 &  $\phantom{+}0.0$ &  $\phantom{+}0.0$ &  1.05078 &  4.94472 &   113.41 &  11 & 128 & $2.344\cdot10^{-2}$ & scattering         & $\cdots$ & $\cdots$ &$   257.3\pm0.6$  \\
1.00 &  $\phantom{+}0.0$ &  $\phantom{+}0.0$ &  1.05078 &  4.96474 &   113.41 &  11 & 128 & $2.344\cdot10^{-2}$ & scattering         & $\cdots$ & $\cdots$ &$   248.5\pm0.8$  \\
1.00 &  $\phantom{+}0.0$ &  $\phantom{+}0.0$ &  1.05078 &  4.98476 &   113.40 &  11 & 128 & $2.344\cdot10^{-2}$ & scattering         & $\cdots$ & $\cdots$ &$   240.7\pm1.0$  \\
1.00 &  $\phantom{+}0.0$ &  $\phantom{+}0.0$ &  1.05079 &  5.10488 &   113.40 &  11 & 128 & $2.344\cdot10^{-2}$ & scattering         & $\cdots$ & $\cdots$ &$   205.7\pm1.1$  \\
1.00 &  $\phantom{+}0.0$ &  $\phantom{+}0.0$ &  1.05079 &  5.20497 &   113.40 &  11 & 128 & $2.344\cdot10^{-2}$ & scattering         & $\cdots$ & $\cdots$ &$   186.1\pm0.5$  \\
1.00 &  $\phantom{+}0.0$ &  $\phantom{+}0.0$ &  1.05079 &  5.30507 &   113.40 &  11 & 128 & $2.344\cdot10^{-2}$ & scattering         & $\cdots$ & $\cdots$ &$   171.1\pm0.5$  \\

\end{tabular}
\end{ruledtabular}
\end{center}
\end{table*}

\begin{table*}[t]
	\caption{\label{tab:runs_q1_spin_q23} Analogous to Table~\ref{tab:runs_q1_nospin}, but for the nonspinning $q=\lbrace 2,3 \rbrace$ and equal mass spinning runs.}
\begin{center}
\begin{ruledtabular}
\begin{tabular}{c c c | c c c | c c c |  c | c | c | c } 
$q$ & $\chi_1$ & $\chi_2$ & $\hat{E}_0$ & $p_\varphi^0$ & $D/M$ & $N_L$ & $N_M^{\rm max}$ & $\delta x_p$ & phenom & $M_f\,[M]$ & $\hat{a}_f$ & $\chi\,[{}^\circ]$ \\
\hline
\hline
2.00 &  $\phantom{+}0.0$ &  $\phantom{+}0.0$ &  1.01609 &  4.15000 &   104.18 &  12 & 128 & $1.172\cdot10^{-2}$ & direct capt.       & $  0.9566 \pm   0.0021$ & $  0.6670 \pm   0.0075$ &    $\cdots$      \\
2.00 &  $\phantom{+}0.0$ &  $\phantom{+}0.0$ &  1.01609 &  4.20000 &   104.18 &  12 & 128 & $1.172\cdot10^{-2}$ & direct capt.       & $  0.9555 \pm   0.0021$ & $  0.6236 \pm   0.0075$ &    $\cdots$      \\
2.00 &  $\phantom{+}0.0$ &  $\phantom{+}0.0$ &  1.01609 &  4.25001 &   104.18 &  12 & 128 & $1.172\cdot10^{-2}$ & double enc.\phantom{*} & $  0.9669 \pm   0.0021$ & $  0.6590 \pm   0.0075$ &    $\cdots$      \\
2.00 &  $\phantom{+}0.0$ &  $\phantom{+}0.0$ &  1.01609 &  4.35001 &   104.18 &  12 & 128 & $1.172\cdot10^{-2}$ & scattering         & $\cdots$ & $\cdots$ &$   308.6\pm1.3$  \\
2.00 &  $\phantom{+}0.0$ &  $\phantom{+}0.0$ &  1.01609 &  4.40001 &   104.18 &  12 & 128 & $1.172\cdot10^{-2}$ & scattering         & $\cdots$ & $\cdots$ &$   274.8\pm0.3$  \\
\hline
2.00 &  $\phantom{+}0.0$ &  $\phantom{+}0.0$ &  1.02014 &  4.25010 &   105.31 &  12 & 256 & $5.859\cdot10^{-3}$ & direct capt.       & $  0.9550 \pm   0.0016$ & $  0.6609 \pm   0.0061$ &    $\cdots$      \\
2.00 &  $\phantom{+}0.0$ &  $\phantom{+}0.0$ &  1.02014 &  4.27510 &   105.31 &  12 & 256 & $5.859\cdot10^{-3}$ & direct capt.       & $  0.9539 \pm   0.0015$ & $  0.6397 \pm   0.0060$ &    $\cdots$      \\
2.00 &  $\phantom{+}0.0$ &  $\phantom{+}0.0$ &  1.02014 &  4.30010 &   105.31 &  12 & 256 & $5.859\cdot10^{-3}$ & direct capt.       & $  0.9598 \pm   0.0015$ & $  0.6079 \pm   0.0058$ &    $\cdots$      \\
2.00 &  $\phantom{+}0.0$ &  $\phantom{+}0.0$ &  1.02014 &  4.32510 &   105.31 &  12 & 256 & $5.859\cdot10^{-3}$ & double enc.\phantom{*} & $  0.9674 \pm   0.0015$ & $  0.6517 \pm   0.0058$ &    $\cdots$      \\
2.00 &  $\phantom{+}0.0$ &  $\phantom{+}0.0$ &  1.02014 &  4.35010 &   105.31 &  12 & 256 & $5.859\cdot10^{-3}$ & double enc.\phantom{*} & $  0.9674 \pm   0.0015$ & $  0.6671 \pm   0.0058$ &    $\cdots$      \\
2.00 &  $\phantom{+}0.0$ &  $\phantom{+}0.0$ &  1.02014 &  4.44010 &   105.31 &  12 & 256 & $5.859\cdot10^{-3}$ & scattering         & $\cdots$ & $\cdots$ &$   298.7\pm0.5$  \\
\hline
3.00 &  $\phantom{+}0.0$ &  $\phantom{+}0.0$ &  1.01714 &  4.32295 &   104.34 &  13 & 256 & $2.838\cdot10^{-3}$ & direct capt.       & $  0.9701 \pm   0.0016$ & $  0.5876 \pm   0.0060$ &    $\cdots$      \\
3.00 &  $\phantom{+}0.0$ &  $\phantom{+}0.0$ &  1.01714 &  4.34794 &   104.34 &  13 & 256 & $2.838\cdot10^{-3}$ & direct capt.       & $  0.9679 \pm   0.0015$ & $  0.5604 \pm   0.0058$ &    $\cdots$      \\
3.00 &  $\phantom{+}0.0$ &  $\phantom{+}0.0$ &  1.01714 &  4.37293 &   104.34 &  13 & 320 & $2.271\cdot10^{-3}$ & double enc.\phantom{*} & $  0.9763 \pm   0.0021$ & $  0.5663 \pm   0.0075$ &    $\cdots$      \\
3.00 &  $\phantom{+}0.0$ &  $\phantom{+}0.0$ &  1.01714 &  4.39792 &   104.34 &  13 & 256 & $2.838\cdot10^{-3}$ & double enc.\phantom{*} & $  0.9778 \pm   0.0021$ & $  0.5880 \pm   0.0075$ &    $\cdots$      \\
3.00 &  $\phantom{+}0.0$ &  $\phantom{+}0.0$ &  1.01714 &  4.42291 &   104.34 &  13 & 256 & $2.838\cdot10^{-3}$ & scattering         & $\cdots$ & $\cdots$ &$   379.7\pm3.0$  \\
\hline
3.00 &  $\phantom{+}0.0$ &  $\phantom{+}0.0$ &  1.02531 &  4.64033 &   106.51 &  13 & 256 & $2.777\cdot10^{-3}$ & scattering         & $\cdots$ & $\cdots$ &$   356.9\pm0.6$  \\
3.00 &  $\phantom{+}0.0$ &  $\phantom{+}0.0$ &  1.02531 &  4.66033 &   106.51 &  13 & 256 & $2.533\cdot10^{-3}$ & scattering         & $\cdots$ & $\cdots$ &$   327.8\pm0.5$  \\
\hline
1.00 &  -0.5 &  -0.5 &  1.01619 &  4.60002 &   104.26 &  11 & 256 & $1.172\cdot10^{-2}$ & direct capt.       & $  0.9564 \pm   0.0021$ & $  0.6043 \pm   0.0075$ &    $\cdots$      \\
1.00 &  -0.5 &  -0.5 &  1.01619 &  4.63002 &   104.26 &  11 & 256 & $1.172\cdot10^{-2}$ & direct capt.       & $  0.9548 \pm   0.0021$ & $  0.5820 \pm   0.0075$ &    $\cdots$      \\
1.00 &  -0.5 &  -0.5 &  1.01619 &  4.66002 &   104.26 &  11 & 256 & $1.172\cdot10^{-2}$ & direct capt.       & $  0.9581 \pm   0.0021$ & $  0.5390 \pm   0.0075$ &    $\cdots$      \\
1.00 &  -0.5 &  -0.5 &  1.01619 &  4.67002 &   104.26 &  11 & 256 & $1.172\cdot10^{-2}$ & double enc.\phantom{*} & $  0.9656 \pm   0.0021$ & $  0.5484 \pm   0.0075$ &    $\cdots$      \\
1.00 &  -0.5 &  -0.5 &  1.01619 &  4.72002 &   104.26 &  11 & 256 & $1.172\cdot10^{-2}$ & double enc.\phantom{*} & $  0.9652 \pm   0.0021$ & $  0.5982 \pm   0.0075$ &    $\cdots$      \\
1.00 &  -0.5 &  -0.5 &  1.01619 &  4.75002 &   104.26 &  11 & 256 & $1.172\cdot10^{-2}$ & scattering         & $\cdots$ & $\cdots$ &$   348.7\pm3.8$  \\
\hline
1.00 &  +0.5 &  -0.5 &  1.01619 &  4.12001 &   104.27 &  11 & 256 & $1.172\cdot10^{-2}$ & direct capt.       & $  0.9446 \pm   0.0021$ & $  0.6756 \pm   0.0075$ &    $\cdots$      \\
1.00 &  +0.5 &  -0.5 &  1.01619 &  4.30002 &   104.27 &  11 & 256 & $1.172\cdot10^{-2}$ & scattering         & $\cdots$ & $\cdots$ &$   303.6\pm2.0$  \\
\hline
1.00 &  +0.5 &  +0.5 &  1.01619 &  3.40001 &   104.28 &  11 & 256 & $1.172\cdot10^{-2}$ & direct capt.       & $  0.9188 \pm   0.0021$ & $  0.8307 \pm   0.0075$ &    $\cdots$      \\
1.00 &  +0.5 &  +0.5 &  1.01619 &  3.50001 &   104.28 &  11 & 256 & $1.172\cdot10^{-2}$ & direct capt.       & $  0.9245 \pm   0.0021$ & $  0.7884 \pm   0.0075$ &    $\cdots$      \\
1.00 &  +0.5 &  +0.5 &  1.01619 &  3.60001 &   104.28 &  11 & 256 & $1.172\cdot10^{-2}$ & double enc.\phantom{*} & $  0.9432 \pm   0.0021$ & $  0.8281 \pm   0.0075$ &    $\cdots$      \\
1.00 &  +0.5 &  +0.5 &  1.01619 &  3.70001 &   104.28 &  11 & 256 & $1.172\cdot10^{-2}$ & double enc.\phantom{*} & $  0.9293 \pm   0.0021$ & $  0.8374 \pm   0.0075$ &    $\cdots$      \\
1.00 &  +0.5 &  +0.5 &  1.01619 &  3.80001 &   104.28 &  11 & 256 & $1.172\cdot10^{-2}$ & double enc.*       & $\cdots$ & $\cdots$ &    $\cdots$      \\
1.00 &  +0.5 &  +0.5 &  1.01619 &  4.07001 &   104.27 &  11 & 256 & $1.172\cdot10^{-2}$ & scattering         & $\cdots$ & $\cdots$ &$   211.1\pm1.2$  \\
1.00 &  +0.5 &  +0.5 &  1.01619 &  4.12001 &   104.27 &  11 & 256 & $1.172\cdot10^{-2}$ & scattering         & $\cdots$ & $\cdots$ &$   202.2\pm1.1$  \\

\end{tabular}
\end{ruledtabular}
\end{center}
\end{table*}

\clearpage
\bibliography{refs20250305, refs_loc20250305}

\end{document}